\documentclass{JHEP3}
\let\ifpdf\relax
\usepackage{ifpdf}
\usepackage{multicol,bbm}
\usepackage{graphicx,psfrag}
\usepackage{dcolumn}
\usepackage{bm}
\usepackage{amsmath,amssymb, bbold}
\usepackage{color}
\let\normalcolor\relax
\newcommand{\eq}{\begin{equation}}
\newcommand{\eqe}{\end{equation}}

\newcommand{\eqa}{\begin{eqnarray}}
\newcommand{\eqae}{\end{eqnarray}}

\title{Accelerated expansion from ghost-free bigravity: a statistical analysis with improved generality}

\author{Yashar Akrami, Tomi Sebastian Koivisto and Marit Sandstad\\
       Institute of Theoretical Astrophysics, University of Oslo\\
       P.O. Box 1029 Blindern, N-0315 Oslo, Norway\\
       E-mail: \email{yashar.akrami@astro.uio.no, t.s.koivisto@astro.uio.no, marit.sandstad@astro.uio.no}}

\received{\today} %%
\accepted{} %% These are for published papers.
\abstract{We study the background cosmology of the ghost-free, bimetric theory of gravity. We perform an extensive statistical analysis of the model using both frequentist and Bayesian frameworks and employ the constraints on the expansion history of the Universe from the observations of supernovae, the cosmic microwave background and the large scale structure to estimate the model's parameters and test the goodness of the fits. We explore the parameter space of the model with nested sampling to find the best-fit chi-square, obtain the Bayesian evidence, and compute the marginalized posteriors and mean likelihoods. We mainly focus on a class of sub-models with no explicit cosmological constant (or vacuum energy) term to assess the ability of the theory to dynamically cause a late-time accelerated expansion. The model behaves as standard gravity without a cosmological constant at early times, with an emergent extra contribution to the energy density that converges to a cosmological constant in the far future. The model can in most cases yield very good fits and is in perfect agreement with the data. This is because many points in the parameter space of the model exist that give rise to time-evolution equations that are effectively very similar to those of the $\Lambda$CDM. This similarity makes the model compatible with observations as in the $\Lambda$CDM case, at least at the background level. Even though our results indicate a slightly better fit for the $\Lambda$CDM concordance model in terms of the $p$-value and evidence, none of the models is statistically preferred to the other. However, the parameters of the bigravity model are in general degenerate. A similar but perturbative analysis of the model as well as more data will be required to break the degeneracies and constrain the parameters, in case the model will still be viable compared to the $\Lambda$CDM.}

\preprint{}
\keywords{Modified Gravity, Massive Gravity, Bigravity, Dark Energy, Background Cosmology, Statistical Analysis}

\begin{document}

\section{Introduction}\label{sec:introduction}
Ever since the first proposal by Fierz and Pauli \cite{Fierz_et_Pauli1939}, giving gravity a mass has seemed like a theoretically appealing extension of gravity. However, the fact that these theories contained ghost instabilities leading to the wrong limits in systems such as the solar system \cite{Boulware_et_Deser1939} rendered them unattractive, and research endeavors into massive gravity theories lay for the most part abandoned for decades.\footnote{For a recent review of massive gravity see \cite{Hinterbichler2011}.}

Recently this has changed as new nonlinear interactions free of ghosts have been discovered, described and explored by de Rham, Gabadadze, Tolley and others \cite{deRham2009, deRham_et_Gabadadze2010a, deRham_et_Gabadadze2010b, deRham_et_al2010, deRham_Gabadadze_et_Tolley2010, deRham_et_al2011, Koyama:2011xz, Hassan_et_Rosen2011a, Hassan:2011hr, Koyama:2011yg, deRham_et_Heisenberg2011, deRham_Gabadadze_et_Tolley2011a, deRham_Gabadadze_et_Tolley2011b, deRham_Gabadadze_et_Tolley2011c, D'Amico_et_al2011, Koyama:2011wx, Berezhiani_et_al2011, Burrage_et_al2011, de_Rham_et_Renaux-Petel2012, De_Felice_Gumrukcuoglu_etMukohyama2012, D'Amico2012, Fasiello_et_Tolley2012, Langlois_et_Naruko2012, Motohashi:2012jd}. This formulation of massive gravity used a fixed extra two-tensor with no dynamics of its own \cite{Hassan_et_Rosen2011a} and does not have stable homogeneous and isotropic solutions \cite{D'Amico_et_al2011, De_Felice_Gumrukcuoglu_etMukohyama2012,Comelli_et_al2011b}.

A generalization to having two free and dynamic metrics in the formulation was then proposed and showed a theoretically viable option by Hassan and Rosen in \cite{Hassan_et_al2012a, Hassan_et_Rosen2012a, Hassan_et_Rosen2012b}. Various theoretical aspects of this interesting theory have been explored for instance in \cite{Comelli_et_al2011a,  Paulos_et_Tolley2012, Hinterbichler_et_Rosen2012, Comelli_et_al2012b, Baccetti_et_al2012a, Baccetti_et_al2012b, Baccetti_et_al2012c,Hassan:2012wr,Hassan:2012gz,Nojiri:2012zu}. In particular the cosmology has been explored in \cite{Volkov2012a} which emphasized the energy contribution of the extra field, in \cite{Comelli_et_al2011b} that described the Friedmann-Lema\^{\i}tre-Robertson-Walker (FLRW) solutions and sorted them into two branches and in \cite{Strauss_et_al2011} that looked at data constraints in some specialized corners of the parameter space. In \cite{Volkov2012c} the cosmology of the theory in the setting where the fundamental metric is homogeneous and isotropic while the other is inhomogeneous was explored. In \cite{Comelli_et_al2012a} the cosmological perturbation equations for this theory were given for the first time and these were reformulated and explored further in \cite{Khosravi_et_al2012} and \cite{Berg_et_al2012}. Using an alternative approach, it was recently claimed that the cosmological solutions exhibit an instability for some parameter combinations \cite{Kuhnel:2012gh}. Such instabilities were however not observed\footnote{This is possibly because the ``backward instabilities" reported in \cite{Kuhnel:2012gh} correspond to decaying modes. In any case, in our comprehensive background analysis, we found no instabilities, which strongly suggests that at least the homogeneous modes are stable in these cosmologies.} in the studies \cite{Khosravi_et_al2012, Berg_et_al2012} that used standard cosmological perturbation theory.

Although the background cosmology of this theory with two homogeneous and isotropic metrics has been explored in \cite{Volkov2012a, Comelli_et_al2011b, Strauss_et_al2011}, none of them has given an exhaustive parameter estimation and exploration of the full parameter space in comparison with data in order to find whether this theory has new dynamics yielding better or comparable fits to those of the standard model of cosmology ($\Lambda$ Cold Dark Matter; $\Lambda$CDM; for an introduction, see e.g. \cite{Weinberg:2008,Mukhanov:2005}). This is what this paper endeavors to do. It is quite interesting to study precisely how viable these models are, as a finite mass of a graviton would provide a well-motivated and theoretically consistent explanation for the enigmatic speed-up of the late-time expansion of the Universe (for a review, see e.g. \cite{Frieman:2008sn}). 
In particular, were the data to prefer the higher order interactions of graviton that give it a mass over the constant term, one could also hope to shed new light on the long-standing cosmological constant problem \cite{Martin:2012bt}.  
For a large collection of other possible dynamical models of dark energy and modified gravity that have been proposed in attempts to address the issue of cosmic acceleration, see e.g. the reviews \cite{Copeland:2006wr, Clifton:2011jh, Amendola:2012ys}.

The paper is organized as follows; In section \ref{sec:TheoreticalIntro} we give a brief theoretical introduction to bigravity theory and in particular its formulation in two homogeneous and isotropic metrics. In section \ref{sec:NumericalSetup} we describe the observational data that we used to compare with predictions of the model in different parts of parameter space. We also present a handy reformulation that we used for the numerical integration of the equation set and how we performed this integration. In addition we describe the statistical methods we used to perform the parameter estimation. In section \ref{sec:Results} we describe the results of our analysis and parameter estimation when a method for numerical integration of the equation set was established. To better understand the dynamics of the solutions we first present the analysis of certain specific parts of the parameter space, and in this way we demonstrate how the system becomes degenerate, how good fits to the data can be obtained, but also how the search for a best-fit model reveals that a theory which mimics the $\Lambda$CDM as closely as possible is preferred, and the closeness\footnote{Which is in fact arbitrarily good in more than one corner of parameter space.} to the $\Lambda$CDM that is obtainable in these theories is just what makes them such good fits to the data.

\section{The model}\label{sec:TheoreticalIntro}
\subsection{Bimetric theory of massive gravity}\label{sec:Bigravity}
In this section, we briefly review the ghost-free, bimetric theory of gravity without restricting the framework to any particular choices of metrics. We start from the action level and present the modified field equations for the dynamical variables of the theory.

The theory contains two space-time metrics. The first one, which we denote by $g_{\mu\nu}$, is assumed to be the physical metric, namely the metric based on which all usual distances in cosmology are defined. The other metric, denoted by $f_{\mu\nu}$, is a dynamical rank-two tensor that is essential for the theory to be ghost-free; this metric when coupled to the physical metric, gives mass to gravitons\footnote{For other recent
developments in bimetric and biconnected spacetimes, see \cite{Hossenfelder:2008bg,Koivisto:2011vq,Tamanini:2012mi,Westman:2012xk,BeltranJimenez:2012sz}.}.

The action for our ghost-free, massive theory of gravity was first presented in \cite{Hassan_et_Rosen2012a}. We will follow the notation of \cite{Strauss_et_al2011} everywhere in the paper, and most of this introductory section (and the next section) will follow that paper quite closely. Let us begin with the action which has the following form:

\begin{eqnarray}
  S & = & - \frac{M^2_g}{2}\int d^4x\sqrt{-\det g}R(g) - \frac{M^2_f}{2}\int
d^4x\sqrt{-\det f}R(f) \nonumber\\
  && + m^2M_g^2\int d^4x\sqrt{-\det
g}\sum_{n=0}^{4}\beta_ne_n\left(\sqrt{g^{-1}f}\right) + \int d^4x\sqrt{-\det
g}\mathcal{L}_m\left(g, \Phi\right)\label{eq:ActionOriginal} .
\end{eqnarray}

Here, the matrix $\sqrt{g^{-1}f}$ is defined such that $\sqrt{g^{-1}f}\sqrt{g^{-1}f}=g^{\mu\lambda}f_{\mu\lambda}$, $R(g)$ and $R(f)$ are the Ricci scalars for the metrics $g$ and $f$, respectively, $M_g$ and $M_f$ denote the Planck masses corresponding to the two metrics, and $\mathcal{L}_m\left(g,\Phi\right)$ shows the Lagrangian for the matter sector. In addition, $e_n(\mathbb{X})$ are elementary symmetric polynomials of the eigenvalues of the matrix $\mathbb{X}$ and have the following forms:

\begin{eqnarray}
  &e_0\left(\mathbb{X}\right)= 1, \qquad e_1\left(\mathbb{X}\right)=
\left[\mathbb{X}\right] \qquad, & e_2\left(\mathbb{X}\right)=
\frac{1}{2}\left(\left[\mathbb{X}\right]^2 - \left[\mathbb{X}^2\right]\right),
\nonumber \\
  &e_3\left(\mathbb{X}\right)= \frac{1}{6}\left(\left[\mathbb{X}\right]^ 3 -
3\left[\mathbb{X}\right]\left[\mathbb{X}^2\right] +
2\left[\mathbb{X}^3\right]\right) \qquad& e_4\left(\mathbb{X}\right)=
\det\left(\mathbb{X}\right),
\end{eqnarray}
where square brackets denote the traces of the matrices $\mathbb{X}$. The small $m$ in eq. (\ref{eq:ActionOriginal}) is the graviton mass and the five quantities $\beta_n$ ($n=0,...,4$) are free parameters that need to be determined observationally\footnote{\label{alphanote}It is important to note that there are other ways of parameterizing the model, in particular one that is used relatively widely in the literature on massive gravity in terms of parameters $\alpha_n$. In this case the action is still written in terms of $e_n$ but as functions of $\mathbb{K}=\sqrt{g^{-1}f}-1$ (see e.g. \cite{Hassan_et_Rosen2011a} for expressions that relate $\alpha_n$ and $\beta_n$). The potentials corresponding to the $e_n$-terms ($n=1,...,4$) however contain constant terms that contribute to the cosmological constant of the model, meaning that $\alpha_0$ does not capture all the cosmological terms. Since one of the main goals of the present paper is to investigate whether the bigravity theory can explain the late-time acceleration of the Universe in absence of an explicit cosmological constant (i.e. when it is set to zero) we adhere to the $\beta$-parameterization where at least the vacuum energy contribution to the cosmological constant that corresponds to the physical metric $g$ is represented by $\beta_0$. This is the quantity we will set to zero in most of the analyses of this paper. Also see the discussion of footnote \ref{CCnote}.}. 

The equations of motion for the two metrics (or the equivalents to the Einstein equations) can be derived by varying the action with respect to the metrics; this gives:

\begin{eqnarray}
  R_{\mu\nu} - \frac{1}{2}g_{\mu\nu} R +
\frac{m^2}{2}\sum_{n=0}^3\left(-1\right)^{n}\beta_n\left[g_{\mu\lambda}Y^{\lambda}_{(n)\nu}\left(\sqrt{g^{-1}
f}\right) + g_{\nu\lambda}Y^{\lambda}_{(n)\mu}\left(\sqrt{g^{-1} f}\right)\right]=
\frac{1}{M_g^2}T_{\mu\nu},\nonumber\\
\label{eq:EoM4g} \\
  \bar{R}_{\mu\nu} - \frac{1}{2}g_{\mu\nu} \bar{R} +
\frac{m^2}{2M^2_\star}\sum_{n=0}^3\left(-1\right)^{n}\beta_{4 -
n}\left[f_{\mu\lambda}Y^{\lambda}_{(n)\nu}\left(\sqrt{g^{-1} f}\right) +
f_{\nu\lambda}Y^{\lambda}_{(n)\mu}\left(\sqrt{g^{-1} f}\right)\right]= 0,\nonumber\\ \label{eq:EoM4f}
\end{eqnarray}
where over-bar denotes quantities corresponding to the $f$ metric, $T_{\mu\nu}$ is the stress-energy-momentum tensor, $M_\star^2 \equiv M_f^2/M_g^2$,  and
\begin{eqnarray}
  Y_{(0)}\left(\mathbb{X}\right) = \mathbb{1}, \quad Y_{(1)}\left(\mathbb{X}\right)=
\mathbb{X} - \mathbb{1}\left[\mathbb{X}\right],\quad
Y_{(2)}\left(\mathbb{X}\right) = \mathbb{X}^2 - \mathbb{X}\left[\mathbb{X}\right]
+ \frac{1}{2}\mathbb{1}\left(\left[\mathbb{X}\right]^2 -
\left[\mathbb{X}^2\right]\right),\nonumber \\  
  Y_{(3)}\left(\mathbb{X}\right) = \mathbb{X}^3 -
\mathbb{X}^2\left[\mathbb{X}\right] +
\frac{1}{2}\mathbb{X}\left(\left[\mathbb{X}\right]^2 -
\left[\mathbb{X}^2\right]\right) -
\frac{1}{6}\mathbb{1}\left(\left[\mathbb{X}\right]^ 3 -
3\left[\mathbb{X}\right]\left[\mathbb{X}^2\right] +
2\left[\mathbb{X}^3\right]\right).
\end{eqnarray}
As in standard gravity, these equations determine the dynamics of the space-time degrees of freedom, i.e. the geometry and time-evolution of the Universe, if the properties of matter and energy are known (through the tensor $T_{\mu\nu}$).

As observed in \cite{Berg_et_al2012}, we can perform the constant rescaling $f_{\mu\nu} \to \frac{M_g^2}{M_f^2}f_{\mu\nu}$ and $\beta_i \to \left(\frac{M_f}{M_g}\right)^i\beta_i$ in order to set $M_{\star}^2$ to unity. This quantity is therefore not a free parameter of the theory; we will drop it in the rest of the paper. We also assume $M_g$ to be the usual (reduced) Plank mass $M_{Pl}$.

In addition to the equations of motion, further constraints are imposed on the dynamics of the two metrics by Bianchi identities and the assumption that the stress-energy-momentum tensor of the matter components is conserved:

\begin{eqnarray}
 \bigtriangledown^\mu \sum_{n=0}^3\left(-1\right)^{n}\beta_n\left[g_{\mu\lambda}Y^{\lambda}_{(n)\nu}\left(\sqrt{g^{-1}
f}\right) + g_{\nu\lambda}Y^{\lambda}_{(n)\mu}\left(\sqrt{g^{-1} f}\right)\right]&=&0,\label{eq:Bianchi_g} \\
  \bar{\bigtriangledown}^\mu \sum_{n=0}^3\left(-1\right)^{n}\beta_{4 -
n}\left[f_{\mu\lambda}Y^{\lambda}_{(n)\nu}\left(\sqrt{g^{-1} f}\right) +
f_{\nu\lambda}Y^{\lambda}_{(n)\mu}\left(\sqrt{g^{-1} f}\right)\right]&=&0.\label{eq:Bianchi_f}
\end{eqnarray}

As we will see in the next section, this gives us an extra piece of information which dramatically simplifies the evolution equations when applied to the entire Universe.

\subsection{The case for an isotropic and homogeneous universe}\label{sec:FRW}

In the previous section, we described the full theory of ghost-free bimetric gravity where we did not assume any particular forms for the metrics. The purpose of the present paper is however to compare the cosmological predictions of the model to real data. A fundamental assumption in the standard concordance model of cosmology, the $\Lambda$CDM, is that the Universe on large scales is spatially isotropic and homogeneous. This assumption restricts the metric of the Universe to be of the FLRW form. Based on the same observational and theoretical reasons, we follow \cite{Strauss_et_al2011} and assume both of the two metrics $g$ and $f$ in our bigravity model exhibit spatial isotropy and homogeneity. As we will see, this assumption leads to a set of generalized Friedmann equations that we will use in comparing the background dynamics of the Universe, predicted by the model, to different types of cosmological measurements. 

The spatially isotropic and homogeneous metrics $g$ and $f$ read

\begin{eqnarray}
  ds^2_g &=& - dt^2 + a^2\left(\frac{dr^2}{1 - kr^2} + r^2\left(d\theta^2 + \sin^2\theta d\phi^2\right)\right),\nonumber\\
  ds^2_f &=& - X^2dt^2 + Y^2\left(\frac{dr^2}{1 - kr^2} + r^2\left(d\theta^2 + \sin^2\theta d\phi^2\right)\right),
\end{eqnarray}
where $a(t)$ and $Y(t)$ are the time-dependent spatial scale factors corresponding to the metrics $g$ and $f$, respectively. $X(t)$ is a time-dependent function that must be determined through the evolution equations given by the model. We have additionally assumed the same spatial curvature for the two metrics ($k=-1,0,+1$) \cite{Comelli_et_al2011b}.

In order to have consistent solutions in this case, where the matter sources are time-dependent, eqs. (\ref{eq:Bianchi_g}) and (\ref{eq:Bianchi_f}) require the following relationship between the functions $X(t)$, $a(t)$ and $Y(t)$ \cite{Strauss_et_al2011}:

\begin{equation}
  X = \frac{\dot{Y}}{\dot{a}} = \frac{dY}{da} ,
\end{equation}
where over-dot denotes derivative with respect to time. As we mentioned earlier, this relation is necessary for the conservation of the stress-energy-momentum tensor through the Bianchi identity of $g$.

As in the $\Lambda$CDM case, we assume that the Universe is filled with dust, with the energy density $\rho_m$, and radiation, with the energy density $\rho_\gamma$ (other components can be added easily). The equations of motion (\ref{eq:EoM4g}) and (\ref{eq:EoM4f}) then give the following generalized Friedmann equations \footnote{Note that eq. (\ref{eq:spacialFieldEqg}) is slightly different from  the equivalent equation in \cite{Strauss_et_al2011}; this could be a misprint in that paper which has been corrected here.}:
\begin{equation}\label{eq:firstFriedmanng}
  3\left(\frac{\dot{a}}{a}\right)^2  + 3\frac{k}{a^2} - m^2\left[\beta_0 +
3\beta_1\frac{Y}{a} + 3\beta_2\frac{Y^2}{a^2} + \beta_3\frac{Y^3}{a^3}\right]= 
\frac{1}{M_g^2}\left(\rho_m + \rho_\gamma\right),
\end{equation}
\begin{equation}\label{eq:spacialFieldEqg}
  -2\frac{\ddot{a}}{a}-\left(\frac{\dot{a}}{a}\right)^2 - \frac{k}{a^2}+
m^2\left[\beta_0 + \beta_1\left(2\frac{Y}{a} + \frac{\dot{Y}}{\dot{a}}\right) +
\beta_2\left(\frac{Y^2}{a^2} + 2\frac{Y\dot{Y}}{a\dot{a}}\right) + \beta_3
\frac{Y^2\dot{Y}}{a^2\dot{a}}\right] = \frac{1}{3M_g^2}\rho_\gamma,
  \end{equation}
  \begin{equation}\label{eq:firstFriedmannf}
  3\left(\frac{\dot{a}}{Y}\right)^2 + 3\frac{k}{Y^2} -
m^2\left[\beta_4 + 3\beta_3\frac{a}{Y} +
  3\beta_2\frac{a^2}{Y^2} + \beta_1\frac{a^3}{Y^3}\right] = 0,
  \end{equation}
  \begin{equation}\label{eq:spacialFieldEqf}
    m^2\left[\beta_4 + \beta_3\left(2\frac{a}{Y} +
\frac{\dot{a}}{\dot{Y}}\right) + \beta_2\left(\frac{a^2}{Y^2} +
2\frac{a\dot{a}}{Y\dot{Y}}\right) + \beta_1
\frac{a^2\dot{a}}{Y^2\dot{Y}}\right] -2\frac{\ddot{a}\dot{a}}{Y\dot{Y}}-\left(\frac{\dot{a}}{Y}\right)^2-\frac{k}{Y^2} = 0.
  \end{equation}

\section{Numerical investigation of the parameter space}\label{sec:NumericalSetup}

This section is about our strategy and methods we use to compare the predictions of the bigravity model to a set of cosmological observations, assess how well they match and then constrain the parameters of the model. We first, in section \ref{sec:constraints}, introduce the dataset we use and the cosmological quantities that have to be computed theoretically in order to make the comparison with observations. Section \ref{sec:evolution} describes all the simplified dynamical equations for the bigravity theory that we employ in our numerical investigation. It also shows more clearly what dynamical variables play the most important roles in the evolution equations. This in addition helps us to understand the physics of the model in comparison to the standard model. We then continue our description of the model in section \ref{sec:initialconditions} with a discussion about the initial conditions we need to set for the evolution equations in our numerical implementation of the model. Finally, in section \ref{sec:statistics} we focus on the statistical aspects of our work and review different statistical frameworks we employ, as well as our strategy for exploring the parameter space of the model. This clarifies how we aim to compare the model's predictions to the real data in practice. Readers who are familiar with or not interested in such statistical and scanning techniques can skip section \ref{sec:statistics} and continue with our results and discussions in section \ref{sec:Results}.

\subsection{Constraints from cosmology}\label{sec:constraints}

Cosmological data that are used in comparing predictions of cosmological models to observations are classified into two main categories: 1) constraints from measuring the geometry and background evolution of the Universe on large scales (expansion history), and 2) constraints from the formation, distribution and evolution of structures in the Universe (growth history). In order to see whether a model is viable observationally, one usually starts with the background cosmology. This is also simpler to study since the background equations are considerably simpler to derive and implement numerically. Studying the structures requires the field equations to be perturbed around the background (FLRW metrics in our case). Here, we only work with the background dynamics and leave the investigation of the model at the perturbation level for future work.

Three main sources of information in cosmology are 1) anisotropies of the Cosmic Microwave Background (CMB) Radiation, 2) Baryon Acoustic Oscillations (BAO), and 3) Type Ia Supernovae (SNe Ia). At the background level, observational constraints on a theoretical model provided by these sources all involve calculations of different types of cosmological distances. In general, such distances depend on the parameters of the model and by comparing them to the measured distances one can constrain the model and determine how successful the model is in describing the Universe. In what follows, we briefly review the distance measures important in CMB, BAO and SNe Ia observations, how they are calculated from a cosmological model and how they are related to the actual cosmological data.
\\

\noindent
$\bullet$ \textbf{Cosmic Microwave Background:} In order to properly extract the information from the tiny fluctuations on the CMB one usually looks at the angular distribution of the fluctuations through the computation of the angular power spectrum. One can on the other hand derive the theoretical power spectrum by solving some Boltzmann codes numerically and fit the model to the data by comparing the two spectra. The latter needs perturbative equations to be solved. There is however one important quantity that can be measured from the observed CMB power spectrum and only depends on the background equations: the position of the first peak on the spectrum (denoted by $l_A$). This represents the angular scale of the sound horizon at the recombination epoch. Since we are working only with the background equations in this paper, we adhere to this quantity to place constraints on our model. One can show that

\begin{equation}
  l_A=\pi \frac{(1+z_r)D_A(z_r)}{r_s(z_r)}, 
\end{equation}
where $D_A(z_r)$ is the angular-diameter distance to the CMB last-scattering surface, i.e. at the redshift of recombination $z_r$. $r_s(z_r)$ is the co-moving sound horizon at $z_r$. Theoretically, $D_A$ and $r_s$ as functions of redshift $z$ can be calculated from the following expressions (we assume a flat universe, i.e. $k=0$):

\begin{eqnarray}
  D_A(z)&=&\frac{1}{(1+z)}\int_0^z c \frac{dz'}{H(z')},\label{eq:D_A}\\
  r_s(z)&=&\int_z^\infty c_s \frac{dz'}{H(z')}.\label{eq:r_s}
\end{eqnarray}

Here, $c$ is the speed of light, $c_s$ is the speed of sound before recombination and $H(z)$ is the Hubble parameter at a given redshift $z$. The latest measurements of the CMB power spectrum by the satellite Wilkinson Microwave Anisotropy Probe (WMAP) \cite{Komatsu:2010fb} has determined the value of $l_A$ to be $302.56\pm0.78$. In addition, the value we assume for $z_r$ is $1091.12$; we neglect the uncertainties about this value \cite{Strauss_et_al2011}.
\\

\noindent
$\bullet$ \textbf{Baryon Acoustic Oscillations:} As in the CMB case, in order to extract full information from the distribution and evolution of large-scale structures in the Universe, one needs to look at the perturbative equations. This involves fitting the theoretical power spectrum of matter to the observed one. There are however simpler ways to work only at the background level. One such possibility is to consider the ratio of the sound horizon at the so-called drag epoch (with the redshift of $z_d$) to another quantity called dilation scale (denoted as $D_V$) at some arbitrary redshifts. $z_d$ is the redshift of an epoch at which the baryon acoustic oscillations are frozen in ($z_d\approx 1020$). The theoretical value of the dilation scale $D_V$ at any redshift $z$ can be obtained from the expression

\begin{equation}
D_V(z)=[\frac{cz(1+z)^2D_A(z)^2}{H(z)}]^{1/3}.\label{eq:D_V}
\end{equation}

In the present analysis, we use the values of this ratio when $D_V$ is measured at redshifts $0.106$, $0.2$, $0.35$, $0.44$, $0.6$ and $0.73$. These measurements have been done by the galaxy surveys 2dFGRS, 6dFGS, SDSS and WiggleZ (\cite{Beutler:2011hx}, \cite{Percival:2009xn}, \cite{Blake:2011en}) and are given in table \ref{tbl:rs2DVdA2DV}.

  \begin{table}[t]
    \begin{center}
      \footnotesize{
      \begin {tabular}{|c|c|c|c|c|c|c|}
        \hline
         $z=0.106$& $z=0.2$ & $z=0.35$ & $z=0.44$ & $z=0.6$ & $z=0.73$\\
	\hline\hline
          \multicolumn{6}{|c|}{$r_s(z_d)/D_V(z):$} \\
	\hline
	$0.336\pm 0.015$& $0.1905\pm 0.0061$& $0.1097\pm 0.0036$ & $0.0916\pm 0.071$ & $0.0726\pm 0.0034$ & $0.0592\pm 0.0032$\\
	\hline\hline
          \multicolumn{6}{|c|}{$(1+z_r)D_A(z_r)/D_V(z):$} \\
	\hline	
	$30.96\pm 1.48$& $17.55\pm 0.64$& $10.11\pm 0.37$ & $8.44\pm 0.67$ & $6.69\pm 0.33$ & $4.46\pm 0.31$\\
	\hline	
      \end{tabular}}
      \caption{Measured values of $\frac{r_s(z_d)}{D_V(z)}$ and $(1+z_r)\frac{D_A(z_r)}{D_V(z)}$ at different redshifts (\cite{Beutler:2011hx}, \cite{Percival:2009xn}, \cite{Blake:2011en}).}
      \label{tbl:rs2DVdA2DV}
    \end{center}
  \end{table}

In applying the CMB and BAO measurements mentioned above to our model, we follow the strategy used in \cite{Strauss_et_al2011} where the BAO measurements of $r_s(z_d)/D_V(z)$ at different redshifts are multiplied by the CMB measurement of $l_A$ in order to reduce the model dependence of the constraints. Assuming the measured value of $1.0451\pm0.0158$ for the ratio $r_s(z_d)/r_s(z_r)$ reported by WMAP collaboration, the combined constraints (from CMB and BAO) used in our analysis are the ones we have given in table \ref{tbl:rs2DVdA2DV}.
\\

\noindent
$\bullet$ \textbf{Type Ia Supernovae:} Luminosity measurements of SNe Ia have proven to be a powerful source of information about the geometry and evolution of the Universe at late times. After the striking discovery of the accelerated expansion in 1998 \cite{Riess:1998cb,Perlmutter:1998np}, they have received much attention as standard candles in observational cosmology. The key quantity in using SNe Ia in cosmological model analysis is the luminosity distance $D_L(z)$ which can be computed theoretically for a model and constrained directly from the SNe observations:

\begin{eqnarray}
  D_L(z)&=&(1+z)\int_0^z c \frac{dz'}{H(z')}.\label{eq:D_L}
\end{eqnarray}

The dataset we use in our analysis is the Union2.1 compilation of SNe \cite{Suzuki:2011hu} which contains $580$ SNe in total. We include the reported systematic uncertainties for the measurements given in that paper.
\\

\noindent
$\bullet$ \textbf{Present value of Hubble parameter ($H_0$):} $H_0$ is a quantity that appears in all expressions used in comparing predictions of a theoretical model to the real data. It is therefore important to know what value it has in reality. We use the value provided by seven-year WMAP observations, i.e. $H_0=71\pm2.5$ km s$^{-1}$ Mpc$^{-1}$. In $\Lambda$CDM cosmology, $H_0$ is a free parameter of the model that is to be constrained observationally. As we will see in the next section, in our bigravity model, it is a prediction of the model when other parameters are fixed.

\subsection{Hubble parameter and evolution equation}\label{sec:evolution}

We saw in the previous section (eqs. (\ref{eq:D_A}), (\ref{eq:r_s}), (\ref{eq:D_V}) and (\ref{eq:D_L})) that all quantities used in constraining a model through cosmological observations can be calculated theoretically only if the Hubble parameter $H$ is known during the evolution of the Universe. For the standard $\Lambda$CDM model, this can be calculated in terms of the usual cosmological parameters such as various density parameters. In our bimetric theory, we have assumed that one metric (we chose it to be $g$) is the physical one and this is the metric that should be used in defining different cosmological quantities including distances. The metric is additionally assumed to be of FLRW form with a time-dependent scale factor ($a$) as the only dynamical quantity which determines the evolution of the Universe. This all means that we can follow the standard recipe and use the Hubble parameter defined based on that scale factor in our calculations of observable quantities (i.e. $H = \dot{a}/a$). Our next task is therefore to find a time evolution equation for the Hubble parameter.

As usual, we define the redshift $z$ as $z = a_0/a-1$, where $a_0$ is the scale factor at present time (which is chosen to be unity). This gives the possibility of calculating $H$ as a function of $z$ and using it directly in the expressions for cosmological distances. In addition, it turns out that working with the variable $y = Y/a$ (where $Y$ is the scale factor corresponding to the metric $f$) significantly simplifies the equations used in numerical calculations. In fact, as we will demonstrate below, all the interesting equations and dynamical variables (including the Hubble parameter) can be written in terms of $y$ (which we expect to be a time-dependent quantity). For numerical integration, as will be discussed below, it is additionally useful to reformulate the equations in terms of the e-folding time $x = \ln a$. This transforms the time derivative $\frac{d}{dt}$ into the e-folding time derivative: $\frac{d}{dt} = \frac{d\ln a}{d t}\frac{d}{dx} = H \frac{d}{dx}$, which we will denote by $\frac{d}{dx} \equiv~'$. 

Before giving the expression for $H$ in terms of $y$ and parameters of our model, we perform one more redefinition, this time on the $\beta_i$ parameters. It is obvious already from the action (eg. (\ref{eq:ActionOriginal})) that the parameter $m^2$ is redundant since it just multiplies all the $\beta$s. We can therefore redefine $\beta$s such that they absorb $m^2$. However, since $m$ is presumably quite small (something of the order of the present value of the Hubble parameter, $H_0$), we further normalize the $\beta$ parameters to the observed value of $H_0$ (i.e. $71\pm2.5$ km s$^{-1}$ Mpc$^{-1}$). This gives us a set of new parameters that we call $B$s: $B_i \equiv m^2\beta_i/H_0^2$. Note that the exact value of $H_0$ is not important here and we use it only for simplicity reasons, i.e to work with a set of parameters that are likely to have values of not more than a few orders of magnitude larger or smaller than unity. As we will see later, the present value of the Hubble parameter is not a free parameter of the theory and will be predicted by the model in terms of the other parameters.

After including all the aforementioned modifications into eq. (\ref{eq:firstFriedmanng}), we get an equation for the Hubble parameter as follows \footnote{Here, and in the rest of this section, we include the curvature term in the equations for a general $k$-value. As we will mention later, in the present paper we however analyze the bigravity model only for the $k=0$ case (flat Universe); this value has been chosen for simplicity reasons and is also a justified assumption based on the current constraints on the curvature of the Universe from analyses of the $\Lambda$CDM concordance model. Including the curvature term is however a very straightforward procedure and we leave it for future work.}:

\begin{equation}\label{eq:firstFriedmanngefolding}
  \frac{H^2}{H_0^2}  = -\frac{k}{H_0^2\exp(2x)} + \frac{B_0}{3} + B_1y  + B_2y^2  + \frac{B_3}{3}y^3 + \Omega_m + \Omega_\gamma,
\end{equation}
where we have defined the density parameters as usual, i.e. $\Omega_i = \rho_i/(3H_0^2M_g^2)$ ($\rho_i$ being the energy density for the component $i$).

It can be seen from this expression that in order to compute the Hubble parameter at any given time during the cosmic evolution, one needs to know the values of the parameters $B_0, ..., B_3$, as well as the dynamical quantities $\Omega_m$, $\Omega_\gamma$ and $y$. Matter and radiation follow the standard evolutions with redshift according to their traditional equations of state, i.e. $\Omega_m=\Omega_m^0(1+z)^3$ and $\Omega_\gamma=\Omega_\gamma^0(1+z)^4$. We have not included an explicit cosmological constant density parameter $\Omega_\Lambda$ in the equation because the parameter $\beta_0$, and correspondingly $B_0$, act as a cosmological constant; including the quantity $\Omega_\Lambda$ is therefore redundant\footnote{\label{CCnote}Even though it is $\beta_0$ that we will call the (explicit) cosmological constant throughout the paper, we should note that in general it may not capture all contributions to the cosmological constant. Strictly speaking, it represents the vacuum energy contribution which is arguably the most interesting type of the cosmological constant. In bimetric theory (the ghost-free massive bigravity that we study in the present paper) the notion of vacuum energy (that receives contributions from the $g$ and $f$ matter loops) is well defined and is given by $\beta_0$ and $\beta_4$ (or correspondingly $B_0$ and $B_4$). On the other hand, the notion of an ``intrinsic cosmological constant" is not as well defined. In General Relativity, ``cosmological constant" and ``vacuum energy" are the same, but not in bimetric theory. Vacuum energy always appears in the form $\Lambda \sqrt{-\det g}$ in the action and this is the quantity that receives large corrections from quantum loops of heavy particles. In the bimetric theory $\beta_0$ measures vacuum energy (same of $\beta_4$ for the $f$ sector). One could also call this an ``explicit" cosmological constant. But the total observed cosmological constant is to be read off from Einstein's equations. In the bimetric setup whenever the equations contain a quantity that is cleanly identifiable as a cosmological constant, that will be a combination of the $\beta$'s. In the specific bigravity model that we are interested in here, i.e. with $g$ and $f$ being FLRW (or FLRW-like) metrics, the total observed cosmological constant must be identified from the Einstein's equations in this case, i.e. the modified Friedmann equation (\ref{eq:firstFriedmanngefolding}). It seems from this equation that the $B_0$-term (or equivalently $\beta_0$-term) is the only term that appears as the cosmological constant for the physical metric $g$ and all other terms (that involve other $\beta$'s) are functions of time since $y$ is not a constant in general. As we will see later, our numerical analysis shows that $y$ is not a constant in order for the model to fit the data. We therefore call $\beta_0$ ($B_0$) the cosmological constant throughout the paper. Even if there is still a hidden cosmological constant lurking within the other $\beta$-terms, we have at least set the vacuum energy contribution to zero by setting $\beta_0$ to zero. An interesting example of bigravity models for which a combination of all $\beta$'s contribute to the cosmological constant is the case of the backgrounds of the type $g=f$ (which are valid backgrounds only when the parameters satisfy some specific constraint; a property that is not satisfied in our case of cosmologically interesting backgrounds). In this case the parameter $\alpha_0$ ($\alpha_0=\beta_0+4\beta_1+6\beta_2+4\beta_3+\beta_4$) corresponds to the cosmological constant (see footnote \ref{alphanote}). The only way of cleanly identifying a cosmological constant is to consider $f=c^2 g$ type backgrounds. Then one obtains the most general expression for the cosmological constant in these models without constraining the parameters in any way (as in \cite{Hassan:2012wr} or \cite{Hassan:2012rq} where the details are given). This gives not $\alpha_0$, but what is called $\alpha^c_0$ in those references, as the cosmological constant. Clearly, even if we set $\alpha_0=0$, we find that $\alpha^c_0$ does not vanish unless we set $c=1$. One may also think of this in the following way: in massive gravity (not bimetric theory) if one decides on $f=\eta$, then $\alpha_0$ becomes an intrinsic cosmological constant that must vanish for consistency. But as soon as $f$ is made dynamical, this privilege is lost.}.

As the next step in solving eq. (\ref{eq:firstFriedmanngefolding}) for $H(z)$, we need to find what the values of $y$ are at different times (or redshifts). Using eqs. (\ref{eq:firstFriedmanng}) - (\ref{eq:spacialFieldEqf}) we obtain the following equation for $y'$ (derivative of the $f$-metric's normalized scale factor $y$ with respect to $x$):

\begin{eqnarray}\label{eq:diffEqFory}
  y' &=& \frac{ y \left[-2 B_0 + B_3 y^3 + 3\Omega_{m}
 + 6\Omega_\gamma + 6 B_2+3 B_3 y\right] + 3B_1 \left(1-y^2\right)}{B_0 +3\Omega_m + 3\Omega_{\gamma} + 6 B_1  y + 3B_2 \left(3 y^2 - 1\right)+ 2B_3y\left(2y^2-3\right) - 3 B_4 y^2} - y.
\end{eqnarray}

This shows that the dynamical variable $y$ has a one-dimensional phase space, meaning that knowing its value at any given time is sufficient for obtaining its values at all other times; as we will see later, we use this differential equation to determine the value of $y$ (and accordingly $H$) at different redshifts. We will discuss in the next section how we can obtain the value of $y$ at a particular redshift to set an initial value for eq. (\ref{eq:diffEqFory}).

Before continuing with setting the initial conditions, let us take a closer look at the equations of motion (\ref{eq:firstFriedmanng}) - (\ref{eq:spacialFieldEqf}) to see whether there is any alternative approach in calculating the Hubble parameter as a function of redshift. The answer ``seems" to be ``yes": the equation set yields another equation for the Hubble parameter (if at least one of the parameters $B_1$, $B_2$, $B_3$ or $B_4$ is nonzero):
\begin{equation}\label{eq:AlgebraicH}
  \frac{H^2}{H_0^2}=-\frac{k\exp(-2x)}{H_0^2}+\frac{B_4}{3}y^2+B_3y+B_2+\frac{B_1}{3}y^{-1}.
\end{equation}

Eqs. (\ref{eq:firstFriedmanngefolding}) and (\ref{eq:AlgebraicH}) are enough to set the values of the system at all times. We can see this by solving eq. (\ref{eq:AlgebraicH}) for $H^2$ and plugging it into eq. (\ref{eq:firstFriedmanngefolding}). This combination of eqs. (\ref{eq:firstFriedmanngefolding}) and (\ref{eq:AlgebraicH}) yields a quartic equation in $y$\footnote{Differentiating this equation with respect to $x$ gives the differential equation (\ref{eq:diffEqFory}).}:
\begin{eqnarray}
 \frac{B_3}{3}y^4 + \left(B_2 - \frac{B_4}{3}\right)y^3 + \left(B_1 - B_3\right)y^2 + \left( \frac{B_0}{3} + \Omega_m + \Omega_{\gamma} - B_2\right)y - \frac{B_1}{3}=0.\label{eq:quarticEquationForY}
\end{eqnarray}

With equation (\ref{eq:quarticEquationForY}) we could in principle calculate the values of $y$ at any redshifts without integrating the system through the entire history of the Universe. However, since a ``quartic'' equation can in general have as many as four real solutions, it is not necessarily easy to choose the right root at each given point: ``is this root the one that the root from the previous point would have evolved to?'' To insure that this is the case, we instead use the differential equation (\ref{eq:diffEqFory}) to solve for $y$. This also guarantees real solutions throughout the cosmic history since starting with a real solution, a differential equation with only real coefficients is guaranteed to yield solutions that will remain real at all times.

By studying eq. (\ref{eq:quarticEquationForY}), we can immediately deduce what form the asymptotic solutions will take. At early times the terms containing $\Omega_m$ and $\Omega_\gamma$ will dominate since they evolve according to $\Omega_m = \Omega_m^0(1+z)^3$ and $\Omega_\gamma = \Omega_\gamma^0(1+z)^4$, respectively. Deep in the past the equation will then approximately read as
\begin{equation}
  \left(\Omega_{m} + \Omega_\gamma\right)y = 0,
\end{equation}
so that $y = 0$, eliminating the contributions from $y$ in eq. (\ref{eq:firstFriedmanngefolding}). Hence, at early times, we will have the usual $\Lambda$CDM model with $\Omega_\Lambda = B_0/3$. As time progresses the other terms become important, and $y$ starts varying with time. At late times, $\Omega_m$, $\Omega_\gamma$ and the curvature terms become negligible, and $y$ will be given in terms of a time-independent quartic equation. This means that the Universe will have transitioned again into a cosmological constant phase, but now with a different value of the constant. An example of this behavior can be seen in fig. \ref{fig:BiGB1plot_y} where we show $y$ as a function of redshift ($z=-1$ corresponds to far in the future). We therefore expect the model to give good fits to the cosmological observations as long as the change in $y$ is not dramatically large and the Hubble parameter evolves in a way similar to that of the $\Lambda$CDM model.

\FIGURE[t]{
      \includegraphics[width=0.7\textwidth]{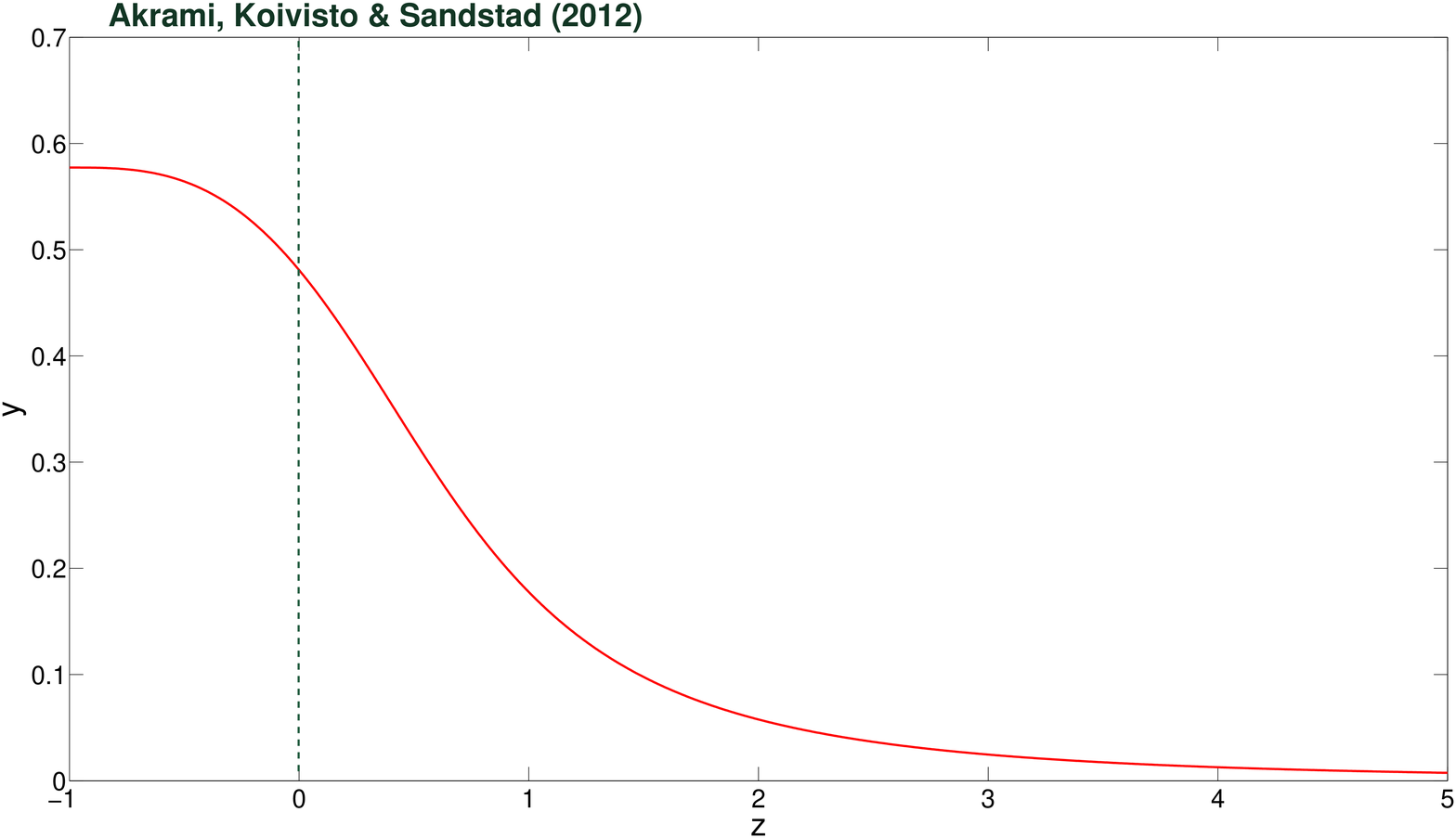}
      \caption{An example of how $y$ evolves with redshift $z$ for a model that is well fitted to the real data. This is obtained for a bigravity model with $B_0=B_2=B_3=B_4=0$, $B_1=1.44$ and $\Omega_m^0=0.307$ (see section \ref{sec:2freeparams}). We see that $y$ starts out at a zero-value fixed-point in the past and transitions to a larger-value fixed-point at late times. The dashed vertical line illustrates the present time ($z=0$).}
      \label{fig:BiGB1plot_y}
    }

\subsection{Setting initial conditions}\label{sec:initialconditions}
To perform the numerical integration of eq. (\ref{eq:diffEqFory}), we start from the present time and integrate into the past. This is done by solving the quartic equation (\ref{eq:quarticEquationForY}) at $z=0$ and then using eq. (\ref{eq:diffEqFory}) to obtain $y$ at all interesting redshifts throughout the evolution of the Universe. One could in principle solve eq. (\ref{eq:quarticEquationForY}) at any other arbitrary redshift instead, but choosing $z=0$ is particularly useful in practice for simplicity reasons, as well as to set the initial conditions stably. In addition, it makes it possible to compute co-moving distances belonging to each redshift of interest in the same integration code as the one used for computing the Hubble parameter. This makes the statistical analysis of the model faster and the numerical calculations more rational. Now that the starting values of $y$ and $H$ (i.e. at $z=0$) are determined for a set of model parameters using eqs. (\ref{eq:quarticEquationForY}) and (\ref{eq:AlgebraicH}), the numerical integration can commence using eq. (\ref{eq:diffEqFory}) and the Hubble parameter can be calculated at all redshifts using eq. (\ref{eq:AlgebraicH}).

Some caveats however need to be considered for this scheme since the quartic equation generically has several solutions: Firstly, all values of $y$ should not be accepted. $y$ is the scale factor of a metric ($f$) normalized to the scale factor of another metric ($g$), and it therefore seems meaningless for $y$ to take on complex or negative values. Therefore only real and positive roots of eq. (\ref{eq:quarticEquationForY}) should be considered; this means that many choices of parameters yield no viable solutions. Secondly, we see from eq. (\ref{eq:firstFriedmanngefolding}) and (\ref{eq:AlgebraicH}) that $H^2$ and not $H$ is the quantity that is directly calculated from $y$; we should therefore ensure that it receives positive values. We will see in more detail in the next section how we implement these conditions numerically. And thirdly, for a single set of parameters of the model, there may be several viable roots for eq. (\ref{eq:quarticEquationForY}). Again, as we will see later, this subtlety is taken care of by treating the initial value of $y$ as an additional parameter in our statistical analysis. This parameter can in general take on $0$, $1$, $2$, $3$ or $4$ different values at each point in the model's parameter space (depending on the number of allowed solutions for the quartic eq. (\ref{eq:quarticEquationForY})). Hence, in exploring the parameter space for the best-fit parameter values, we should consider all the allowed initial values of $y$, perform the analysis and then choose the initial $y$ that gives the best fit.

\subsection{Statistical framework, scanning strategy and comparison to observation}\label{sec:statistics}
So far, we have set up our theoretical model: we know how to solve the dynamical equations of the model, how to compute necessary observable quantities to be used in comparison of the theoretical predictions to various cosmological observables, and which data to use. What we intend to discuss in this section is how to confront the model with observation in practice, i.e. how to explore the parameter space of the model to test how fit it is to the data and what values of the parameters are preferred. Let us recap what we have found and discussed so far:

\begin{itemize}
    \setlength{\itemsep}{1pt}
    \setlength{\parskip}{0pt}
    \setlength{\parsep}{0pt}
    \item The full model has the following six free parameters:
\begin{equation} \label{allparams}
\Theta = (\beta_0, \beta_1, \beta_2, \beta_3, \beta_4, \Omega_m^0).
\end{equation}
Here, as we stated earlier, we have assumed a flat Universe (i.e. $k=0$). In addition, we have neglected the radiation contribution to the evolution equation and background observables, i.e. we have assumed $\Omega_\gamma^0\approx 0$. This latter assumption can be justified by the fact that in the standard $\Lambda$CDM cosmology, the present energy density of radiation is vanishingly small ($\Omega_\gamma^0=\mathcal{O}(10^{-5})$) compared to the matter contribution ($\Omega_m^0 = \mathcal{O}(1)$). Its value then remains negligible up to the redshift of matter-radiation equality which goes well beyond the redshift range we are using in the present analysis. We expect a similar ratio between $\Omega_\gamma^0$ and $\Omega_m^0$ in our bigravity model which enables us to neglect radiation in the entire period of interest (from the present time to the epoch of recombination) as radiation and matter scale with redshift as $(1+z)^4$ and $(1+z)^3$, respectively.
    \item By fixing the values of all parameters $\Theta$, eqs. (\ref{eq:quarticEquationForY}) and (\ref{eq:diffEqFory}) tell us how $y$, corresponding to that chosen point in the parameter space, evolves with time (or with redshift).
    \item Eq. (\ref{eq:AlgebraicH}) can then be used to calculate the Hubble parameter $H$ as a function of $z$.
    \item Knowing $H(z)$ at a given point in the parameter space is then enough for calculating all the quantities we need in order to compare the model (at that point) with the observations we described in section \ref{sec:constraints}.
\end{itemize}

In order to constrain the model and test how fit it is to the real data, we need to perform a statistical analysis. This requires a scanning of the parameter space of the model and this can be done in various ways depending on which statistical framework one chooses to work with. In statistical parameter estimation and model selection, there are in general two different types of statistical inference that are fundamentally very different: ``Bayesian'' and ``frequentist'' statistics (for a general introduction to statistical inference, see e.g. \cite{Cowan:1998}, and for some discussions about the differences between the two approaches in data analysis, see \cite{Akrami:2009hp,Akrami:2010cz,Akrami:2011vh} and references therein). In presence of sufficient observational data where the parameters of the model are strongly constrained, both statistics are proven to yield the same results. Any discrepancy between the inferences therefore shows that either we need more constraining data, or the numerical methods employed in exploring the parameter space are not accurate enough. In what follows we first briefly review the two approaches and their main ingredients used for both parameter estimation and model selection. We then discuss why it is essential to use both approaches in order to ensure that the results of the statistical inference are reliable and robust. We use both approaches in our analysis of the bigravity model in this paper.
\\

\noindent
$\bullet$ \textbf{Bayesian inference:} This approach has now become among the standard tools in cosmological data analysis (for applications of Bayesian statistics in physics, see e.g. \cite{D'Agostini:1995fv}, and for reviews of its applications in cosmology, see e.g. \cite{Trotta:2005ar,Trotta:2008qt,Liddle:2009xe,Hobson:2010}). Here, one can assign probabilities to the parameters of the model under consideration. Let us first define the following quantities: 1) $P(\Theta;H)\equiv \pi(\Theta)$, which denotes our ``prior'' probability density function (PDF) and reflects our knowledge or prejudices about the values of the parameters before comparing the model with observations. 2) $P(D|\Theta;H)\equiv \mathcal{L}(\Theta)$, which is called the ``likelihood'', and is the probability of obtaining the data $D$ from the model parameters $\Theta$ when it is considered as a function of $\Theta$. 3) $P(\Theta|D;H)$, or the ``posterior'' PDF, which represents the probability of the model parameters $\Theta$ when the data $D$ are used. Finally, 4) $P(D;H)\equiv \mathcal{Z}$, which is called the Bayesian ``evidence'' and is the probability of observing the data $D$ when we integrate over the entire parameter space. Here, we have assumed $H$ to be the model hypothesis under consideration which we have parameterize by $\Theta$. Bayes' Theorem then relates these quantities in the following way:

\begin{equation}\label{eq:BayesTheorem}
P(\Theta|D;H)=\frac{P(D|\Theta;H)P(\Theta;H)}{P(D;H)}.
\end{equation}

This expression simply shows how our prior degree of belief about different values of model parameters is updated when new observational data are used. Our updated knowledge, given in terms of the multi-dimensional PDF, $P(\Theta|D;H)$, is the quantity of interest in Bayesian statistics through which various statistical statements can be made. One can for example construct different (``credible'') intervals and regions in the parameter space corresponding to certain levels of confidence. In this framework, one-, two-, ... and $(N-1)$-dimensional (where $N$ is the total number of free parameters of the model) credible regions for any sub-space of the parameter space can then be easily constructed by integrating (or `marginalizing') the full posterior PDF, $P(\Theta|D;H)$, over all other parameters; this procedure then results in joint ``marginal'' PDFs for the parameters of interest.

In Bayesian parameter estimation, one reports the ``posterior means'' of the parameters (the expectation values of the parameters according to the marginalized posterior) as their most-favored values. Uncertainties about the favored values (at a given confidence level) are then estimated by ``probability mass'' surfaces containing corresponding percentages of the total marginalized posterior PDF.

In Bayesian model selection on the other hand, the quantity of interest is the evidence

\begin{equation}\label{eq:EvidenceIntegral}
\mathcal{Z}\equiv P(D;H)=\int_V\mathcal{L}(\Theta)\pi(\Theta)d\Theta,
\end{equation}
where $V$ is the entire $N$-dimensional parameter space of the model. Assume that we want to select the best-fit model between two hypotheses $H_0$ and $H_1$. This can be done by looking at the ratio

\begin{equation}\label{eq:modelselection}
\frac{P(H_1|D)}{P(H_0|D)}=\frac{P(D|H_1)P(H_1)}{P(D|H_0)P(H_0)}=\frac{\mathcal{Z}_1}{\mathcal{Z}_0}\frac{P(H_1)}{P(H_0)},
\end{equation}
where $P(H_1|D)$ and $P(H_0|D)$ are the posterior PDFs for $H_1$ and $H_0$ to be the true hypotheses given the observed data, respectively, and $P(H_1)$ and $P(H_0)$ are the prior PDFs for the hypotheses before observing the data. The ratio $P(H_1)/P(H_0)$ can be set to unity in case we have no preference for any of the two models a priori. In that case, one observes from eq. (\ref{eq:modelselection}) that in order to see whether a model is preferred compared to another one one needs to evaluate Bayesian evidences for the two cases and look at their ratio.

This is an interesting recipe for selecting between different theoretical models, but with two caveats: 1) The evidence ratio is useful only if the models at hand are equally motivated theoretically or based on previous observational constraints; this is not always true though. 2) One usually calculates the evidences numerically, meaning that the exact value of the integral (\ref{eq:EvidenceIntegral}) cannot be evaluated. In practice, one needs to choose some ranges for the parameters in the parameter space to scan over and this is effectively equivalent to imposing a prior. In most cases (at least when the likelihood has a Gaussian or nearly Gaussian shape), choosing larger ranges will give a smaller evidence (this can be seen also from the observation that the evidence is nothing but the average of the likelihood over the parameter space, at least when flat priors are used). In cases where the alternative hypothesis $H_1$ differs from the null hypothesis $H_0$ in that it is only an extension of the $H_0$ parameter space by adding new parameters (i.e. $H_1$ contains $H_0$ as a sub-space), the evidence comparison can be a very powerful method. The reason is that increasing the dimension of the parameter space yields a smaller evidence (see integral (\ref{eq:EvidenceIntegral})) unless the likelihood values improve in such a way that the increase in the volume of the parameter space is compensated (the evidence recipe therefore automatically implements Occam's razor). For completely different models however, one needs to know prior probabilities for the models, as well as the parameter ranges within each model.
\\

\noindent
$\bullet$ \textbf{Frequentist inference:} In this approach to statistical data analysis, probabilities are assigned only to the data and cannot be assigned to model parameters. This means that the posterior PDF, $P(\Theta|D)$, on which Bayesian parameter estimation is based, is completely meaningless to frequentists. In this framework, one instead works only with the likelihood which by definition is the probability of the observed data for a fixed set of model parameters and is well-defined in the frequentist framework.

Knowing the full, $N$-dimensional likelihood of the model, one can infer various statistical properties of the model by constructing ``confidence'' intervals and regions (as opposed to the credible intervals and regions in Bayesian statistics). Perhaps the most common method for such constructions, that is numerically feasible enough, is the ``profile likelihood'' procedure \cite[and references therein]{2005NIMPA.551..493R}. Here, instead of marginalizing over unwanted parameters, one maximizes (or profiles) the full likelihood along those parameters and obtains one-, two-, ... and $(N-1)$-dimensional profile likelihoods. The most-favored values for the parameters are then given by the values that maximize the full likelihood, and uncertainties upon those values (at a given confidence level) are constructed by iso-likelihood contours in the model parameter space around the best-fit point. For Gaussian-like likelihoods (which are always the case if sufficient data are available), the profile likelihood method is a good approximation to the exact frequentist construction of confidence intervals and regions proposed by Neyman~\cite{Neyman}. One method that provides exact confidence regions and intervals even for complex (non-Gaussian) parameter spaces (although harder to implement numerically), is the ``confidence belt'' construction scheme \cite{1998PhRvD..57.3873F}.

In frequentist inference, in order to assess how fit a theoretical model is to the observed data, one again works with the full likelihood. Let us assume that the number of data points (measurements) used in the analysis is $N(D)$ and the model has $N(\Theta)$ free parameters. In addition, we assume that each measured data point follows a Gaussian distribution (an assumption that is approximately true for the cosmological data we use in the present paper). One then expects the $\chi^2$ ($\equiv-2\ln{\mathcal{L}}$) at the best-fit (highest-likelihood) point to follow a chi-squared distribution with $N(D)-N(\Theta)$ degrees of freedom if one repeats the measurements ideally an infinite number of times. Calculating the $p$-value (the probability of obtaining a $\chi^2$ as large as, or larger than the one actually observed, assuming that the model is true) corresponding to the observed $\chi^2$ then provides a powerful tool to assess the goodness of fit of the model to the data.
\\

\noindent
$\bullet$ \textbf{Scanning of the parameter space:} Our discussions so far about the statistical frameworks have made it clear that no matter which strategy we use, one quantity that we need to estimate accurately is the likelihood of the model as a function of the free parameters. Mapping of the likelihood correctly is therefore the first essential step in both Bayesian and frequentist statistics. Especially in the latter approach, it is all we need to know: the globally maximum likelihood value provides us with a goodness-of-fit test of the model (through calculating the $p$-value at the best-fit point), as well as the favored values of parameters and uncertainties around them (through the profile likelihood construction and iso-likelihood contours). For the Bayesian framework however, one needs to take one step further and calculate the evidence (for model comparison) and the posterior PDF (for parameter estimation).

Simple grid scans are obviously the most straightforward methods in mapping the posteriors, but they are notoriously slow when implemented numerically for high-dimensional parameter spaces. An alternative approach that has become highly popular in cosmological data analysis is based on the Markov Chain Monte Carlo (MCMC) methods (for an introduction and overview of MCMC methods, see \cite{Gamerman:2006}, and for their applications in cosmology, see e.g. \cite{Lewis:2002ah}) which revolve around the idea that the density of the points obtained in the numerical exploration of the parameter space be proportional to the actual probability function that one aims to map. This provides a simple way of marginalizing over any set of parameters (required by Bayesian statistics) just by counting the number of sample points corresponding to a point in the selected sub-set of the full parameter space. MCMCs provide a largely improved scanning efficiency in comparison to grid searches.

More recently, another scanning algorithm based on the framework of nested sampling \cite{2004AIPC..735..395S,SkillingNS2}, called {\sf {MultiNest}} \cite{Feroz:2007kg,Feroz:2008xx}, has attracted considerable attention in both particle physics and cosmology. The primary purpose of this method is to calculate the Bayesian evidence for a model, but it also provides the posterior distribution as a byproduct. It has been claimed that the results of the {\sf {MultiNest}} and the MCMC parameter estimations are identical (up to numerical noise), while the former is two orders of magnitude faster than the latter. The technique is clearly optimized for Bayesian inference, but in the absence of a competitive alternative, it can be used also for frequentist analyses if the model parameter space is not far too complex with many spike-like, fine-tuned regions. We do not expect our cosmological model to be of this sort and therefore choose {\sf {MultiNest}} as the statistical tool in our present analysis (for an alternative algorithm optimized for frequentist statistics, see e.g. \cite{Akrami:2009hp}).

As we mentioned earlier, the results of Bayesian and frequentist statistics do agree if the information provided by data is so strong that the likelihood dominates over prior contributions to the posterior distribution. In this case, the credible and confidence regions will coincide providing unique statistical conclusions about the model parameters. In Bayesian statistics, the set of prior assumptions play an important role in the final inference. In cases where such effects are dominant, the statistical conclusions cannot be trusted. This feature of Bayesian statistics is however very interesting because it provides a good measure of the robustness of a fit; one should not consider the fit definitive if it strongly depends upon priors. In this case the data are not constraining enough and/or the model under consideration is so complex that a more detailed analysis of its parameter space is required (for a detailed discussion of the impacts of priors, but in a different context, see \cite{Trotta:2008bp}). One other way to check whether the results of a fit are independent of priors is to compare them with the results of a frequentist analysis. In cases where the frequentist measures (such as the profile likelihood) do not point to similar preferred values for parameters and uncertainties around them, we can conclude that the model is not constrained properly with the data, and priors have strong impacts on the results.

The other interesting reason why one should consider both Bayesian and frequentist measures, such as marginal posteriors and profile likelihoods, respectively, in any scan, is that they probe different properties of the parameter space. Marginal posteriors are sensitive only to the total posterior probability mass in the sub-space that has been marginalized over, and they consequently do not probe fine-tuned regions with spike-like likelihood surfaces. Frequentist measures, such as profile likelihoods, are on the other hand (by construction) strongly sensitive to such regions. Considering both marginal posteriors and profile likelihoods therefore enables us to acquire a complete picture of the favored parameter regions and statistical properties of the model.

As we discussed above, most of the state-of-the-art scanning algorithms that are widely used in cosmological data analysis are based on either MCMCs or nested sampling techniques (as we use in the analysis of the present paper). All such methods are designed and optimized for Bayesian inference as their main objectives are to calculate either the posterior PDFs or the Bayesian evidence (or both). Since there are no competitive algorithms (in terms of speed, efficiency and accuracy) that are optimized for the frequentist approach, it is very common to use those Bayesian algorithms also to map the likelihood of the model, the quantity that is needed for frequentist inference. This gives another reason why one has to deal with Bayesian inference (though implicitly) even if s/he is interested only in frequentist statistics. In principle, results of frequentist inference are independent of prior assumptions. This is however the case only if an accurate construction of the likelihood function is available, which in turn depends on how powerful the employed scanning algorithm is. Using algorithms that are optimized for Bayesian exploration of the parameter space then implies that any mapping of the likelihood function based on such scans can in principle be strongly impacted by prior effects (for more discussions, see e.g. \cite{Akrami:2009hp,Akrami:2010cz}).

Ideally, one wants to compute both marginal posteriors and profile likelihoods by sampling the parameter space. Our discussion in the previous paragraph illustrates that MCMC-based sampling techniques would give a good estimate of the former, but they may not be sufficiently accurate in providing the latter. One may be able to get a better estimate of the maximum values of the likelihood at each point in the parameter sub-space of interest by generating a much larger number of samples, but with typical numbers of Monte-Carlo samples generated by standard algorithms, the estimated values may be too far from the actual ones. The reason simply is that the algorithms may miss the spike-like maxima as they are not sensitive to the fine-tuned regions. One can however obtain sufficiently accurate values for the ``mean" values of the likelihood along the unwanted parameters (as defined in \cite{Lewis:2002ah}), even from a small number of samples. Since in the present paper we will give our estimates of the parameter values only in terms of marginal posteriors, and frequentist measures are used only to assess the degree of dependence of the Bayesian results on the priors (we do not aim to compute frequentist confidence regions and intervals), we plot the mean likelihoods, instead of the profile likelihoods, on top of the posteriors. We still expect a good agreement between marginal posteriors and mean likelihoods for well constrained parameters \cite{Lewis:2002ah}. If the posterior PDF is Gaussian or is separable with respect to the sub-space of the parameter space for which the posterior has been marginalized, the mean likelihood will be proportional to the marginal posterior. Mean likelihoods are however not purely frequentist quantities, as the posterior PDF is used in their definition. They can additionally help us acquire more information about the shape of the distribution along the directions over which we marginalize the distribution. The marginalization procedure looses all such information in particular about the skewness with respect to the marginalized dimensions, which is important for understanding possible correlations between the parameters; this information can be to a great extent recovered by plotting the mean likelihoods in addition to the marginal posteriors.

We end this section by a discussion about the construction of the model likelihood that is used in constraining the model. The scanning algorithm we use in our analysis (the {\sf {MultiNest}} algorithm) works based on the calculation of the likelihood (or the $\chi^2$) value at each point in the parameter space. This is done by summing over all $\chi^2$ values for different contributions  from cosmological measurements of section \ref{sec:constraints} when Gaussian distributions are assumed for uncertainties around the measured values. In cases where a point in the parameter space, i.e. a set of parameters, gives unacceptable values for a computed quantity (such as negative values for $y$ or $H^2$ as discussed in section \ref{sec:initialconditions}), we assume a very large value for the $\chi^2$ (or a very small value for the likelihood); this automatically disfavors such points. Also, in cases where the parameters give more than one solution for the initial value of $y$ (at $z=0$), we compare the $\chi^2$ values obtained for all the solutions and retain the one that gives the lowest value.

\section{Results and discussion}\label{sec:Results}

In the previous sections, we introduced the bimetric theory of gravity and the specific model that we aim to constrain using cosmological data. In addition, we presented the data set, the parameter space of the model, statistical methods and measures, as well as the scanning algorithm that we employ in our analysis. In the following sections, we present our results for particular sub-spaces of the full parameter space, as well as the full theory. We also discuss our findings and some subtleties that must be taken into account when interpreting the results. At the end of this section, we summarize our statistical results for all different models in table \ref{tbl:ResultsDifferentParmaeterRegimes}.

\subsection{$\Lambda$CDM as the reference model}\label{sec:LCDM}

In order to assess how well the bigravity model fits the data, it is useful to compare the results to those of a reference model, with the obvious choice of the standard $\Lambda$CDM model. We know that the $\Lambda$CDM model is in perfect agreement with all existing cosmological constraints, in particular at the background level. This makes the model phenomenologically very interesting and so far the best way of parameterizing the evolution of the Universe. We therefore expect any alternative models that agree with observations to effectively mimic the $\Lambda$CDM at late times; our bigravity model is no exception.

In order to be consistent with the assumptions in the bigravity case, we define our $\Lambda$CDM model in terms of two free parameters $\Omega_m^0$ (present value of the matter density parameter) and $H_0$ (present value of the Hubble parameter). Assuming a flat universe, $\Omega_\Lambda$ will be a derived quantity and not a free parameter ($\Omega_\Lambda=1-\Omega_m^0$). We scan this two-dimensional parameter space using the data and methods we described in the previous sections. The flat (top-hat) prior ranges we choose for the parameters are $[0,1]$ and $[50,100]$ km s$^{-1}$ Mpc$^{-1}$, for $\Omega_m^0$ and $H_0$, respectively.

\begin{figure}[t]
\begin{center}
\includegraphics[width=0.49\textwidth]{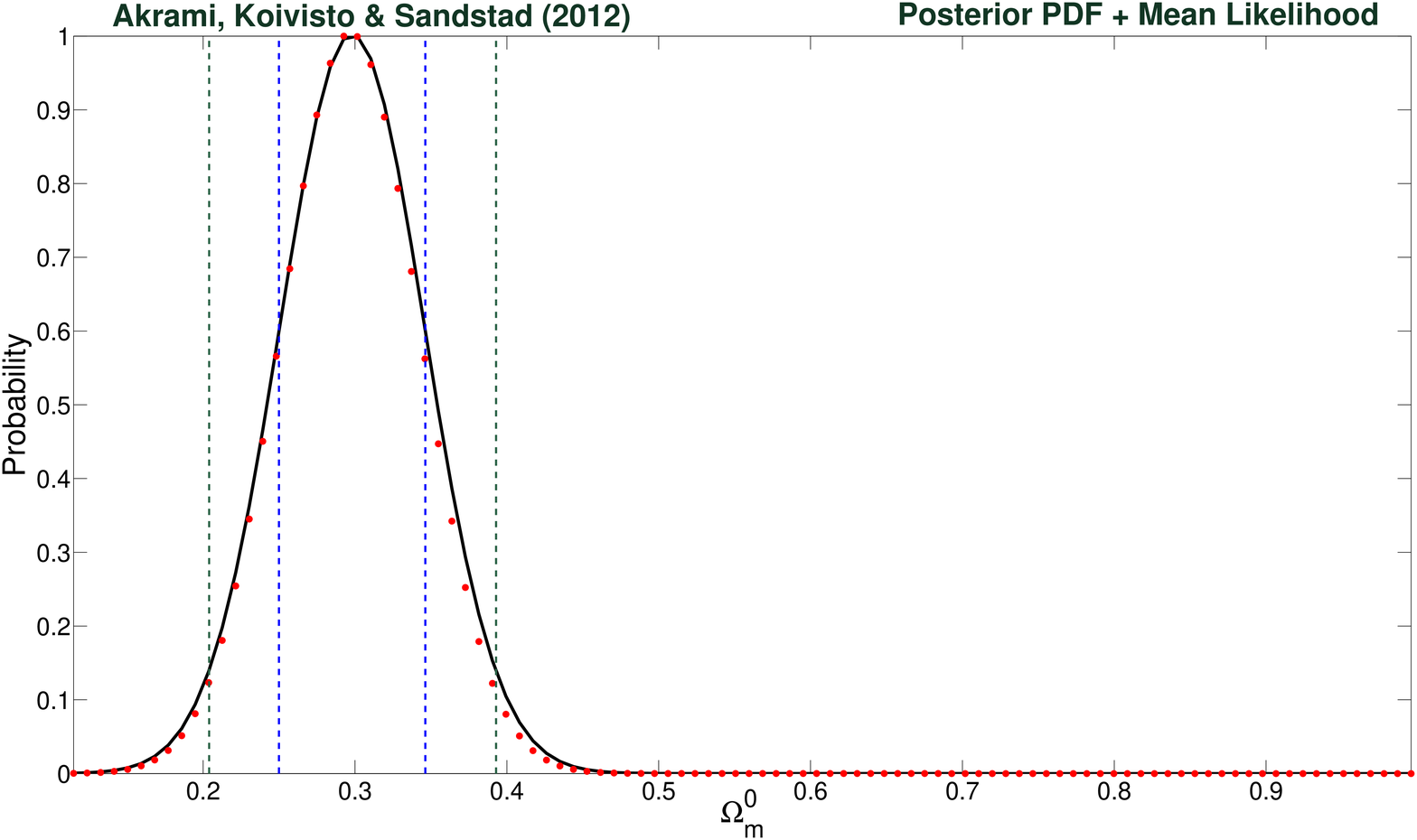}
\includegraphics[width=0.49\textwidth]{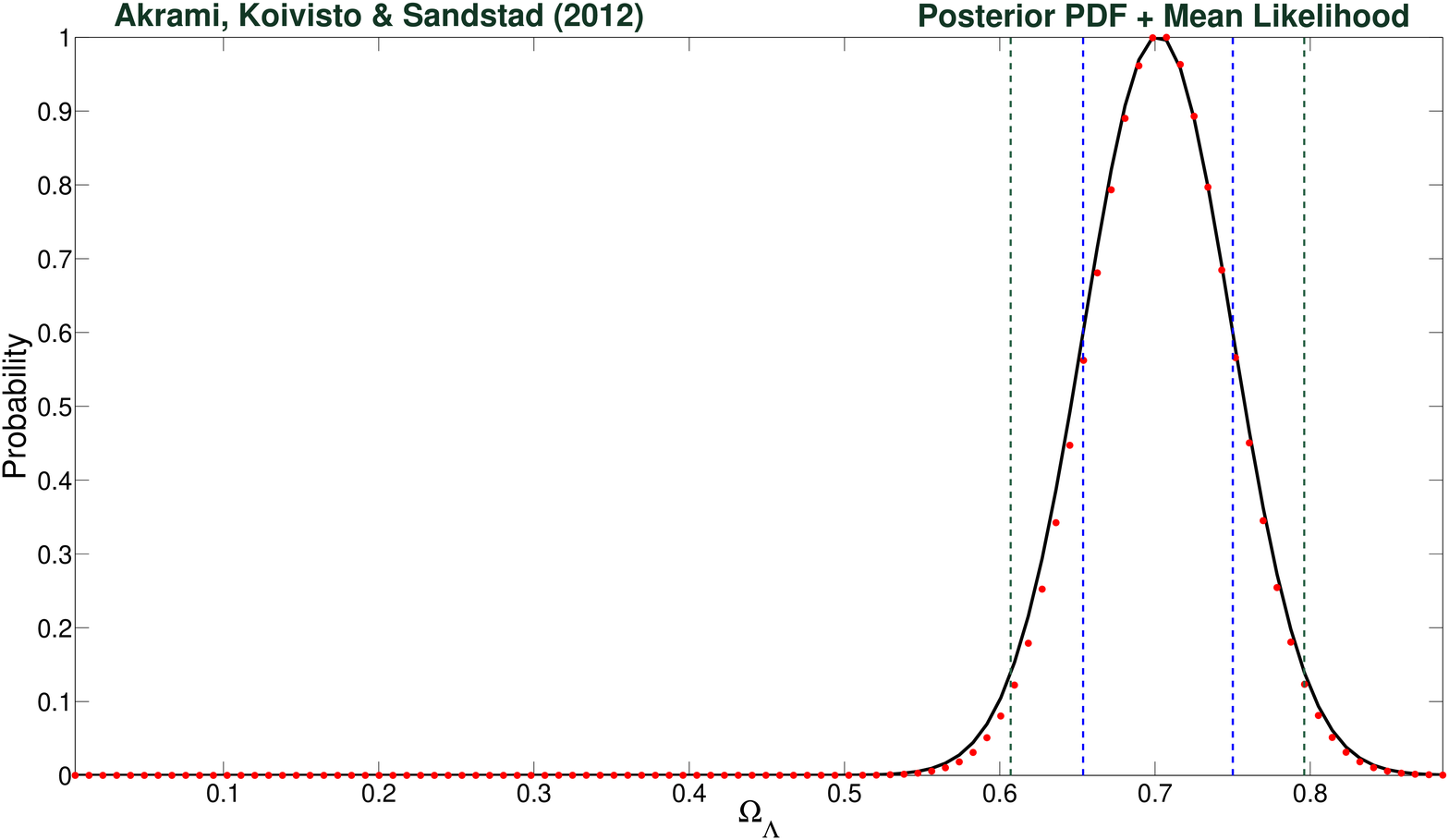}
\caption[aa]{One-dimensional marginal posterior PDFs (black solid curves) and mean likelihoods (red dots) for the parameters $\Omega_m^0$ (left panel) and $\Omega_\Lambda$ (right panel) in the (flat) $\Lambda$CDM concordance model fitted to the cosmological data. $68\%$ ($1\sigma$) and $95\%$ ($2\sigma$) credible intervals for the parameters are shown by blue (inner) and green (outer) vertical dashed lines, respectively.}
\label{fig:LCDMplots}
\end{center}
\end{figure}

The $\chi^2$ value that we find at the best-fit point is $546.54$, which corresponds to a $p$-value of $0.8709$. This value shows, as expected, that the observed $\chi^2$ is less than $1\sigma$ away from the predicted value, an indication that the model is very well consistent with the data. The value of $log \mathcal{Z}$ ($\mathcal{Z}$ being the Bayesian evidence) we have obtained for the model is $-278.50$. In addition, fig. \ref{fig:LCDMplots} shows the one-dimensional marginal posterior PDFs (black solid curves) and mean likelihoods (red dots) for $\Omega_m^0$ and $\Omega_\Lambda$ as the result of our analysis. The blue (inner) and green (outer) vertical dashed lines indicate $68\%$ ($1\sigma$) and $95\%$ ($2\sigma$) credible intervals, respectively. Both marginal posterior and mean-likelihood curves are normalized to their values at their peaks. Both curves have Gaussian forms and perfectly match, another indication that the model fits the data very well and its parameters are observationally well constrained.

\subsection{No explicit cosmological constant ($B_0 = 0$)}

Let us now turn to our bigravity model. Looking again at the model's action, eq. (\ref{eq:ActionOriginal}), we see that the parameter $\beta_0$ (and correspondingly $B_0$) is nothing but a cosmological constant term for the physical metric $g$ (with $\Omega_\Lambda=B_0/3$)\footnote{Strictly speaking, $\beta_0$ represents the vacuum energy contribution to the cosmological constant; see the discussions in footnotes \ref{alphanote} and \ref{CCnote} for more details.}. In fact, by setting all other $\beta$s to zero, it is obvious that our bigravity model will reduce to the standard $\Lambda$CDM model (it can be seen also from eq. (\ref{eq:firstFriedmanngefolding}) for the Hubble parameter). We know from the observational constraints on the standard model that one gets a very good fit to the data if such a term is present. This constant term has been proposed as the standard source of cosmic acceleration at late times. As we stated in section \ref{sec:introduction}, this assumption has been strongly challenged by both cosmology and particle physics considerations, mainly because of the difficulty of explaining its small but not zero value compared to the theoretical predictions without the need for a huge fine-tuning. Finding an alternative to the cosmological constant that explains the late-time acceleration of the Universe has therefore been of great interest. In order to see whether one can get an accelerated universe from our bimetric modification of gravity without an explicit cosmological term (i.e. through a self-acceleration mechanism), we set the parameter $B_0$ to zero. We then try to fit the other parameters of the model to the data examining how well the model can explain the cosmic acceleration.

In addition, before we analyze the full system, it is helpful to study its sub-systems where only one or a couple of the parameters $B_i$ are nonzero and free to vary. This may help us to understand the dynamics of the model better, for instance, which terms are necessary for self-acceleration giving a good fit to the data, which parameters can be constrained observationally, and which ones remain unconstrained. In addition, such a step-by-step analysis will be helpful in understanding possible correlations between the parameters.

\subsubsection{Sub-models with two free parameters:  ($B_1,\Omega_m^0$), ($B_2,\Omega_m^0$), ($B_3,\Omega_m^0$)}\label{sec:2freeparams}

We start with sub-sets of the parameter space (\ref{allparams}) where all $\beta$s (or $B$s) are set to zero except one of them. We have already assumed $B_0$ is fixed to zero (again, it is obvious that a model with only nonzero $B_0$ and $\Omega_m^0$ will yield good fits to observations, as this is just the $\Lambda$CDM model that we studied in section \ref{sec:LCDM}). Keeping $\Omega_m^0$ a free parameter, we therefore have four choices for the reduced two-dimensional sub-space, i.e. cases with $B_1$, $B_2$, $B_3$, or $B_4$ being free. Clearly having only $B_4$ non-zero will not yield an accelerated universe and hence a good fit. It is easy to see this by observing that with only $B_4 \neq 0$, eq. (\ref{eq:firstFriedmanngefolding}) is nothing but the standard Hubble equation containing only matter and radiation. Such a model does not fit the set of cosmological observations. The interesting cases to study are therefore those of free $B_1$, $B_2$ or $B_3$ (when the matter density parameter $\Omega_m^0$ is also allowed to vary).
\\

\begin{figure}[t]
\begin{center}
\includegraphics[width=0.49\textwidth]{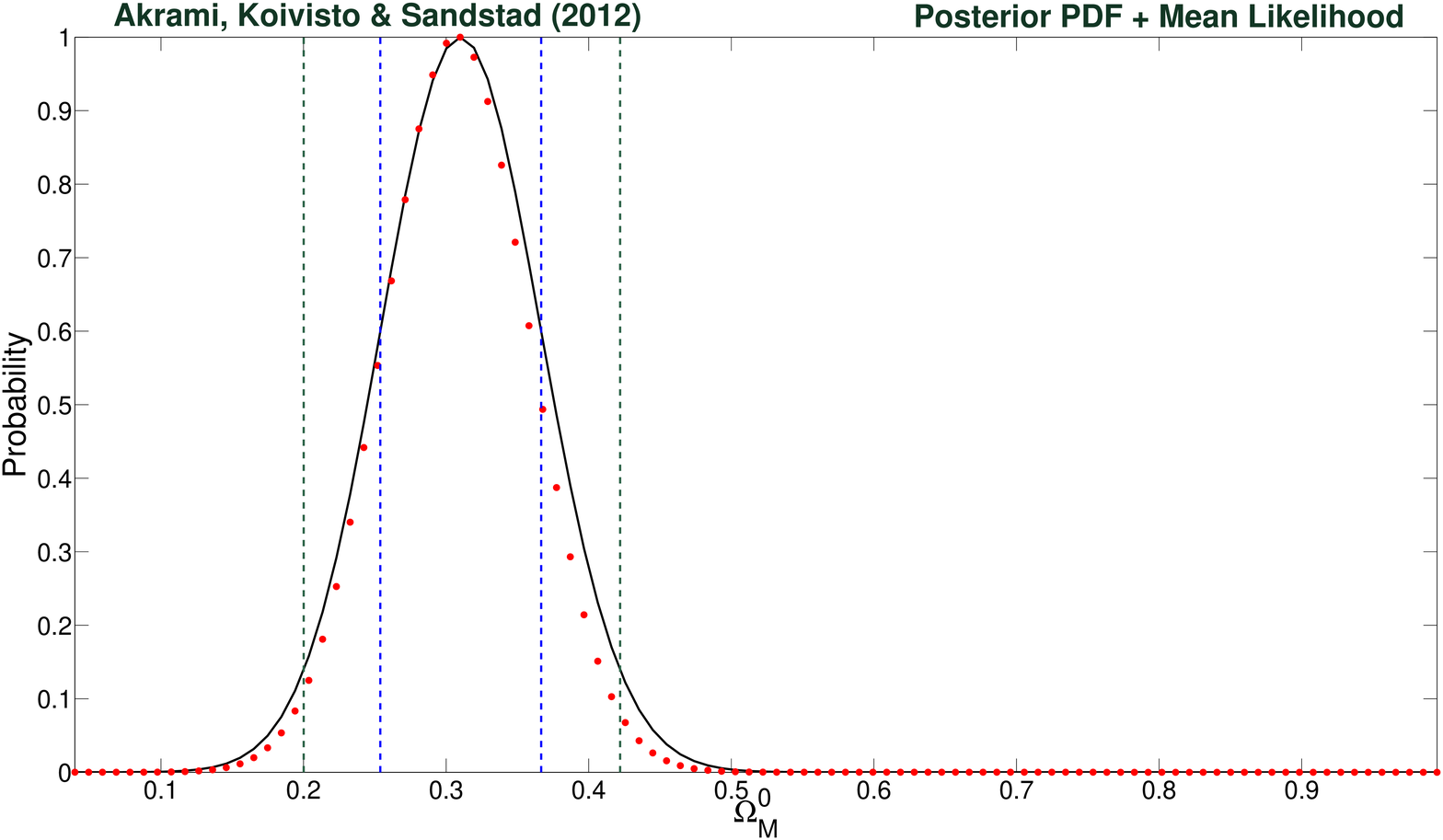}
\includegraphics[width=0.49\textwidth]{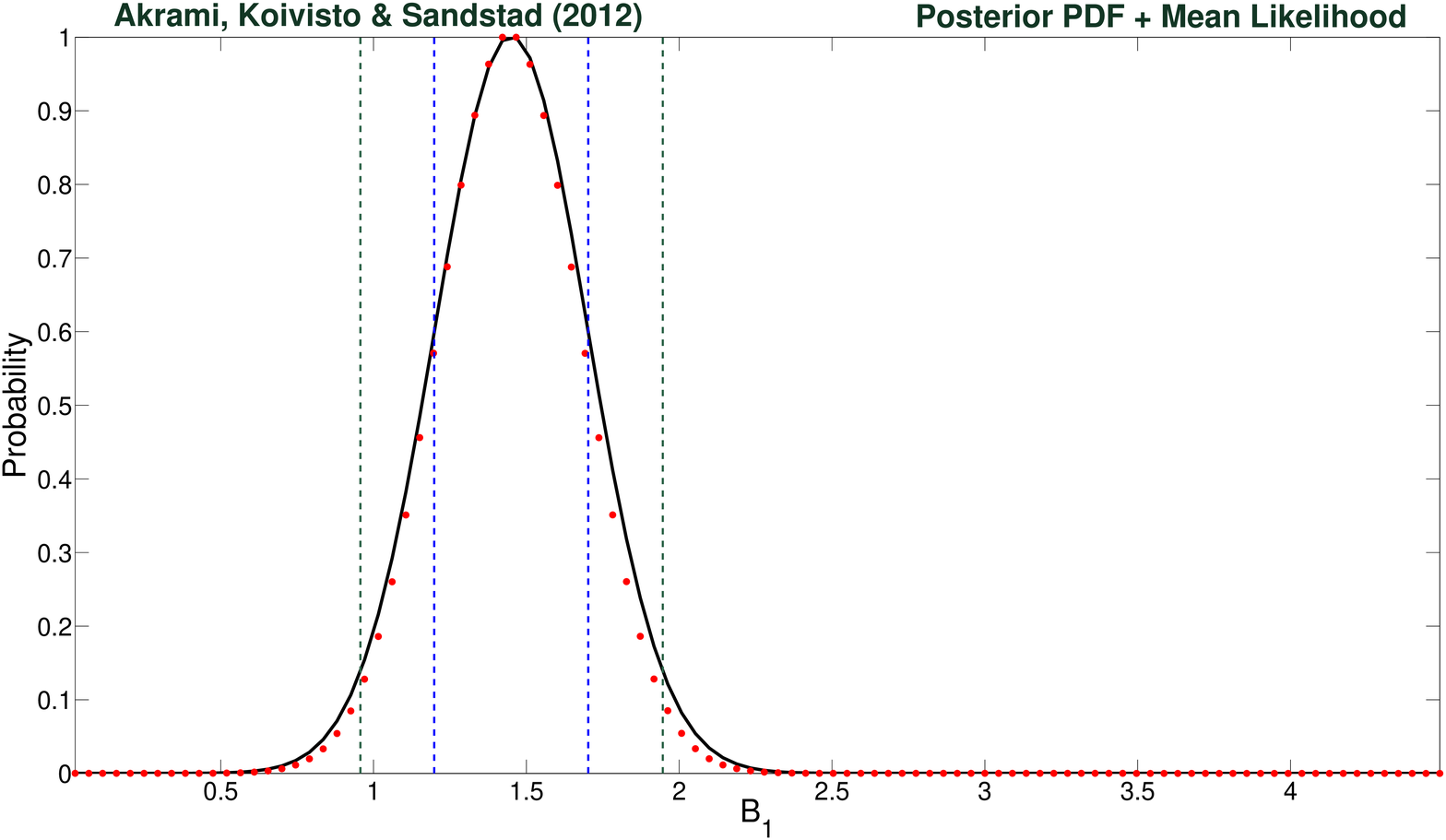}
\caption[aa]{As in fig. \ref{fig:LCDMplots}, but for the bigravity model with only $B_1$ and $\Omega_m^0$ varying (other parameters are set to zero). The constraint on the value of graviton mass ($m$) in this model is $m^2=(1.45\pm0.25) H_0^2$, where $H_0=71\pm2.5$ km s$^{-1}$ Mpc$^{-1}$ (right panel).}
\label{fig:BiGB1plots1D}
\end{center}
\end{figure}

\begin{figure}[t]
\begin{center}
\includegraphics[width=0.7\textwidth]{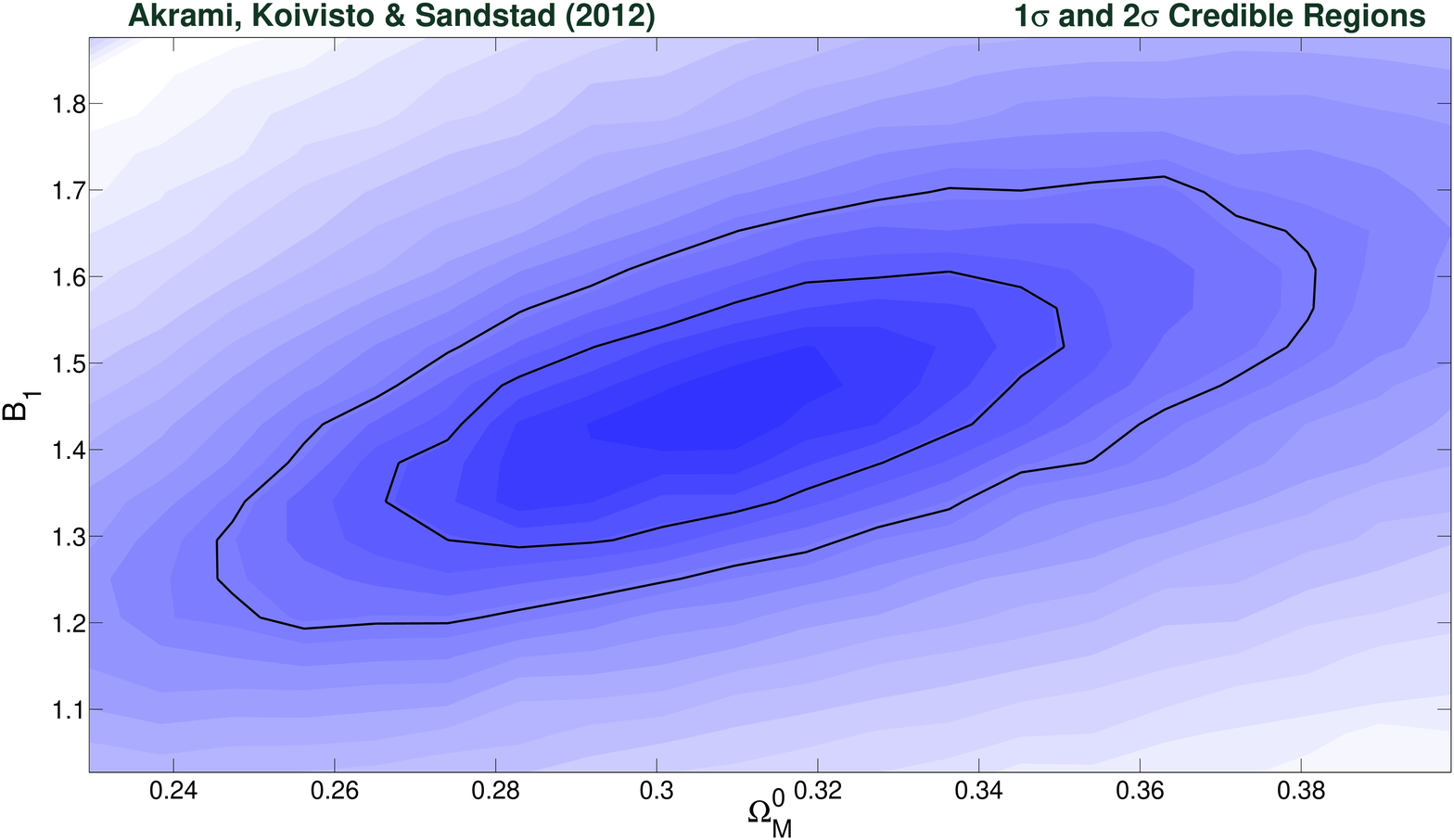}
\caption[aa]{Two-dimensional marginal posterior PDF for the bigravity model with only $B_1$ and $\Omega_m^0$ varying (other parameters are set to zero). The inner and outer contours represent $68\%$ ($1\sigma$) and $95\%$ ($2\sigma$) credible regions, respectively.}
\label{fig:BiGB1plot2D}
\end{center}
\end{figure}

\noindent
$\bullet$ \textbf{$B_1$ and $\Omega_m^0$ nonzero:} The one-dimensional posterior PDFs and mean likelihoods for this case are depicted in fig. \ref{fig:BiGB1plots1D}, where the left (right) panel corresponds to $\Omega_M^0$ ($B_1$). As for the $\Lambda$CDM case, the black solid curves (red dots) show the posterior PDFs (mean likelihoods), and the blue (green) vertical dashed lines indicate $68\%$ ($95\%$) credible intervals. The two-dimensional posterior PDF in the $\Omega_m^0$-$B_1$ plane is given in fig. \ref{fig:BiGB1plot2D}, where the inner (outer) contours correspond to $68\%$ ($95\%$) credible regions. The Gaussian shapes of the curves in fig. \ref{fig:BiGB1plots1D}, as well as the observation that the posteriors and the mean likelihoods perfectly match indicate that the model is well constrained. Here, we have used flat priors in our scans with prior ranges of $[0,1]$ and $[-5,5]$ for $\Omega_m^0$ and $B_1$, respectively. The impact of increasing the scanning range for $B_1$ on the plots is negligible, which is another indication that the analysis is statistically robust and the constraints upon the model parameters are reliable. 

The $\chi^2$ value at the best-fit point is $551.60$, with the corresponding $p$-value of $0.8355$. The $\chi^2$ in this case is slightly bigger than that of the $\Lambda$CDM model, but the $p$-value shows that the observed $\chi^2$, assuming that the ($B_1,\Omega_m^0$)-bigravity is the true model, is still less than $1\sigma$ away from the predicted value. We conclude that the model is perfectly consistent with observations.

With the mentioned chosen parameter ranges for the scans, the value of $log \mathcal{Z}$ we get is $-281.73$, which is smaller than that of the $\Lambda$CDM. An immediate conclusion might be that the $\Lambda$CDM is favored compared to our bigravity model, since the $\Delta\mathcal{Z}$ (evidence difference between the two models) is more than $3$. One should however be cautious in this, as first of all the Bayesian model selection procedure we described in section \ref{sec:statistics} is based on not only the evidence ratio (or log-evidence difference), but also on our priors about the models before observing the data (see eq. (\ref{eq:modelselection})). This includes all previous observational constraints on the models as well as the theoretical preferences. We do not consider any previous observational prejudices about the models here, but theoretically one may argue that the bigravity model is preferred to the $\Lambda$CDM model because the latter (although very well consistent with observations) strongly suffers from purely theoretical problems (the measured value for $\Omega_\Lambda$ is not technically natural from a particle physics point of view). In addition, the prior ranges one chooses for the parameters to scan over can change the calculated value of the evidence. We can reduce the range for $B_1$ in our scans and get better evidence values. For example, as we will see below, there are reasons why we expect $B_1$ to possess positive values for the considered sub-set of the full model to agree with observations; this reduces the prior range by a factor of two. Other possible theoretical reasons may reduce the prior range even further, giving rise to an even larger evidence. For these reasons we do not use the Bayesian model selection approach in comparing the bigravity model with the $\Lambda$CDM in the present paper. We will however use it later when we investigate how our fits may improve by allowing more parameters of the model to vary. This would tell us whether increasing the dimensionality of the parameter space could help the model to explain the data better or not (Occam's razor).

In order to better understand why the model gives a good fit to the data in this particular case, let us look at the dynamics of the scale factor ratio $y$ at the best-fit point of the model. Fig. \ref{fig:BiGB1plot_y} that we used as an example in section \ref{sec:evolution} actually corresponds to this particular sub-space of the parameter space. We observe from that plot that $y$ starts off from the asymptotic value of zero in the far past (high redshifts) and evolves toward another constant value of $y\approx5.8$ at the far future. We obtained this value from our statistical analysis of the model, but it can be understood also from analytical studies of the model, in particular by analyzing eq. (\ref{eq:quarticEquationForY}) at $z=-1$ (far in the future). The redshift dependence of $\Omega_m$ and $\Omega_\gamma$ tells us that the energy densities of both matter and radiation will vanish at $z=-1$, and consequently for the ($B_1,\Omega_m^0$)-model, the quartic equation (\ref{eq:quarticEquationForY}) reduces to the following asymptotic form and analytical solution for $y$:

\begin{equation}
  B_1y^2 - \frac{B_1}{3} = 0 \qquad \Longrightarrow \qquad y = \sqrt{\frac{1}{3}}\approx0.577.
\end{equation}

This value perfectly matches the one we obtained numerically. Moreover, plugging this asymptotic value into eq. (\ref{eq:firstFriedmanngefolding}) gives an equation for the Hubble parameter that resembles that of the $\Lambda$CDM cosmology with only a cosmological constant contribution to the energy density of the Universe (i.e. an asymptotically de Sitter universe). The energy density parameter in this case will become $B_1/\sqrt{3}$. If $y$ were constant over the entire evolution of the Universe, this would mean that the model could give a good fit to the observations with the value of $\sqrt{3}\Omega_\Lambda^{\Lambda CDM}\approx 1.2$, where $\Omega_\Lambda^{\Lambda CDM}$ is the measured value of $\Omega_\Lambda$ within the $\Lambda$CDM cosmology ($\sim 0.7$). We however know that $y$ is not constant and for all redshifts larger than zero (which our measurements actually correspond to) has a value smaller than its asymptotic limit. Therefore, in order for the model to compensate for the smallness of $y$, we expect a good fit to the data with a somewhat larger value for $B_1$ at the best-fit point. This is exactly what we observe as the result of our statistical analysis, i.e. a fit comparable to the $\Lambda$CDM with $B_1\approx1.44$.

Obviously, a more rigorous analysis is needed in order to analytically justify the obtained value for $B_1$ at the best-fit point (i.e. $\approx 1.44$), and to understand why this actually gives good fits for the entire time of the cosmic evolution. However, we can get a better understanding of this by considering eq. (\ref{eq:quarticEquationForY}) once again for our particular sub-set of the model where $B_0$, $B_2$, $B_3$ and $B_4$ are set to zero (and the contributions from radiation and curvature are neglected). The equation in this case becomes quadratic in $y$ with the following solution:

\begin{equation}\label{eq:yWithJustB1}
  y = \frac{-\Omega_{m} \pm \sqrt{{\Omega_m}^2 + \frac{4}{3}{B_1}^2}}{2B_1}.
\end{equation}

The model is expected to give a relatively better fit to the data if instead of $z=-1$ (far in the future), it matches the $\Lambda$CDM model at $z=0$ (the present time). Since the effective $\Omega_\Lambda$ in this case is $B_1y$ (from eq. (\ref{eq:firstFriedmanngefolding})), eq. (\ref{eq:yWithJustB1}) can be written in the following form at $z=0$:

\begin{equation}\label{eq:yWithJustB1_2}
 2\Omega_\Lambda=-\Omega_m^0 \pm \sqrt{{\Omega_m^0}^2 + \frac{4}{3}{B_1}^2} \qquad \Longrightarrow \qquad B_1=\pm\sqrt{\frac{3}{4}[{(1+\Omega_\Lambda)}^2-{\Omega_m^0}^2]}.
\end{equation}

First of all, we must chose a positive value for $B_1$. The reason is the following: As we saw, the effective $\Omega_\Lambda$ has the value of $B_1y$ with a positive value favored by observations (to yield cosmic acceleration). $y$ is the ratio of the two scale factors of the theory and has to be positive. We therefore observe that positive $B_1$ is favored by data. In order to compute the value of $B_1$, we plug the values of $\Omega_\Lambda$ and $\Omega_m^0$ preferred by the data in the case of the $\Lambda$CDM model (i.e. $\Omega_\Lambda\approx0.7$ and $\Omega_m^0\approx0.3$) into eq. (\ref{eq:yWithJustB1_2}). This gives $B_1\approx 1.45$ which is very close to the value we have obtained numerically.

Furthermore, by increasing the redshift (moving into the past), $\Omega_m$ increases and eq. (\ref{eq:yWithJustB1}) shows that $y$ always remains positive and real, and it smoothly transitions into $y = 0$ in the extreme past. This means that the energy contributed by $y$ to the dynamics of the Universe becomes smaller and smaller compared to the matter contribution at higher redshifts and the Universe becomes matter-dominated. The model therefore has a well-behaved solution and resembles the $\Lambda$CDM model at larger redshifts. This is another reason why the model yields good fits to the data.

Finally, let us mention one more interesting implication of our results. We have seen that assuming only one of the $B$s to vary (i.e. $B_1$), we get a robust constraint on its value (see the right panel of fig. \ref{fig:BiGB1plots1D}). We have already given the value at the best-fit point, i.e the highest-likelihood value ($B_1\approx 1.44$). Marginalizing the full posterior PDF over the unwanted parameters ($\Omega_m^0$ in this case), gives the Bayesian preferred value for $B_1$ and uncertainties upon it: $B_1=1.45\pm0.25$. In the absence of other $B$s (or $\beta$s), eq. (\ref{eq:ActionOriginal}) indicates that the graviton mass is purely determined by $B_1$. Using the definition of $B_1$ (i.e. $m^2\beta_1/H_0^2$) and the obvious choice of $\beta_1=1$ (there is in fact no need to define $\beta_1$ as an independent parameter in this case; it can be absorbed into $m^2$), we can conclude that $m^2=(1.45\pm0.25) H_0^2$, where $H_0=71\pm2.5$ km s$^{-1}$ Mpc$^{-1}$. Obviously, this constraint on the graviton mass is true only if our particular construction of the massive gravity theory is correct, no explicit cosmological term exists and all $\beta$ parameters (except $\beta_1$) are zero. These are strong assumptions that need to be tested observationally or justified theoretically.
\\

\noindent
$\bullet$ \textbf{$B_2$ and $\Omega_m^0$ nonzero:} The results of our statistical analysis of the model in this case shows that the best-fit $\chi^2$ has the value of $894.0$, which corresponds to a $p$-value smaller than $0.0001$. This clearly means that the model is strongly excluded by observations (with more than $5\sigma$ confidence). This is confirmed by the very low value obtained for the Bayesian evidence: $log\mathcal{Z}=-450.25$ (compare this with the $\Lambda$CDM or ($B_1,\Omega_m^0$) models). Let us see if we can analytically explain why one does not get a good fit in this case.

As for the ($B_1,\Omega_m^0$) case, we use eq. (\ref{eq:quarticEquationForY}) to study how $y$ behaves in the ($B_2,\Omega_m^0$) case. The equation becomes a cubic one with the following form:

\begin{equation}\label{eq:yWithJustB2}
  y\left(B_2\left(y^2 - 1\right) + \Omega_m\right) = 0.
\end{equation}

The trivial solution $y = 0$ is clearly not interesting, and will not yield a good fit, as it does not produce any term in eq. (\ref{eq:firstFriedmanngefolding}) that mimics the cosmological constant (which is essential for giving a good fit). The model in this case is nothing but the CDM model (with $\Omega_\Lambda=0$). The acceptable solution for $y$ is therefore:

\begin{equation}
y = \sqrt{1 - \frac{\Omega_m}{B_2}}.\label{eq:yforB2}
\end{equation}

We again chose positive $y$ in order to have a physically meaningful model. We observe from eq. (\ref{eq:firstFriedmanngefolding}) that at late times we have an effective cosmological term $\Omega_\Lambda=B_2y^2$. To have any hope of obtaining good fits this term must be positive. This implies that $B_2$ must be positive.

We immediately see from eq. (\ref{eq:yforB2}) that as $\Omega_m$ gets bigger than $B_2$, $y$ turns imaginary, which is unphysical. Since $\Omega_m$ scales as $(1+z)^3$ with redshift, we expect this to always happen for redshifts bigger than some $z^*$, where

\begin{equation}
 z^*=(\frac{B_2}{\Omega_m^0})^{\frac{1}{3}}-1.
\end{equation}

In case a value of $B_2$ could be chosen such that $z^*$ becomes larger than the maximum redshift we have considered in our analysis (i.e. $z\approx1100$) we would still be able to get a good fit, even though the model would need to be modified at redshifts higher than $z^*$. We can check this by trying to find an estimate for $B_2$ at the best-fit point. As in the $B_1$ case, this can be done by assuming that $B_2y^2 = \Omega_\Lambda^{\Lambda CDM}$ today. Eq. (\ref{eq:yWithJustB2}) then reads

\begin{equation}
  B_2 = \Omega_m^0 + \Omega_\Lambda^{\Lambda CDM}=1.
\end{equation}

Therefore, as soon as $\Omega_m > B_2=1$ the model becomes unphysical. Assuming a value of $\sim 0.3$ for $\Omega_m^0$, this corresponds to $z^*\approx0.5$, which is well within the range of the cosmological data we have used. We therefore conclude that the ($B_2,\Omega_m^0$)-model cannot yield good fits.
\\

\noindent
$\bullet$ \textbf{$B_3$ and $\Omega_m^0$ nonzero:} As for the previous case, our statistical analysis shows that the model is excluded observationally. The reason is that the best-fit $\chi^2$ in this case is $1700.5$ ($p$-value $<0.0001$), which is even larger than that of the ($B_2,\Omega_m^0$)-model. The Bayesian log-evidence is $-850.26$ confirming that the model does not fit observations. Again, in order to understand the reason analytically, we proceed in the same way as in the previous cases, starting from eq. (\ref{eq:quarticEquationForY}). The equation becomes:

\begin{equation}
  y\left(\frac{B_3}{3}y^3 - B_3 y + \Omega_m\right) = 0.\label{eq:yWithJustB3}
\end{equation}

Again, it is the nontrivial cubic equation in brackets that is a possible solution. To see how this behaves we consider the discriminant of the equation:

\begin{equation}
  \Delta = \frac{4}{3}B_3^2\left(B_3^2 - \frac{9}{4}\Omega_m^2\right).
\end{equation}

When $\Delta$ becomes negative, as it will when $\frac{3}{2}\Omega_m>B_3$, only one root of the cubic equation will be real. Hence, it is natural to only consider this solution, which is:

\begin{equation}
  y = -\frac{1}{B_3}\sqrt[3]{\frac{3}{2}B_3^2}\left[\sqrt[3]{\Omega_m + \sqrt{\Omega_m^2 - \frac{4}{9}B_3^2}} + \sqrt[3]{\Omega_m - \sqrt{\Omega_m^2 - \frac{4}{9}B_3^2}}\right].\label{eq:yforB3}
\end{equation}

The effective cosmological constant in this case is (see eq. (\ref{eq:firstFriedmanngefolding})) $\Omega_\Lambda=B_3 y^3/3$. A good fit to the data in this case means a positive cosmological constant, which given that $y$ must be positive, implies a positive value for $B_3$. Eq. (\ref{eq:yforB3}) is however not consistent with both $B_3$ and $y$ being positive. This means that the model cannot fit the data if $\Delta$ becomes negative within the range of the data. The redshift $z^*$ beyond which this happens can be estimated in a similar manner as for the $B_2$ case:

\begin{equation}
 z^*=\left(\frac{2B_3}{3\Omega_m^0}\right)^{\frac{1}{3}}-1.\label{eq:z*B3}
\end{equation}

At the present time, eq. (\ref{eq:yWithJustB3}) reads

 \begin{equation}
  B_3y = \Omega_m^0 + \Omega_\Lambda^{\Lambda CDM}=1 \qquad \Longrightarrow \qquad y=\frac{1}{B_3}.
\end{equation}
 
Plugging this value for $y$ in terms of $B_3$ into the relation $\Omega_\Lambda^{\Lambda CDM}=B_3 y^3/3$, we get

 \begin{equation}
B_3\approx (3\Omega_\Lambda^{\Lambda CDM})^{-\frac{1}{2}}\label{eq:B_3approx}
\end{equation}
as an approximation for $B_3$ at the best-fit point. Assuming a value of $\approx 0.7$ for $\Omega_\Lambda^{\Lambda CDM}$, eqs. (\ref{eq:B_3approx}) and (\ref{eq:z*B3}) give $z^*\approx 0.15$, which is clearly within the range of the data. This implies that the model cannot give good fits to the data and is therefore excluded.

\subsubsection{Three-parameter sub-models: $\Omega_m$ and two $B$s}

\begin{figure}[t]
\begin{center}
\includegraphics[width=0.49\textwidth]{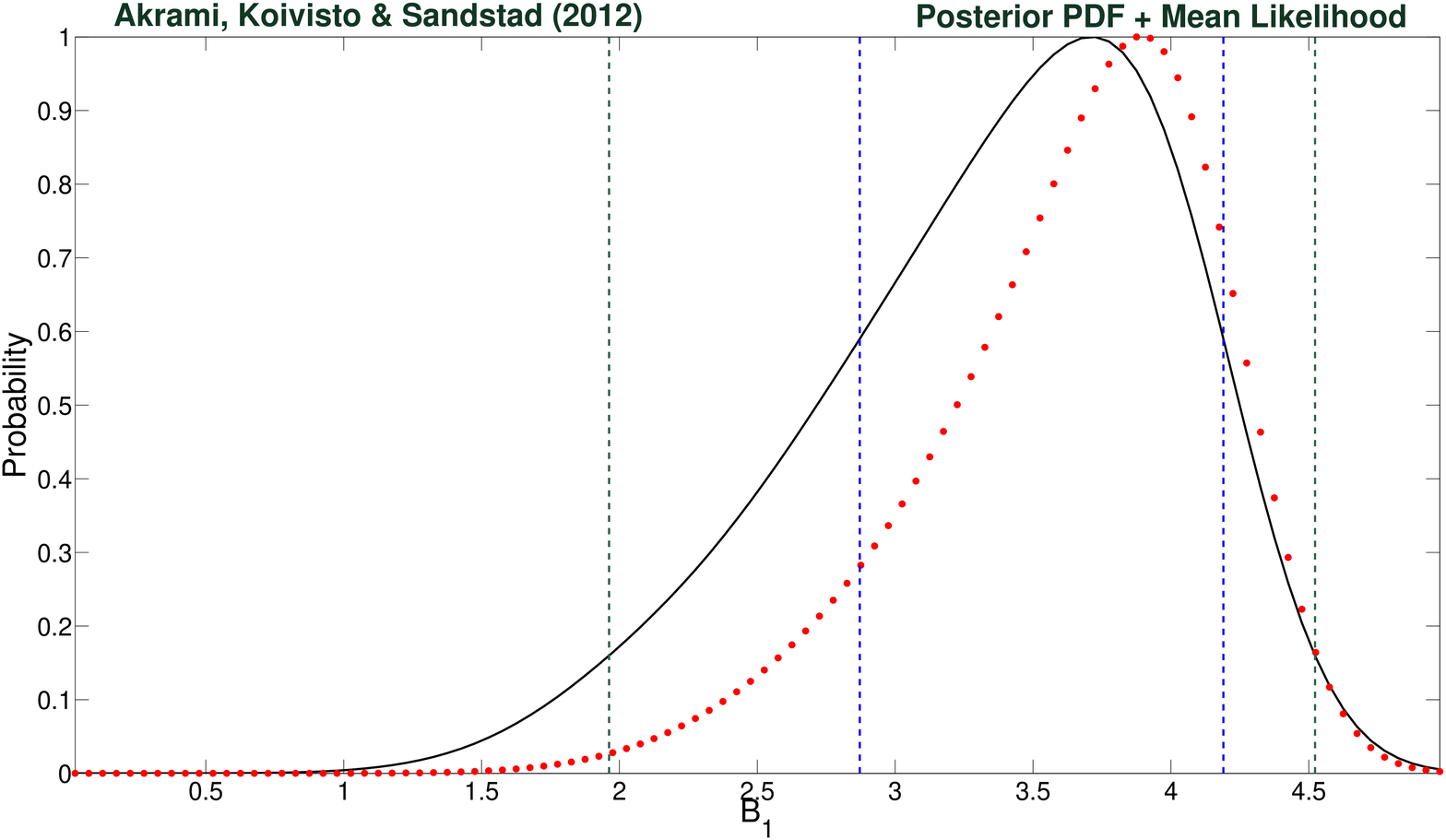}
\includegraphics[width=0.49\textwidth]{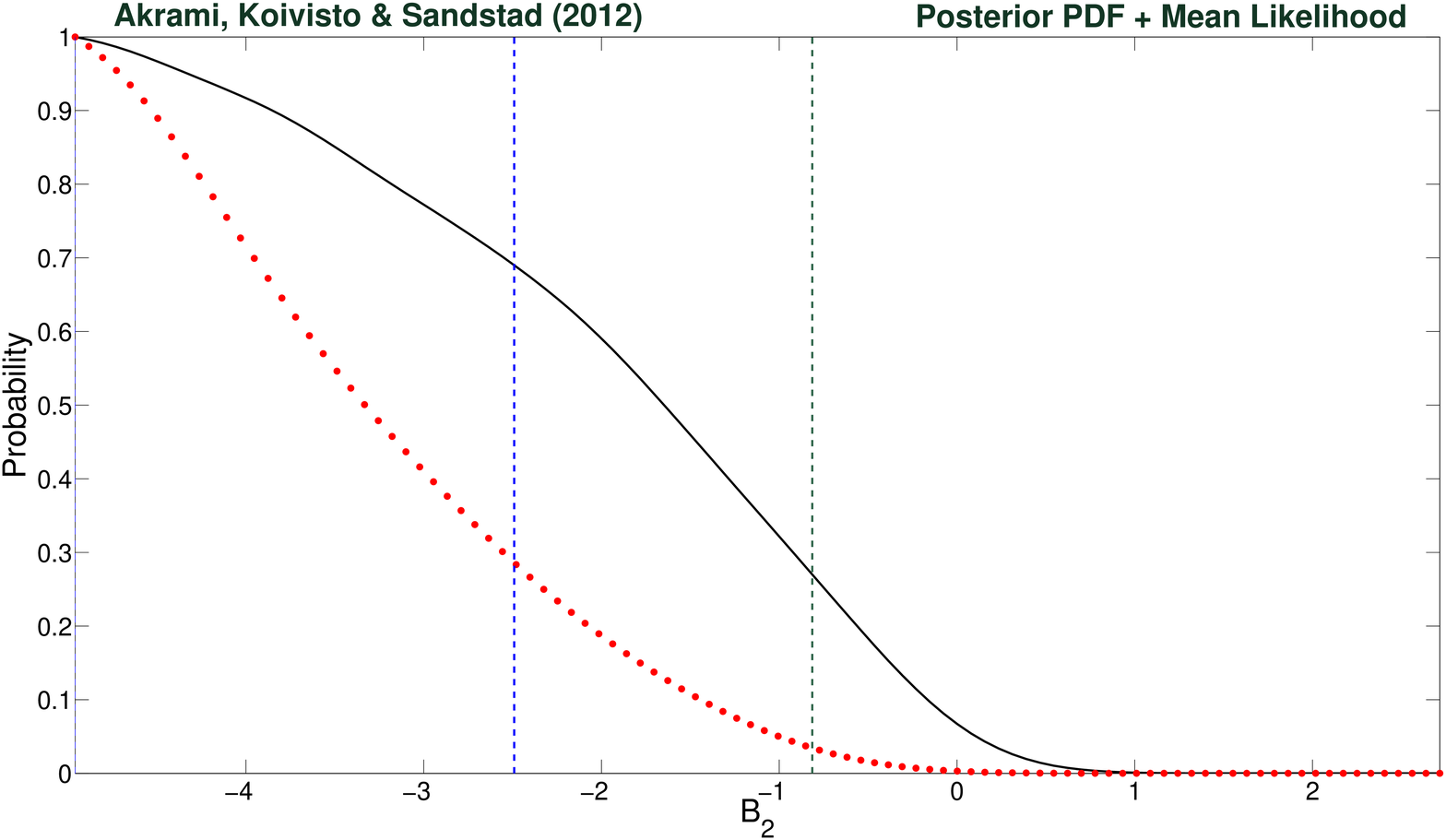}
\includegraphics[width=0.49\textwidth]{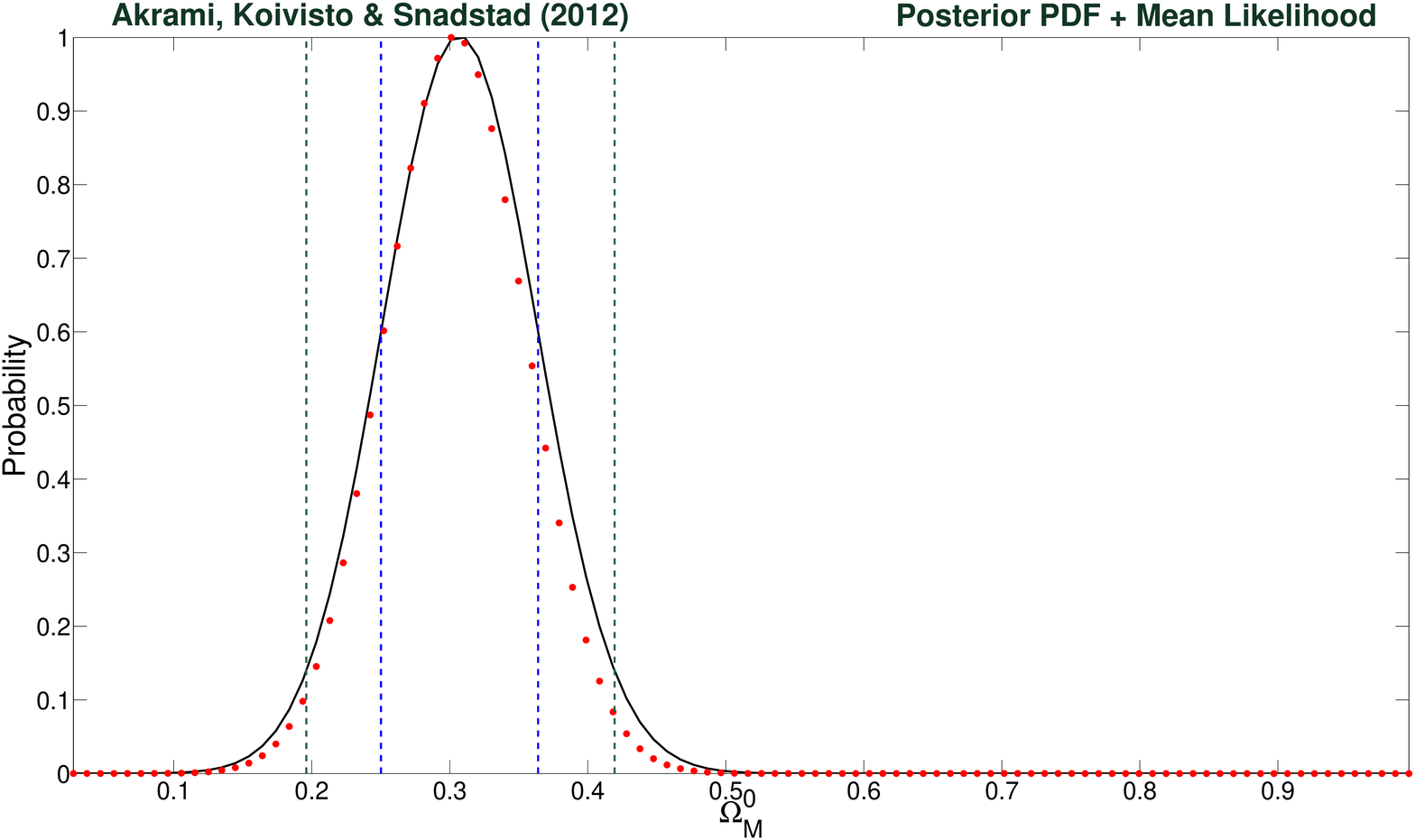}
\caption[aa]{As in figs. \ref{fig:LCDMplots} and \ref{fig:BiGB1plots1D}, but for the bigravity model with only $B_1$, $B_2$ and $\Omega_m^0$ varying (other parameters are set to zero). Here, (flat) prior ranges for both $B_1$ and $B_2$ are $[-5,5]$, and for $\Omega_m^0$ is $[0,1]$.}
\label{fig:BiGB1B2plots1D_r5}
\end{center}
\end{figure}

Now that we have some hold on what the different mass terms can do, it is interesting to study how two terms may interact to give even better fits. This is especially important since the interactions are non-trivial. As we will see, by adding more and more free parameters the model shows some degeneracy patterns that lead to a behavior which makes parameter fits elusive, while at the same time reveals important facts about the physics of the model.

We start with the ($B_1,B_2,\Omega_m^0$)-model where all $B$s, except $B_1$ and $B_2$, are fixed to zero and flat priors are imposed on all free parameters. Scanning over the parameter space with prior ranges $[-5,5]$ for both $B_1$ and $B_2$, and $[0,1]$ for $\Omega_m^0$ results in the best-fit $\chi^2$ of $546.52$ corresponding to the $p$-value of $0.8646$; the log-evidence is $-279.77$. The improvement in both $p$-value and evidence, compared to the ($B_1,\Omega_m^0$)-model, indicates that opening up a new dimension in the parameter space helps the theory fit the data better. The improvement in the evidence is of particular interest since adding more free parameters to a model can in general lower the evidence, as the prior volume increases, unless this effect is compensated by significant increase in the number of high-likelihood points.

\begin{figure}[t]
\begin{center}
\includegraphics[width=0.49\textwidth]{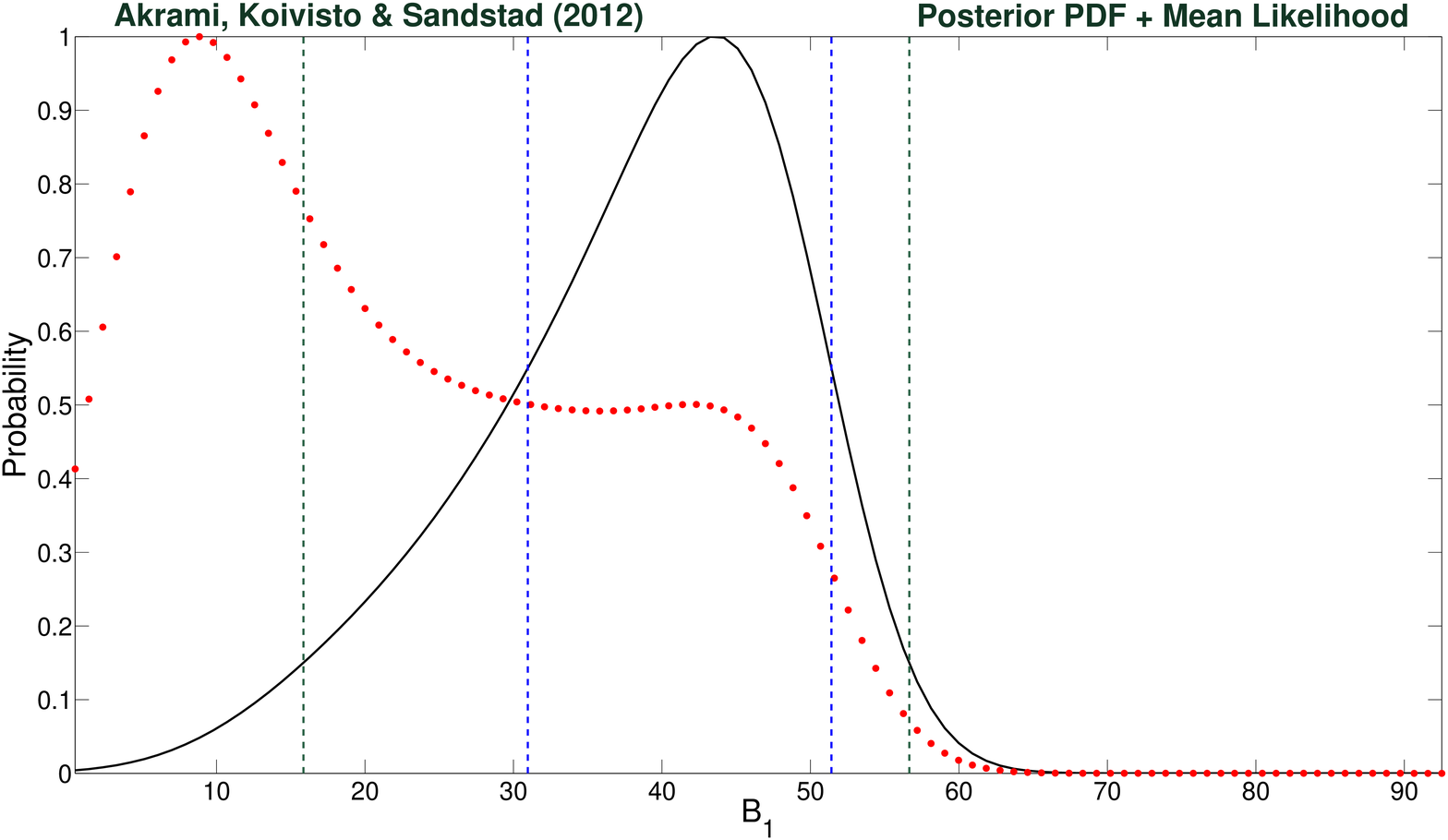}
\includegraphics[width=0.49\textwidth]{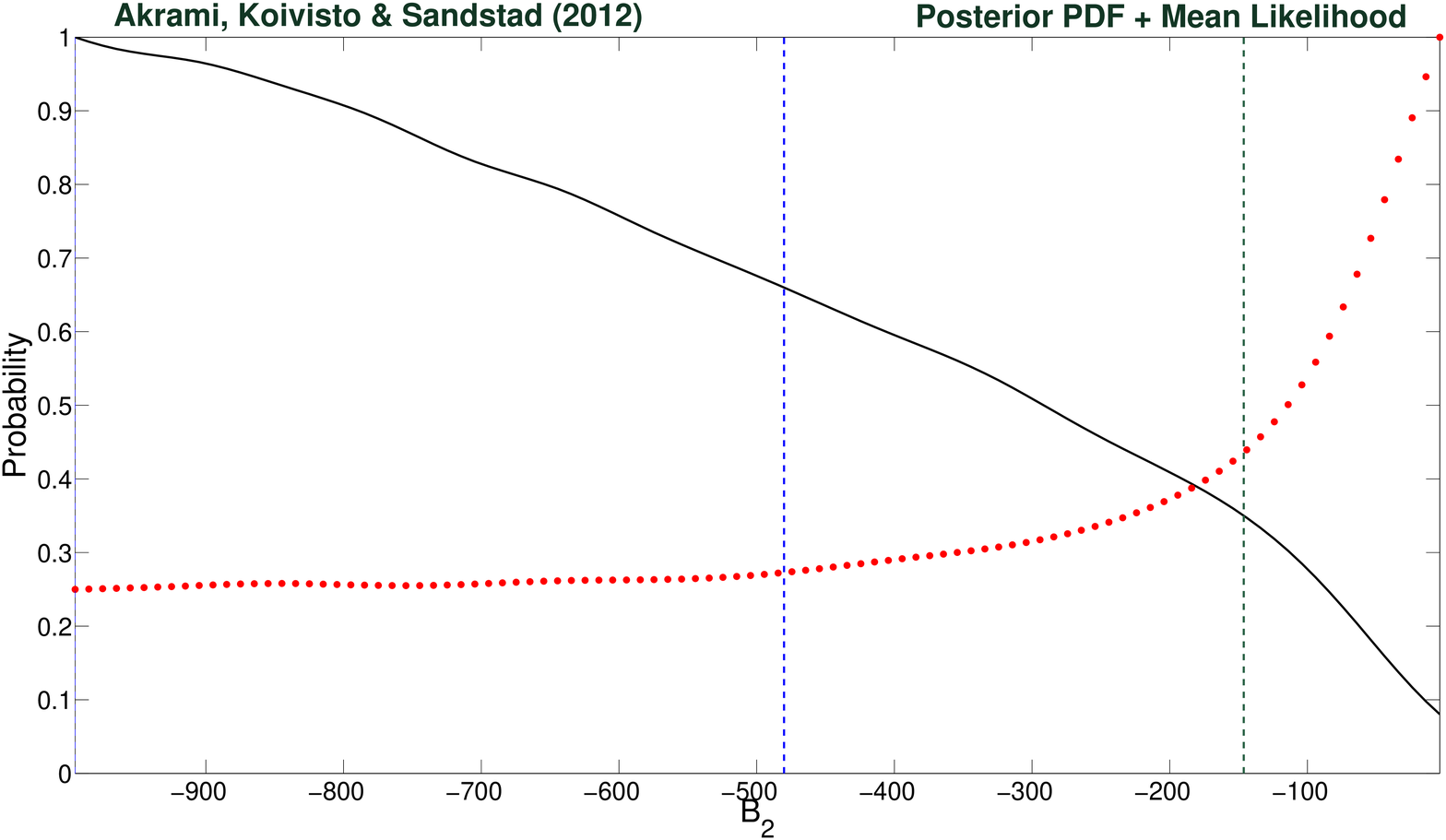}
\includegraphics[width=0.49\textwidth]{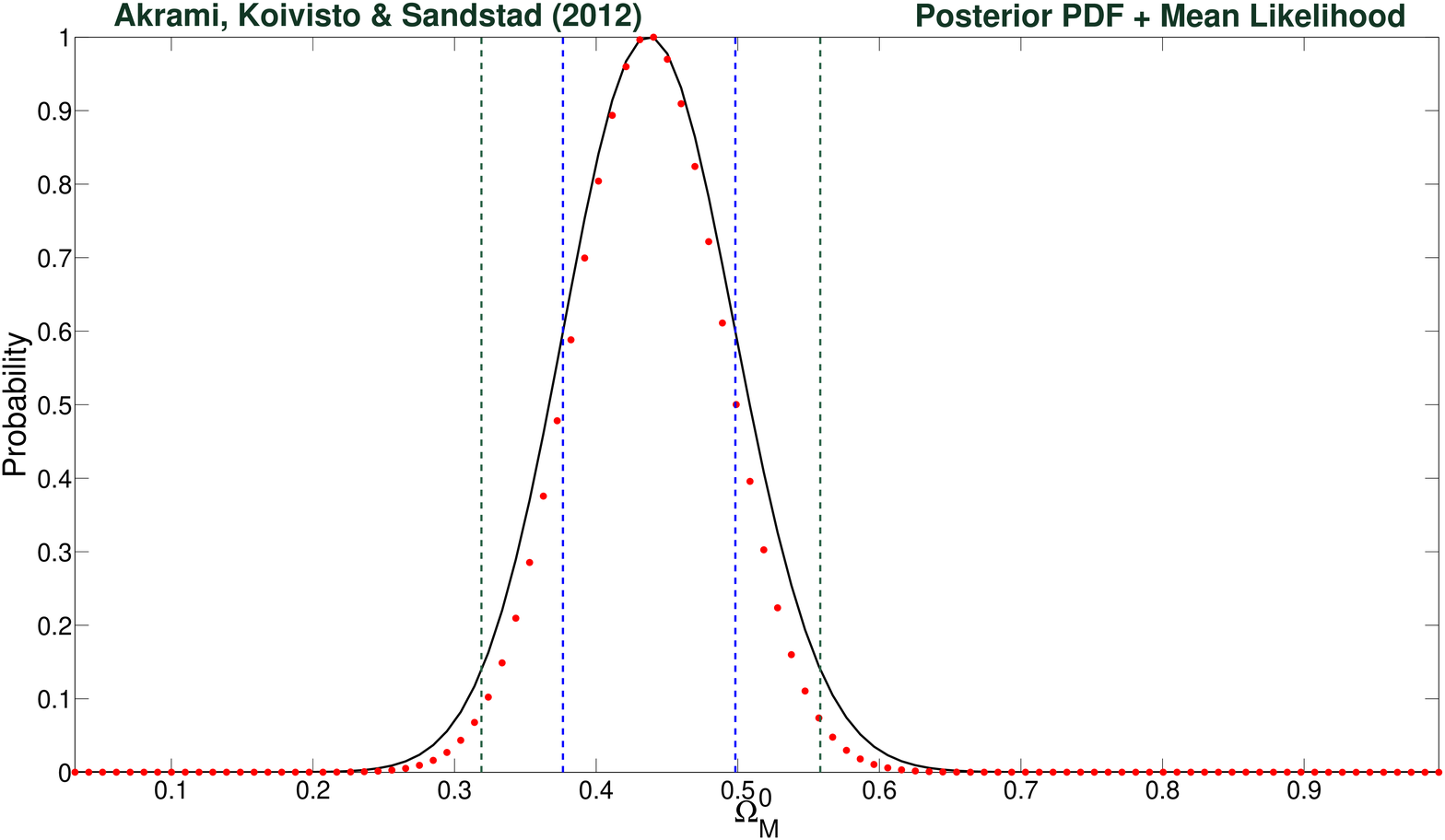}
\caption[aa]{As in fig. \ref{fig:BiGB1B2plots1D_r5}, but for larger prior ranges for $B_1$ and $B_2$ ($[-1000,1000]$). The prior range for $\Omega_m^0$ is kept the same ($[0,1]$).}
\label{fig:BiGB1B2plots1D_r1000}
\end{center}
\end{figure}

However, the one-dimensional posterior PDFs and mean likelihoods for the model parameters in this case tell us that the model is not well constrained by the data (see fig. \ref{fig:BiGB1B2plots1D_r5}). $\Omega_m^0$ seems to be well constrained. Even though $B_1$ also seems to be constrained, the mismatch between the posterior PDF and mean-likelihood curves, as well as their non-Gaussian shapes are a sign that the constraint may not be trusted. For $B_2$, both curves show that negative values are preferred, although they do not match very well. One may think that this is because the prior ranges, at least for $B_2$, have not been chosen large enough and by increasing them Gaussian curves around preferred values of the parameters would show up. Fig. \ref{fig:BiGB1B2plots1D_r1000} however demonstrates that this is not the case. Here, we have enlarged the prior ranges for both $B_1$ and $B_2$ to $[-1000,1000]$ and we still see a similar pattern for the $B_2$ posterior PDF, now extended to even more negative values. The posterior PDF and mean-likelihood curves for both $B$s now show completely different patterns, an indication that the full posterior PDF of the model is not a Gaussian distribution; in addition it indicates that the posterior is not separable with respect to the $B_1$ and $B_2$ sub-spaces \cite{Lewis:2002ah}. The differences in the $B_1$ curves indicate non-Gaussianity, which can be due to the fact that at least one of the marginalized parameters ($B_2$ in this case) is skewing the distribution in its direction. If we fixed $B_2$ at its maximum likelihood value, the marginalized posterior distribution would change in the direction of its mean-likelihood curve, a feature that we observed in the ($B_1,\Omega_m^0$)-model. The different posterior and mean-likelihood curves for $B_2$ similarly show that the distribution is skewed in the direction of another parameter of the model ($B_1$). In other words, the model parameters are correlated.

\begin{figure}[t]
\begin{center}
\includegraphics[width=0.49\textwidth]{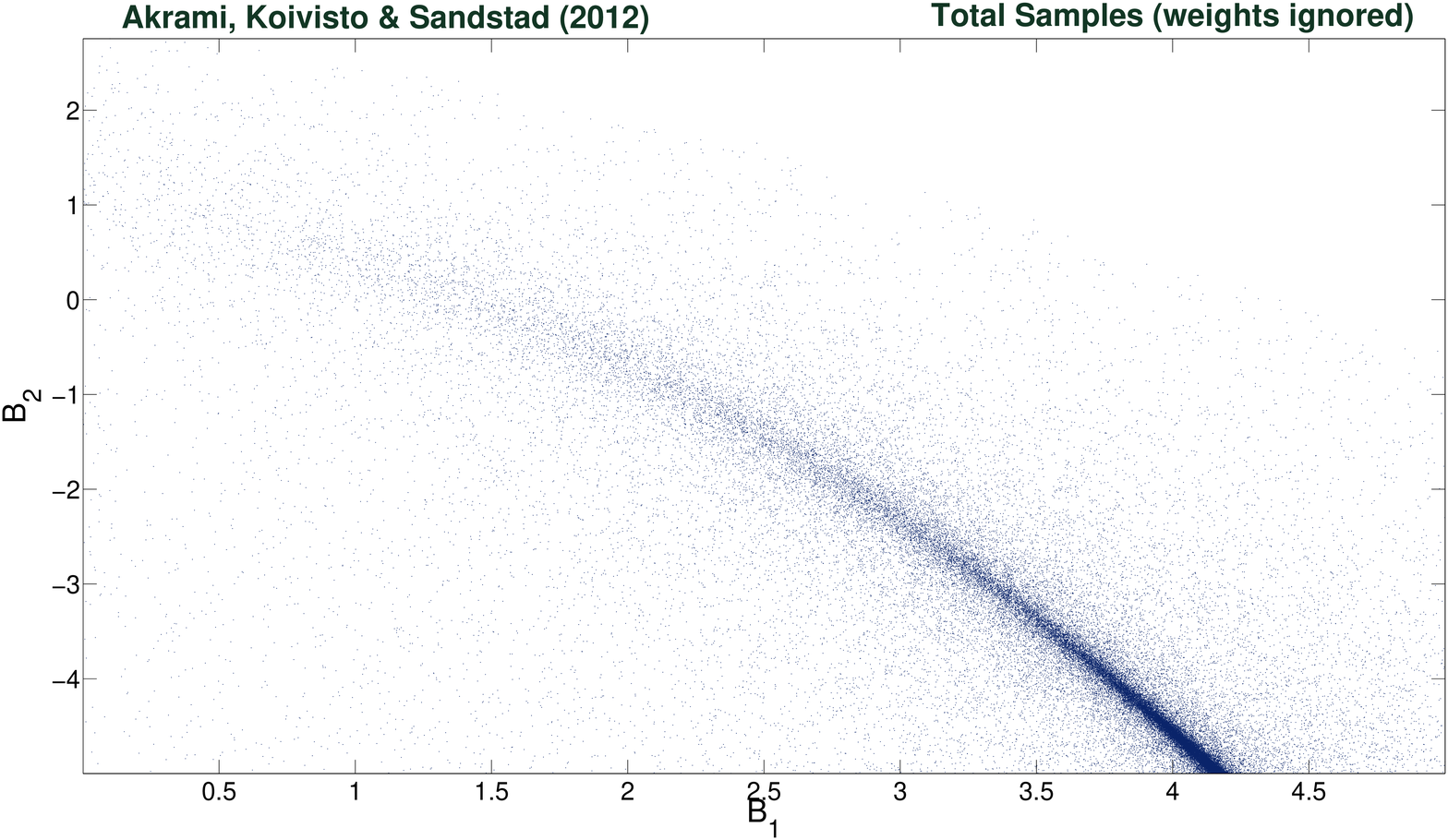}
\includegraphics[width=0.49\textwidth]{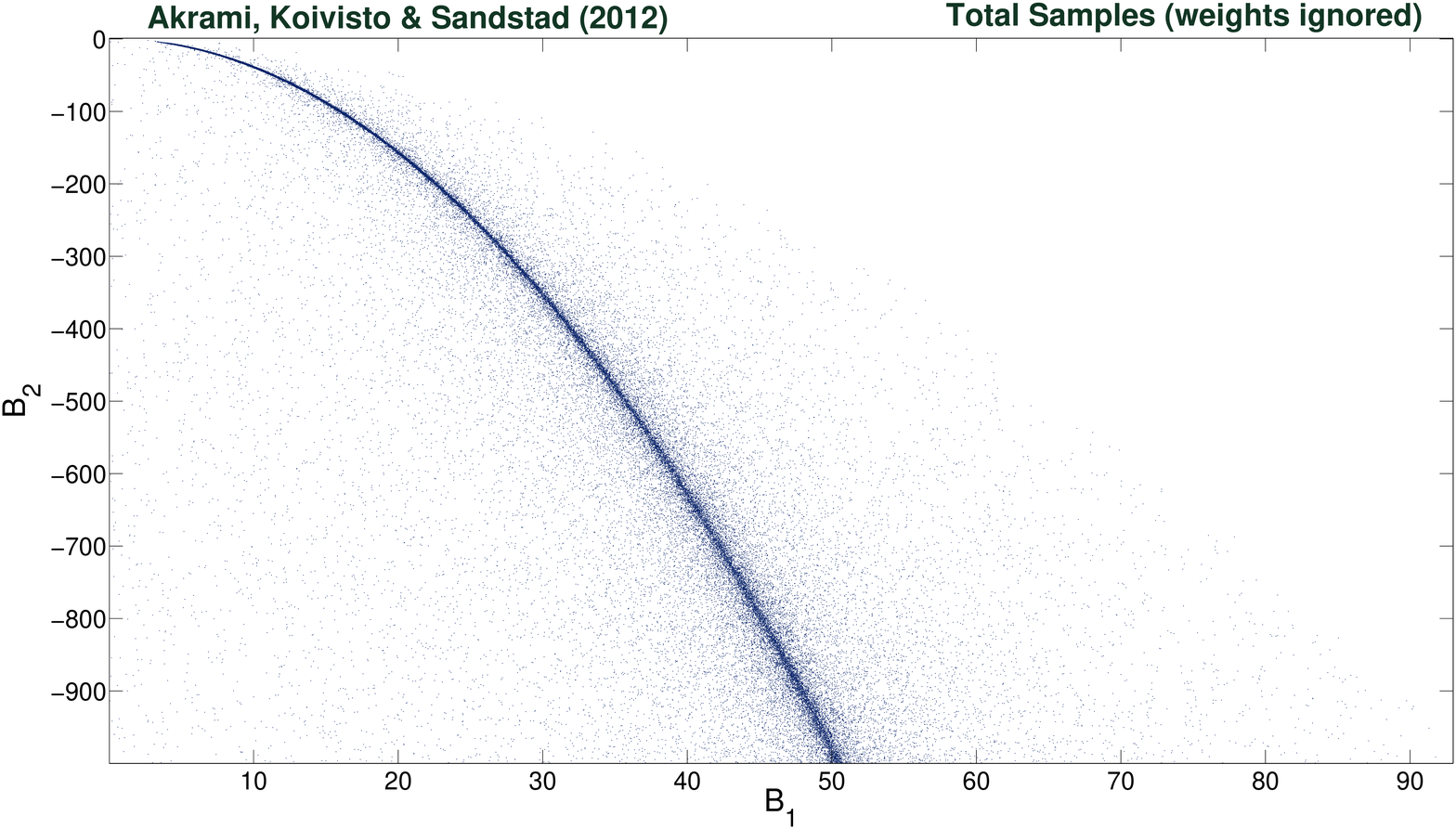}
\caption[aa]{Two-dimensional unweighted distributions of the sample points in the $B_1$-$B_2$ plane for the bigravity model with only $B_1$, $B_2$ and $\Omega_m^0$ varying (other parameters are set to zero) and with different choices of prior ranges. These illustrate how the two parameters are strongly correlated.}
\label{fig:BiGB1B2plots2D}
\end{center}
\end{figure}

In order to see the correlation between $B_1$ and $B_2$ more clearly, we show in fig. \ref{fig:BiGB1B2plots2D} the distribution of the obtained sample points in the $B_1$-$B_2$ plane, for both scans with $[-5,5]$ and $[-1000,1000]$ prior ranges; the strong correlation can be observed from these plots. This simply means that in order for the model to yield good fits to the data, the point ($B_1,B_2$) must be located on the narrow correlation region given by fig. \ref{fig:BiGB1B2plots2D}. This is a clear example of the fact that one-dimensional marginalized curves do not always reflect all the interesting properties of model parameters, in particular when non-trivial correlations exist between the parameters.

The parameter $\Omega_m^0$ is much more robust under the change of the priors, with Gaussian-like shapes for both marginalized posterior and mean-likelihood curves; the curves also look very much the same. A small correlation between $\Omega_m^0$ and the $B$s however makes the distributions slightly move if we change the prior ranges, as we observe from figs. \ref{fig:BiGB1B2plots1D_r5} and \ref{fig:BiGB1B2plots1D_r1000}.
 
\begin{figure}[t]
\begin{center}
\includegraphics[width=0.49\textwidth]{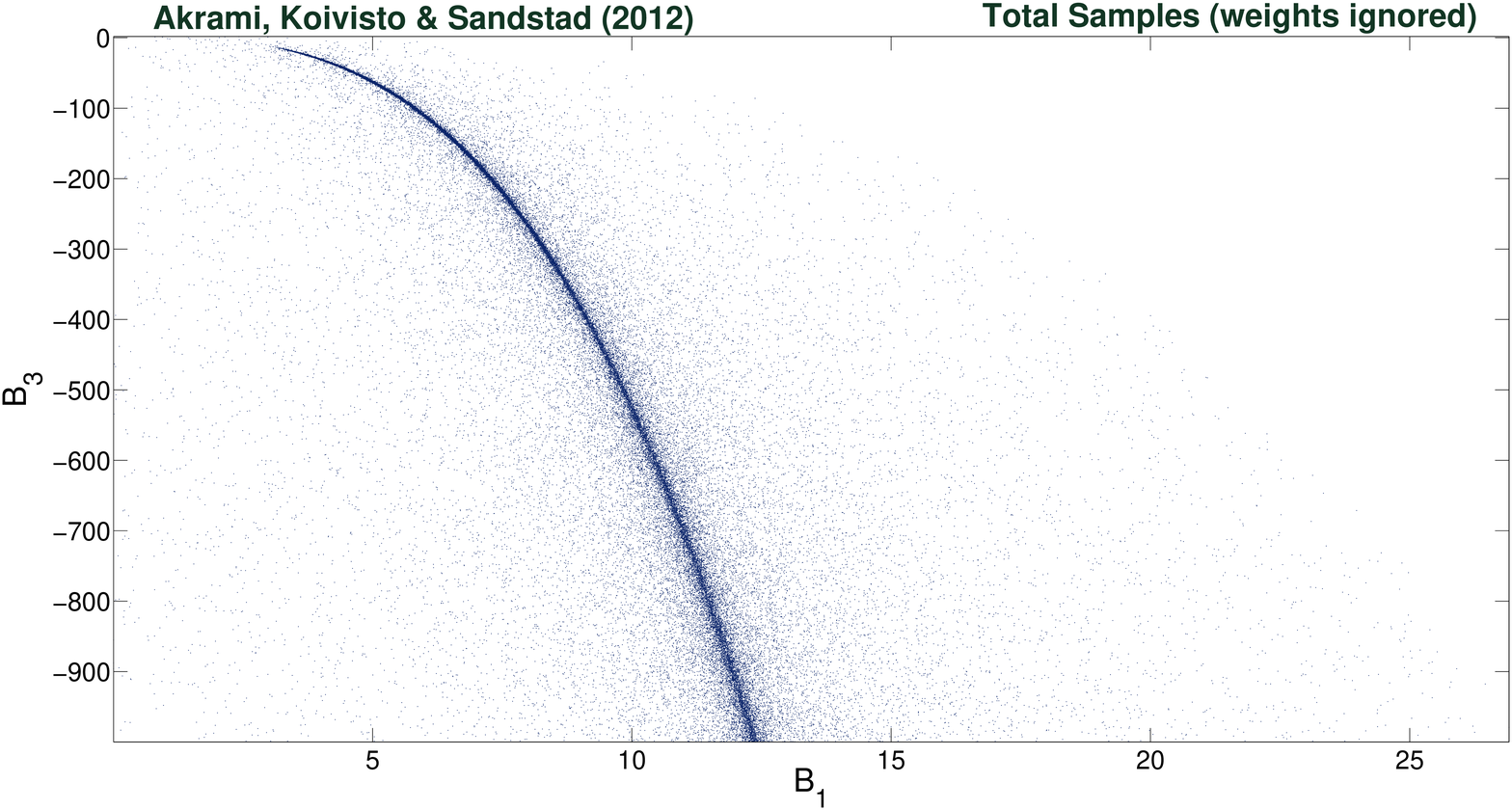}
\includegraphics[width=0.49\textwidth]{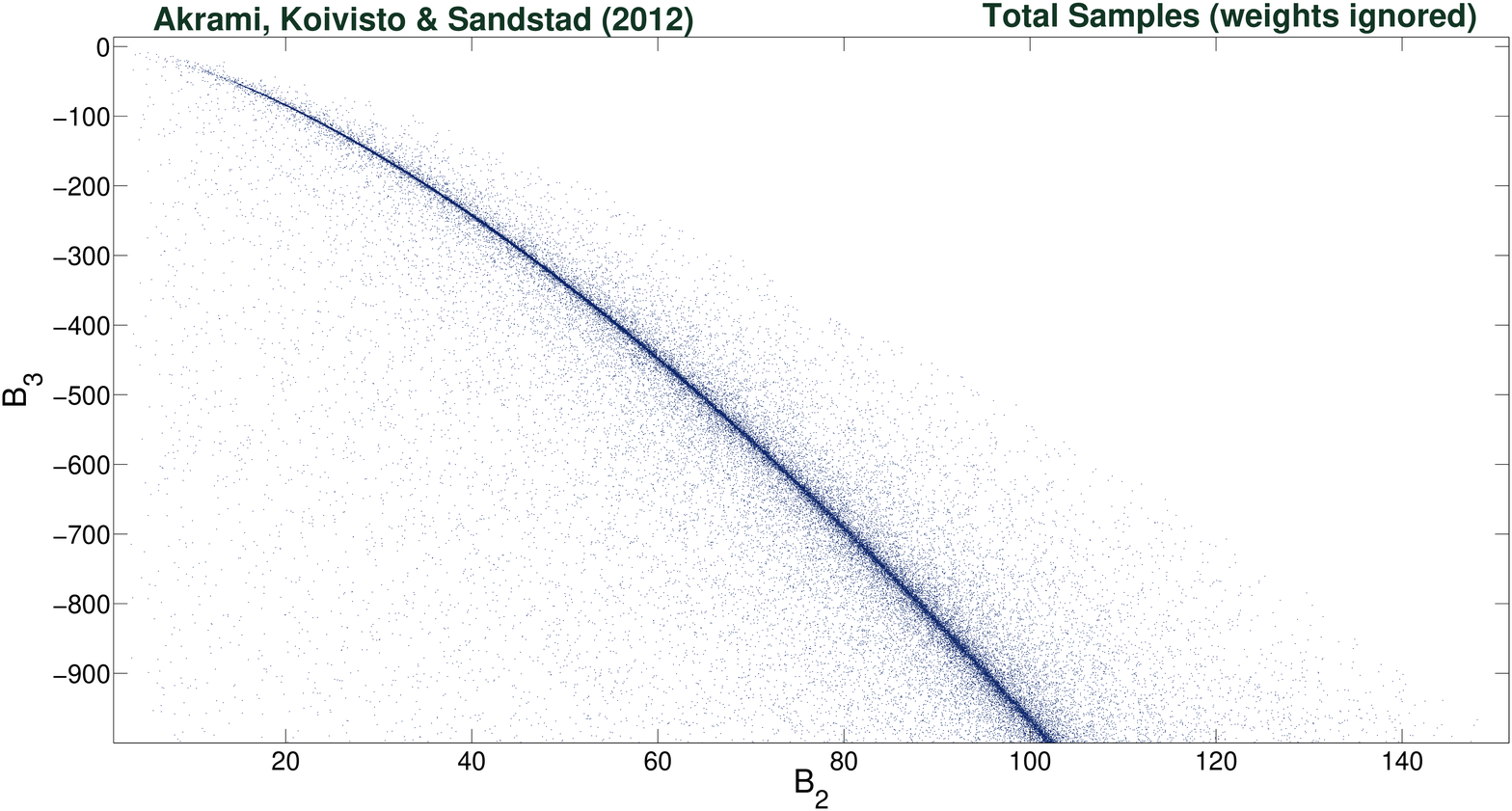}
\caption[aa]{{\it Left:} Two-dimensional unweighted distribution of the sample points in the $B_1$-$B_3$ plane for the bigravity model with only $B_1$, $B_3$ and $\Omega_m^0$ varying. {\it Right:} Similar distribution in the $B_2$-$B_3$ plane for the model with only $B_2$, $B_3$ and $\Omega_m^0$ varying.}
\label{fig:BiGB1B3_B2B3plots2D}
\end{center}
\end{figure}

For the ($B_1,B_3,\Omega_m^0$)-model, similar conclusions can be obtained. The best-fit $\chi^2$ for the $B_1$ and $B_3$ prior ranges of $[-5,5]$ (flat priors) is $542.82$, with the $p$-value of $0.8878$; the log-evidence is $-280.10$. Note that in terms of the $p$-value, the model is fit to the data even better than the $\Lambda$CDM (with the $p$-value of $0.8709$). In addition, the improvement in the evidence compared to the ($B_1,\Omega_m^0$), even though it has more free parameters, is an indication that the ($B_1,B_3,\Omega_m^0$)-model is observationally favored to the ($B_1,\Omega_m^0$)-model. In addition, our statistical analysis shows that, as in the previous case of the ($B_1,B_2,\Omega_m^0$)-model, the parameters of the model are again correlated in a very similar way. Therefore, for brevity reasons, here we only show the two-dimensional distribution of the sample points in the $B_1$-$B_3$ plane for prior ranges of $[-1000,1000]$ (left panel of fig. \ref{fig:BiGB1B3_B2B3plots2D}), where a strong correlation between the two parameters can be seen.

For the ($B_2,B_3,\Omega_m^0$)-model we get the best-fit $\chi^2$ of $548.04$ ($p$-value of $0.8543$) and the log-evidence of $-280.91$ (with prior ranges of $[-5,5]$); these show a very good fit to the data. This is a particularly interesting case as our analysis showed (see the previous section) that the ($B_2,\Omega_m^0$) and ($B_3,\Omega_m^0$) models did not fit the data individually and were therefore excluded. The parameters of the model in this case are also correlated. Fig. \ref{fig:BiGB1B3_B2B3plots2D} (right panel) illustrates this correlation in the $B_2$-$B_3$ plane.

Let us now try to analytically understand the correlation patterns observed in figs. \ref{fig:BiGB1B2plots2D} and \ref{fig:BiGB1B3_B2B3plots2D} for the models ($B_1,B_2,\Omega_m^0$), ($B_1,B_3,\Omega_m^0$) and ($B_2,B_3,\Omega_m^0$). The complexity of our bimetric model increases considerably when we increase the number of model parameters and a thorough analytical consideration of each of the cases (as we did in the previous section for the two-parameter models) is beyond the scope of this work. Here, we therefore only consider one of them (the ($B_1,B_2,\Omega_m^0$)-model) in order to get an understanding of what causes the parameters to be strongly correlated with those specific correlation patterns. The other cases can be studied in similar ways.

As for the discussions of the previous section, the key expression to consider here is eq. (\ref{eq:quarticEquationForY}) for the dynamical variable $y$. This equation for the ($B_1,B_2,\Omega_m^0$)-model reduces to the following one:

 \begin{equation}\label{eq:yWithB1B2}
   B_2y^3 + B_1y^2 +\left(\Omega_m - B_2\right)y - \frac{B_1}{3} = 0.
 \end{equation}
 
 This is a cubic equation for which we again consider the discriminant. This turns out to be:
 
 \begin{equation}
   \Delta = - 8B_1^2B_2\Omega_m + 4B_1^2B_2^2 + \frac{4}{3}B_1^4 + B_1^2\Omega_m^2 - 4B_2\Omega_m^3 + 12B_2^2\Omega_m^2 - 12B_2^3\Omega_m  + 4B_2^4.
 \end{equation}

We see that if $B_2$ is negative, the discriminant will be positive, and eq. (\ref{eq:yWithB1B2}) will always have three roots for $y$. In the opposite case (i.e. positive $B_2$), we will always be able to move back to a time when the discriminant was negative and there was only one root. In this case, consideration of the real root reveals that it will at some point in the past (high redshifts) become negative if $B_1$ is negative, whereas this will not happen if the theory has positive $B_1$ in addition to the positive $B_2$. In addition, looking at the equation for the Hubble parameter (\ref{eq:firstFriedmanngefolding}), we see that to get acceleration (required for a good fit to the data), both constants cannot be negative simultaneously (since $y$ has to be positive). Therefore, the possible theories are the ones with either $B_2$ negative and $B_1$ positive, or both $B_1$ and $B_2$ positive. Hence the first conclusion is that $B_1$ must always be positive.

As in the previous section, we now assume that the model is effectively equivalent to the $\Lambda$CDM model with an effective $\Omega_\Lambda^{\Lambda CDM}$ of $yB_1 + y^2B_2$. This should give a good estimate for the best-fit parameters given that $y$ does not vary significantly with time. This gives an estimate for $y$ in terms of the best-fit $B_1$ and $B_2$, and the effective $\Omega_\Lambda^{\Lambda CDM}$:

 \begin{equation}
   y = \frac{-B_1 \pm \sqrt{B_1^2 + 4\Omega_\Lambda^{\Lambda CDM} B_2}}{2B_2}.\label{eq:yeffectiveB1B2}
 \end{equation}

In addition, eq. (\ref{eq:yWithB1B2}) in this case becomes

 \begin{equation}
   \Omega_\Lambda^{\Lambda CDM} y +\left(\Omega_{m} - B_2\right)y - \frac{B_1}{3} = 0.\label{eq:yWithB1B2effective}
 \end{equation}

Let us now insert $y$ from eq. (\ref{eq:yeffectiveB1B2}) into eq. ({\ref{eq:yWithB1B2effective}}) and solve for $B_1$ as a function of $B_2$. Assuming that $y$ is calculated at present ($z=0$), we can also use the relation $\Omega_{\Lambda}^{\Lambda CDM}  + \Omega_m^0 = 1$\footnote{This assumption simplifies the final relation significantly, while does not affect the general features of the relation. In order to obtain an even better approximation, one could relax this assumption and retain both $\Omega_{\Lambda}^{\Lambda CDM}$ and $\Omega_m^0$ explicitly. In this case, the expression will clearly be a function of redshift. By changing the redshift one can then tune the function such that it perfectly matches the numerically obtained correlation curve. This redshift represents a slightly earlier time than today around which the measured data points cluster most.}. We therefore obtain the following relation between $B_1$ and $B_2$\footnote{In the calculations that lead to this expression, we also make use of our previously acquired knowledge of the positivity of $B_1$. This is important in choosing the correct branch of the equation.}:

\begin{equation}
   B_1 = \frac{3\sqrt{\Omega_{\Lambda}^{\Lambda CDM}}\left(1 - B_2\right)}{\sqrt{3 - 2 B_2}}.
 \end{equation}

Here, the reality and positivity conditions for $B_1$ imply that $B_2<1$. This function very well explains the particular correlation pattern observed in fig. \ref{fig:BiGB1B2plots2D}. The overall shape of the function shows a very similar behavior and the only difference is in its normalization. The two curves remarkably match if we tune the value of the effective cosmological constant $\Omega_{\Lambda}^{\Lambda CDM}$ to something about $0.57$, which is slightly different from the best-fit value in the $\Lambda$CDM case (i.e. $\sim0.7$). This is however not a big surprise, given that the obtained analytical expression for $B_1$ versus $B_2$ is the result of approximating a time-dependent quantity, i.e. $y$, with a constant.   

\begin{figure}[t]
\begin{center}
\includegraphics[width=0.49\textwidth]{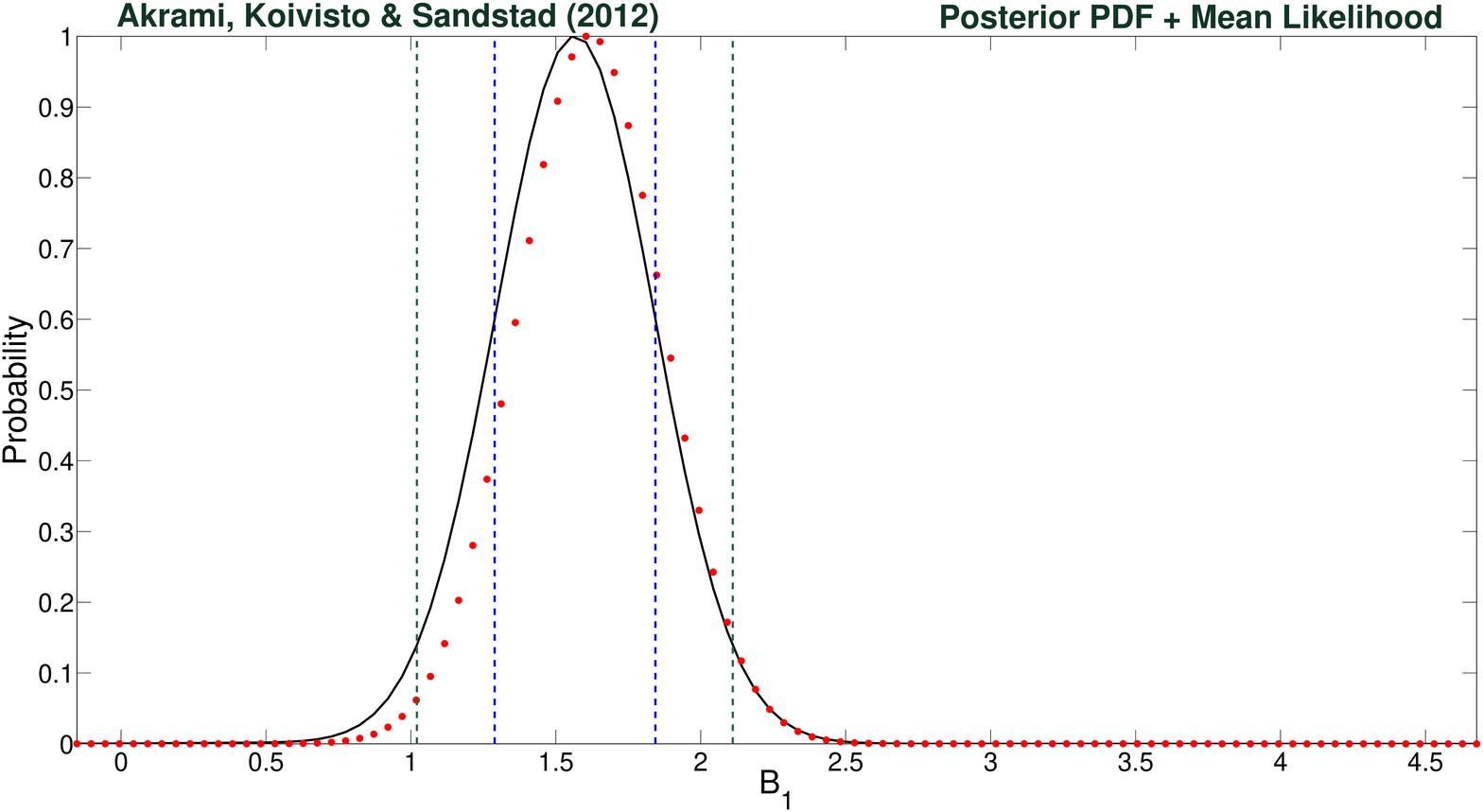}
\includegraphics[width=0.49\textwidth]{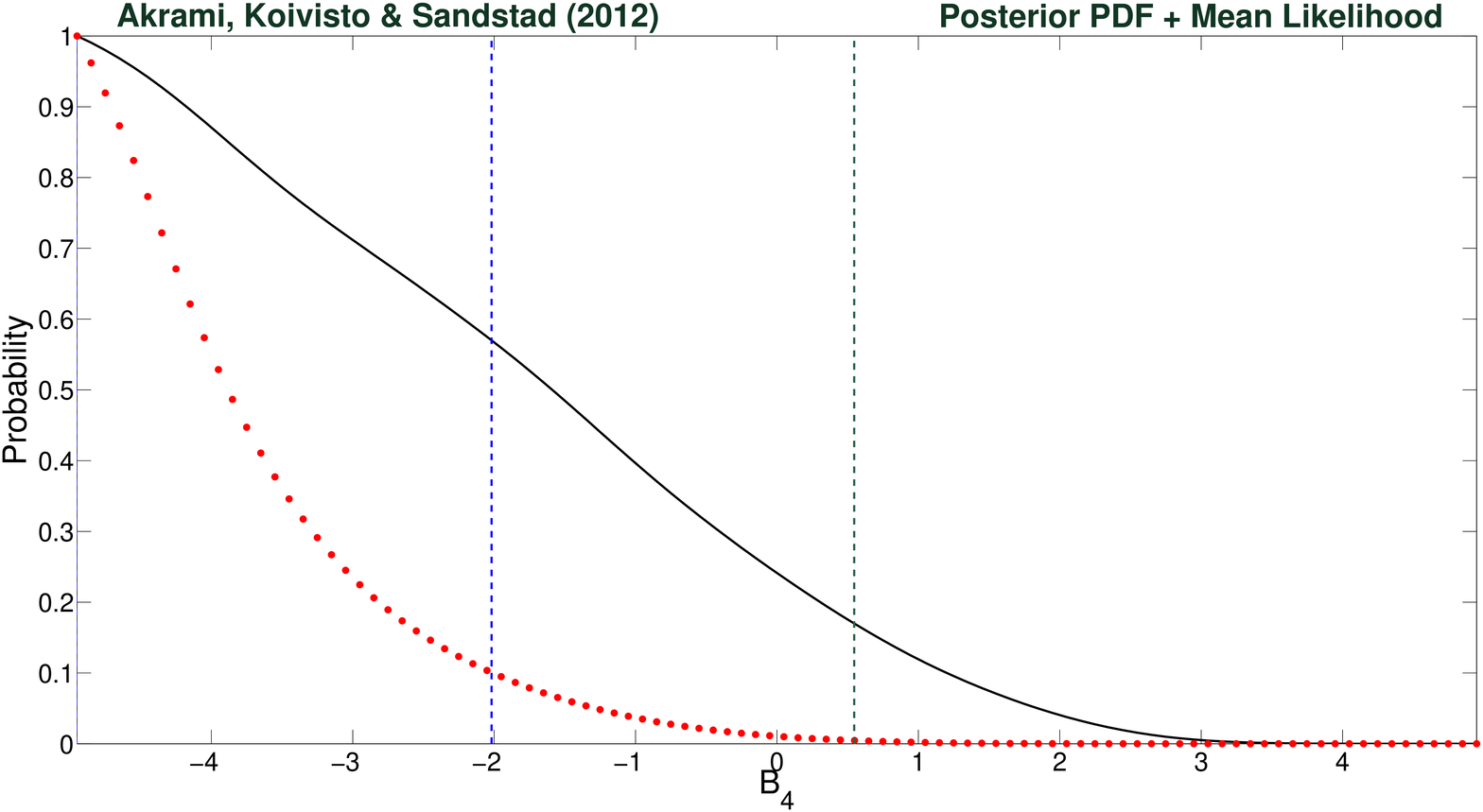}
\includegraphics[width=0.49\textwidth]{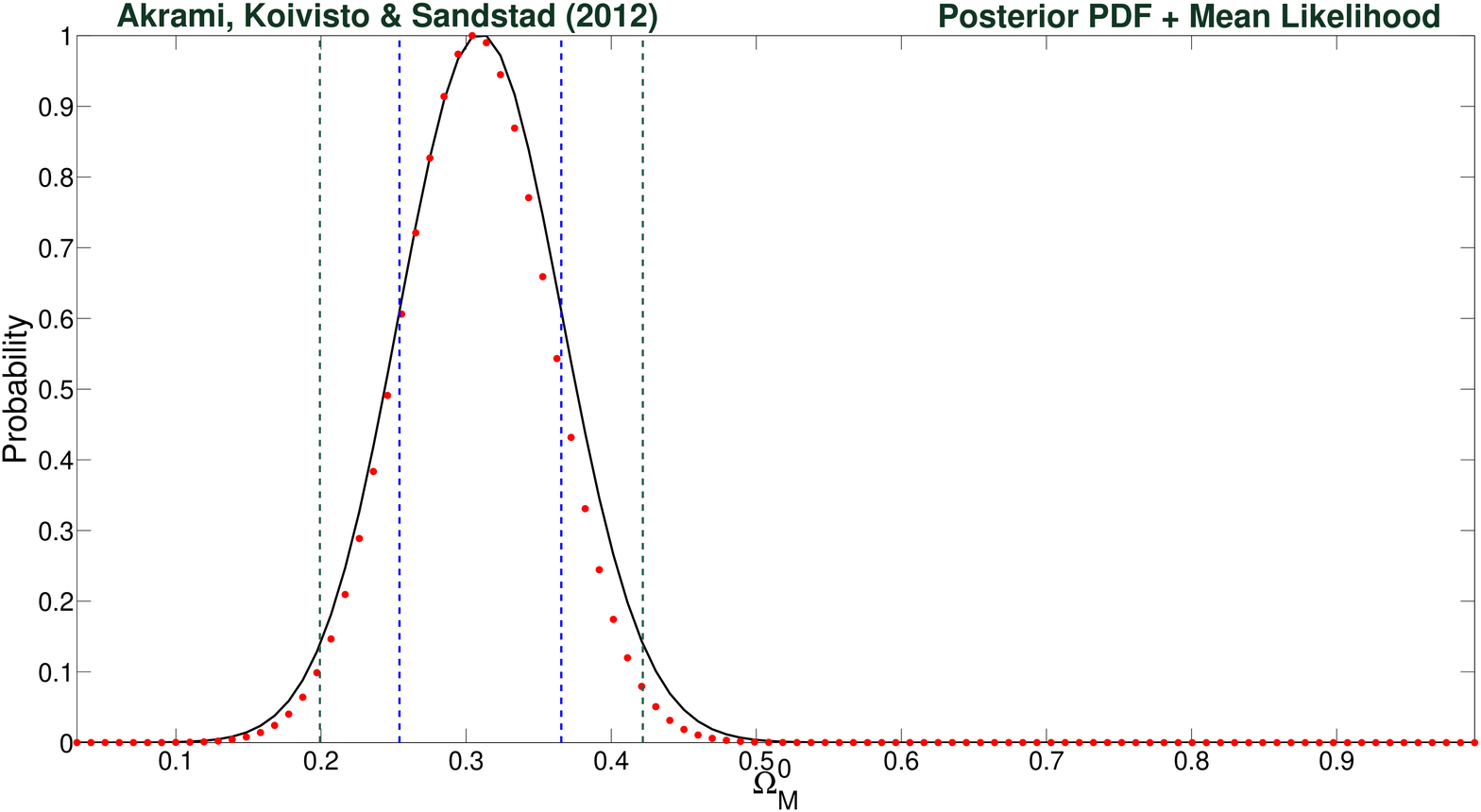}
\caption[aa]{As in figs. \ref{fig:LCDMplots}, \ref{fig:BiGB1plots1D} and \ref{fig:BiGB1B2plots1D_r5}, but for the bigravity model with only $B_1$, $B_4$ and $\Omega_m^0$ varying (other parameters are set to zero). Here, (flat) prior ranges for both $B_1$ and $B_4$ are $[-5,5]$, and for $\Omega_m^0$ is $[0,1]$.}
\label{fig:BiGB1B4plots1D_r5}
\end{center}
\end{figure} 

So far in this section, we have studied three-parameter sub-models where in each case two of the three parameters $B_1$, $B_2$ and $B_3$ are allowed to vary (in addition to $\Omega_m^0$). Three more sub-models with the same dimensionality exist: cases where $B_4$ is considered as a free parameter together with $\Omega_m^0$ and one of the $B_1$, $B_2$ and $B_3$. We argued in the previous section that the two-dimensional model ($B_4,\Omega_m^0$) is equivalent to the CDM model and cannot give acceleration. The situation may however change if we switch on one of the $B_1$, $B_2$ or $B_3$ terms together with the $B_4$ term. For example, we have already seen that the ($B_1,\Omega_m^0$)-model, which is a sub-model of ($B_1,B_4,\Omega_m^0$) can perfectly fit the data, meaning that we expect a good fit from the ($B_1,B_4,\Omega_m^0$)-model too. Our scan for this case with prior ranges of $[-5,5]$ for $B_1$ and $B_4$ gives a best-fit $\chi^2$ of $548.86$, corresponding to the $p$-value of $0.8485$. Compared to the ($B_1,\Omega_m^0$) case, this indicates a slightly better fit. In terms of the Bayesian evidence, the log-evidence we get in this case is $-281.42$ which is again better than the one for the ($B_1,\Omega_m^0$) case. The improvement in the evidence ($\Delta log\mathcal{Z}$) is however not sufficiently large ($\sim 0.3$) for the $B_4$ term to be considered essential in providing a better fit to the data. For the ($B_2,B_4,\Omega_m^0$) and ($B_3,B_4,\Omega_m^0$) models, our results show that they are observationally excluded, although there is improvement in both $p$-value and log-evidence in these cases compared to ($B_2,\Omega_m^0$) and ($B_3,\Omega_m^0$). The ($B_2,B_4,\Omega_m^0$)-model gives the best-fit $\chi^2$ of $806.82$ (with a $p$-value smaller than $0.0001$) and the log-evidence of $-420.87$. The best-fit $\chi^2$ for the ($B_3,B_4,\Omega_m^0$)-model is $685.30$ (corresponding to the $p$-value of $0.0023$) and the log-evidence is $-351.14$. Finally, in terms of the marginalized posteriors and mean likelihoods, all these models again show some degree of correlation between the parameters, although the correlations are much weaker than for the models we studied earlier in this section. The ($B_1,B_4,\Omega_m^0$)-model is clearly more interesting since it yields a good fit to the data, and in this case, the one-dimensional plots (fig. \ref{fig:BiGB1B4plots1D_r5}) illustrate that $B_1$ and $\Omega_m^0$ are constrained to some positive values, while for $B_4$ the observations prefer negative values. By studying the case with larger prior ranges and the two-dimensional plots a weak correlation between the parameters can be seen; here we do not present the plots for such cases for brevity reasons.

\subsubsection{Further generalizations}

So far, our studies have been limited to the sub-models of the full bimetric model where only two or three model parameters (out of six) vary and the rest are fixed to zero. This was however an essential step in understanding the structure of the model, what the roles of each parameter are in providing good fits to the observations and how they interact. We can in principle continue our studies to higher-dimensional models by using the same procedure as before and try to constrain the model parameters in each case. We however saw that even for the three-dimensional models, our data seem to be insufficient in placing strong constraints upon the parameters. We demonstrated this by plotting one-dimensional marginalized posteriors and mean likelihoods and comparing the two when the prior ranges were changed. We observed that some parameters were only constrained to be positive or negative and there were strong correlations between many of them. This all means that adding more and more parameters to such unconstrained models only increases the degrees of correlations and the larger models will most probably become even less constrained. For these reasons, we do not investigate such models in detail and restrict our studies to only reporting the $p$-value and log-evidence in each case. These quantities, together with the ones for the previous models are presented in table \ref{tbl:ResultsDifferentParmaeterRegimes}. These include the four-parameter models ($B_1,B_2,B_3,\Omega_m^0$), ($B_1,B_2,B_4,\Omega_m^0$), ($B_1,B_3,B_4,\Omega_m^0$) and ($B_2,B_3,B_4,\Omega_m^0$), as well as the five-parameter model ($B_1,B_2,B_3,B_4,\Omega_m^0$). Our results show that all these models are in very good agreement with observations, as far as the $p$-values are concerned, and in some cases are more favored compared to the lower-dimensional models from the Bayesian point of view (see the log-evidence values in table \ref{tbl:ResultsDifferentParmaeterRegimes}).

  \begin{table}
    \begin{center}
    \footnotesize{
      \begin {tabular}{|c|c|c|c|c|c|c|c|c|c|}
        \hline
        {\bf Model} & ${\bf B_0}$& ${\bf B_1}$ & ${\bf B_2}$ & ${\bf B_3}$ & ${\bf B_4}$ & ${\bf \Omega_m}$& ${\bf \chi^2_{min}}$& {\bf p-value}&{\bf log-evidence}\\
	\hline
        \hline
	${\bf \Lambda}${\bf CDM}& free& 0 & 0 & 0 & 0 & free& 546.54 & 0.8709 &-278.50\\
	\hline
	${\bf (B_1,\Omega_m^0)}$&0& free & 0 & 0 & 0 & free & 551.60 & 0.8355 &-281.73\\
	\hline
	${\bf (B_2,\Omega_m^0)}$&0& 0 & free & 0 & 0 & free & 894.00 & $<0.0001$ & -450.25\\
	\hline
	${\bf (B_3,\Omega_m^0)}$&0& 0 & 0 & free & 0 & free & 1700.50 & $<0.0001$ & -850.26\\
	\hline
        ${\bf (B_1,B_2,\Omega_m^0)}$&0& free & free & 0 & 0 & free & 546.52 &0.8646 & -279.77\\
	\hline
        ${\bf (B_1,B_3,\Omega_m^0)}$&0& free & 0 & free & 0 & free & 542.82 & 0.8878 & -280.10\\
	\hline
        ${\bf (B_2,B_3,\Omega_m^0)}$&0& 0 & free & free & 0 & free & 548.04 & 0.8543 & -280.91\\
	\hline
        ${\bf (B_1,B_4,\Omega_m^0)}$&0& free & 0 & 0 & free & free & 548.86 & 0.8485 & -281.42\\
	\hline
        ${\bf (B_2,B_4,\Omega_m^0)}$&0& 0 & free & 0 & free & free & 806.82 & $<0.0001$ & -420.87\\
	\hline
        ${\bf (B_3,B_4,\Omega_m^0)}$&0& 0 & 0 & free & free & free & 685.30 & 0.0023 & -351.14\\
	\hline
        ${\bf (B_1,B_2,B_3,\Omega_m^0)}$&0& free & free & free & 0 & free & 546.50 & 0.8582 & -279.61\\
	\hline
        ${\bf (B_1,B_2,B_4,\Omega_m^0)}$&0& free & free & 0 & free & free & 546.52 & 0.8581 & -279.56\\
	\hline
        ${\bf (B_1,B_3,B_4,\Omega_m^0)}$&0& free & 0 & free & free & free & 546.78 & 0.8563 & -280.00\\
	\hline
        ${\bf (B_2,B_3,B_4,\Omega_m^0)}$&0& 0 & free & free & free & free & 549.68 & 0.8353 & -282.89\\
	\hline	
        ${\bf (B_1,B_2,B_3,B_4,\Omega_m^0)}$&0& free & free & free & free & free & 546.50 & 0.8515 & -279.60\\
	\hline
        {\bf full bigravity model}&free& free & free & free & free & free & 546.50 & 0.8445 &-279.82\\
	\hline
      \end{tabular}}
      \caption{Best-fit $\chi^2$, $p$-value and log-evidence for different bigravity sub-models when the models are constrained by SNe, CMB, BAO and $H_0$ measurements at the background level. For each model, parameters that are allowed to vary are marked as ``free''; the non-varying parameters are fixed to zero. The models are named by reduced parameters $B$s, defined as $B_i = m^2\beta_i/H_0^2$. Prior ranges used in the statistical scans are chosen to be $[-5,5]$ for $B$s and $[0,1]$ for $\Omega_m^0$. The sub-model in the uppermost row with only $B_0$ and $\Omega_m^0$ being free is equivalent to the $\Lambda$CDM concordance model of cosmology. The ``full bigravity model'', i.e. the one in the lowermost row, correspond to the full parameter space where all model parameters (including the cosmological term $B_0$) are allowed to vary. In all models, except the $\Lambda$CDM, the present value of the Hubble parameter, $H_0$, is a prediction of the model and is determined in terms of the other parameters. For the $\Lambda$CDM case, $H_0$ is a free parameter which has been constrained by cosmological data.}
      \label{tbl:ResultsDifferentParmaeterRegimes}
    \end{center}
  \end{table}

\subsection{Including a cosmological constant: the full theory}

Our main objective in this paper has been to investigate whether the ghost-free, massive, bimetric gravity, represented by the action (\ref{eq:ActionOriginal}), can give rise to a late-time accelerated universe without any need to an explicit cosmological constant term (or vacuum energy)\footnote{See footnotes \ref{alphanote} and \ref{CCnote} for more details.}. For this reason, we fixed the free parameter $B_0$ to zero in all the analyses we have performed so far. In order to be as comprehensive as possible, here we briefly study the full model, i.e. when the $B_0$ term is also allowed to vary. We expect a very good fit in this case, since the previously considered models that are well fit to the data, including the $\Lambda$CDM model, are basically all sub-models of this six-dimensional model.

Table \ref{tbl:ResultsDifferentParmaeterRegimes} includes the results of our analysis for this case (lowermost row). The best-fit $\chi^2$ value we have obtained for the case where a prior range of $[-5,5]$ is used for all $B$s (while the prior range for $\Omega_m^0$ is chosen to be $[0,1]$) is $546.50$, which corresponds to a $p$-value of $0.8445$. Compared to the $\Lambda$CDM model, we have obtained a better $\chi^2$, even though the $p$-value is smaller. This is natural however regarding the fact that the full bigravity model has more free parameters than the $\Lambda$CDM. Our results show that there is at least one sub-model (the ($B_1,B_3,\Omega_m^0$) case) with a better best-fit $\chi^2$ ($=542.82$). Since this model is obviously a sub-model of the full model, we expect this value to be found even in the latter case. We believe that the reason for observing the contrary is simply that the particular best-fit point in the ($B_1,B_3,\Omega_m^0$)-model is highly fine-tuned so that our scanning algorithm has not been able to probe it when the parameter space has become much larger. We discussed this issue in section \ref{sec:statistics} where we introduced our scanning technique. We mentioned that the technique is optimized for Bayesian inference where the value of the global maximum likelihood is not important as long as a huge number of such high-likelihood points do not exist and the posterior mass is not affected by them. In addition, the log-evidence obtained for the full model is $-279.82$, which compared to the value for the $\Lambda$CDM model (i.e. $-278.50$) indicates that the model is less favored. This is not a big surprise though because we know that the $\Lambda$CDM, as a sub-model of our full bigravity model, is in perfect agreement with background observations. In the absence of any theoretical preference of a model over the other, Bayesian inference naturally selects the lower-dimensional one.

\section{Conclusions and outlook}
In this paper, we have performed a thorough and extensive parameter estimation and model comparison for the ghost-free, bimetric theory of massive gravity by comparing the predictions of the model to various cosmological measurements at the background level. To our knowledge, this has thus far been the most exhaustive statistical analysis of the model.

We assumed the two metrics of the model to be spatially flat, homogeneous and isotropic, and only one metric was considered as the physical one coupled to matter, while both metrics were dynamical. We broadly discussed different observational data used in the analysis (SNe, CMB and BAO), dynamical equations and initial conditions used in numerically computing various cosmological distances, and statistical frameworks employed in constraining the model. We used nested sampling to explore the model's parameter space, find the maximum likelihood points, calculate the Bayesian evidence, compare different sub-models and constrain the model parameters. In many places, we complemented the results of our statistical studies with detailed analytical explanations; this helped us understand various interesting features observed in the results.

The main objective of our analysis has been to investigate the possibility of obtaining a late-time acceleration of the Universe, within the bigravity framework, that is consistent with observations. For this reason, we mainly focused on the interesting case where no explicit cosmological constant (or vacuum energy) was assumed. In the absence of any theoretical restrictions on the possible values of the free parameters of the model, we chose somewhat arbitrary, but justified, ranges for our scans. In order to understand the roles that the different parameters play in the dynamics of the model, we have extensively studied various sub-models of the full model where only some of the free parameters have been allowed to vary.

Our results show that the bimetric model can in general yield very good fits to the observed data at a very high confidence level. In other words, the model can produce the cosmic acceleration without the need to resort to an explicit cosmological constant term. The reason is that the parameter space of the model contains many points for which the model behaves very much like the $\Lambda$CDM, i.e. effectively gives a cosmological constant, at least at the background level. This similarity to the $\Lambda$CDM makes the model be in agreement with the observations. Although the $\Lambda$CDM model remains to be slightly favored by observations, both models are statistically perfectly consistent with the data. Even though the value of the Bayesian evidence for the $\Lambda$CDM model compared to the values for the bigravity model (and its sub-models) suggest that the former is preferred from the Bayesian point of view, we argued that this should be viewed with caution. Our argument was mainly based on the impacts our prior choices of parameter ranges and our prior preferences for different models over the others could have upon such conclusions.

We additionally observe that some particular sub-models with only some parameters varying are ruled out by the data. The first model in this class is the one with only the quartic ($\beta_4$) mass term in the action being allowed to vary and the other parameters are fixed to zero. Other similarly excluded sub-models are the ones where only the quadratic ($\beta_2$) or cubic ($\beta_3$) mass terms in the action are present; turning on both of the terms at the same time however brings the theory back into the game. Models with only quadratic and quartic, or cubic and quartic terms are also excluded. Finally, it turns out that the most important term for obtaining a good fit is the linear ($\beta_1$) term which can individually produce a cosmic evolution in perfect agreement with the cosmological observations.

As far as the observational constraints on the values of the parameters are concerned, one-dimensional marginalized posterior probabilities and mean likelihoods reveal that, except the sub-model with only the linear term existing, in all other viable sub-models the free parameters are correlated. The correlations are in some cases so strong that the one-dimensional probability plots are not reliable. Although in these cases we cannot determine the values of the parameters, the correlation patterns indicate that some parameters are preferred to be positive or negative. In the linear-term-only sub-model, where we have reliable constraints, the graviton mass can be determined and turns out to be of the order of the present value of the Hubble parameter (which is not a big surprise).

In cases where our prior ranges for the parameters are wide enough, our results show that they can take on arbitrarily large or small values (respecting the constraints on their signs) while giving very good fits. In this case, the model tends to become as similar to the $\Lambda$CDM as possible where the time evolution of the effective dynamical variable of the theory ($y$) becomes negligible. Since for a good-fit model, $y$ starts off with an asymptotically zero value in the far past, its value must remain vanishingly small in the far future if its variation with time is to be tiny. This explains why the absolute values of the parameters are required to be infinitely large in those cases: the parameters are always multiplied by different powers of $y$ in the evolution equations and their large values compensate for the smallness of $y$ in order to give a cosmological constant value compatible with observations. Working only with the background dynamics of the model and using only the geometrical measurements of the cosmic evolution (as we have done in the present work) cannot determine how small the time variation of $y$ must be. Currently, the model is consistent with the data even in cases where $y$ changes considerably with time. Additional cosmological or astrophysical constraints on the model are required to break the degeneracy between the parameters and determine how different the model wants to be from the $\Lambda$CDM.

Perhaps the most natural extension of the present work would be to repeat the analysis performed here for the case where perturbative equations are used. One can in this case use a wealth of data available from the measurements of the CMB anisotropies and the growth of large-scale structure of the Universe to further constrain the model. This could tell us whether the bimetric model with nontrivial dynamics could match the data as well as or better than the $\Lambda$CDM without preferring the most $\Lambda$CDM-like corners of the parameter space. In particular, any deviation from a constant equation of state parameter of dark energy observed by existing or forthcoming cosmological experiments, could potentially favor the model over the standard $\Lambda$CDM and place interesting constraints on its parameters. Other astrophysical tests of the model, in particular on scales much smaller than the cosmological ones, such as the solar system or black hole observations, can provide highly important additional pieces of information to the analysis to either rule out the theory or to constrain its properties.

And finally, one can generalize the model to the cases where the flatness assumption for the curvature of the Universe is relaxed, other more nontrivial types of metrics are considered or the physical metric is not identical to only one metric (our $g$) but is defined in terms of a combination of the two metrics ($g$ and $f$). In addition, one very interesting generalization of the model to study would be the case where the second metric ($f$) is also coupled to matter. We leave the investigation of such possibilities for future work.

\begin{acknowledgments}
We thank Tessa Baker, Robert Crittenden, Jonas Enander, Hans Kristian K. Eriksen, Farhan Feroz, Pedro G. Ferreira, Juan Garcia-Bellido, S. F. Hassan, Antony Lewis, Edvard M\"{o}rtsell, David F. Mota, Sigurd K. N{\ae}ss, Claudia de Rham, Rachel A. Rosen, Mikael von Strauss and Andrew J. Tolley for enlightening and helpful discussions. We are particularly grateful to S. F. Hassan and Edvard M\"{o}rtsell for useful comments on a previous version of the manuscript. We also thank the anonymous referee for constructive and helpful comments. YA thanks the Astrophysics Group of the Department of Physics at the University of Oxford and the Centro de Ciencias de Benasque Pedro Pascual for their hospitality during the completion of this work. YA is supported by the European Research Council (ERC) Starting Grant StG2010-257080.
\end{acknowledgments}

% bibliography:
\bibliographystyle{JHEP}
\bibliography{sources}

\providecommand{\href}[2]{#2}\begingroup\raggedright\begin{thebibliography}{10}

\bibitem{Fierz_et_Pauli1939}
M.~Fierz and W.~Pauli, {\it On relativistic wave equations for particles of
  arbitrary spin in an electromagnetic field},  {\em Proceedings of the Royal
  Society of London. Series A, Mathematical and Physical Sciences} (1939)
  211--232.

\bibitem{Boulware_et_Deser1939}
S.~Deser and D.~G. Boulware, {\it Can gravitation have a finite range?},  {\em
  Phys. Rev.} {\bf D6} (1972) 3368--3382.

\bibitem{Hinterbichler2011}
K.~Hinterbichler, {\it Theoretical aspects of massive gravity},  {\em Rev. Mod.
  Phys.} {\bf 84} (2012) 671--710,
  [\href{http://xxx.lanl.gov/abs/1105.3735}{{\tt arXiv:1105.3735}}].

\bibitem{deRham2009}
C.~de~Rham, {\it {Massive gravity from Dirichlet boundary conditions}},  {\em
  Phys. Lett.} {\bf B688} (2010) 137--141,
  [\href{http://xxx.lanl.gov/abs/0910.5474}{{\tt arXiv:0910.5474}}].

\bibitem{deRham_et_Gabadadze2010a}
C.~de~Rham and G.~Gabadadze, {\it {Selftuned Massive Spin-2}},  {\em Phys.
  Lett.} {\bf B693} (2010) 334--338,
  [\href{http://xxx.lanl.gov/abs/1006.4367}{{\tt arXiv:1006.4367}}].

\bibitem{deRham_et_Gabadadze2010b}
C.~de~Rham and G.~Gabadadze, {\it {Generalization of the Fierz-Pauli Action}},
  {\em Phys. Rev.} {\bf D82} (2010) 044020,
  [\href{http://xxx.lanl.gov/abs/1007.0443}{{\tt arXiv:1007.0443}}].

\bibitem{deRham_et_al2010}
C.~de~Rham, G.~Gabadadze, L.~Heisenberg, and D.~Pirtskhalava, {\it {Cosmic
  Acceleration and the Helicity-0 Graviton}},  {\em Phys. Rev.} {\bf D83}
  (2011) 103516, [\href{http://xxx.lanl.gov/abs/1010.1780}{{\tt
  arXiv:1010.1780}}].

\bibitem{deRham_Gabadadze_et_Tolley2010}
C.~de~Rham, G.~Gabadadze, and A.~J. Tolley, {\it {Resummation of Massive
  Gravity}},  {\em Phys. Rev. Lett.} {\bf 106} (2011) 231101,
  [\href{http://xxx.lanl.gov/abs/1011.1232}{{\tt arXiv:1011.1232}}].

\bibitem{deRham_et_al2011}
C.~de~Rham, G.~Gabadadze, D.~Pirtskhalava, A.~J. Tolley, and I.~Yavin, {\it
  {Nonlinear Dynamics of 3D Massive Gravity}},  {\em JHEP} {\bf 1106} (2011)
  028, [\href{http://xxx.lanl.gov/abs/1103.1351}{{\tt arXiv:1103.1351}}].

\bibitem{Koyama:2011xz}
K.~Koyama, G.~Niz, and G.~Tasinato, {\it {Analytic solutions in non-linear
  massive gravity}},  {\em Phys.Rev.Lett.} {\bf 107} (2011) 131101,
  [\href{http://xxx.lanl.gov/abs/1103.4708}{{\tt arXiv:1103.4708}}].

\bibitem{Hassan_et_Rosen2011a}
S.~F. Hassan and R.~A. Rosen, {\it {On Non-Linear Actions for Massive
  Gravity}},  {\em JHEP} {\bf 07} (2011) 009,
  [\href{http://xxx.lanl.gov/abs/1103.6055}{{\tt arXiv:1103.6055}}].

\bibitem{Hassan:2011hr}
S.~Hassan and R.~A. Rosen, {\it {Resolving the Ghost Problem in non-Linear
  Massive Gravity}},  {\em Phys.Rev.Lett.} {\bf 108} (2012) 041101,
  [\href{http://xxx.lanl.gov/abs/1106.3344}{{\tt arXiv:1106.3344}}].

\bibitem{Koyama:2011yg}
K.~Koyama, G.~Niz, and G.~Tasinato, {\it {Strong interactions and exact
  solutions in non-linear massive gravity}},  {\em Phys.Rev.} {\bf D84} (2011)
  064033, [\href{http://xxx.lanl.gov/abs/1104.2143}{{\tt arXiv:1104.2143}}].

\bibitem{deRham_et_Heisenberg2011}
C.~de~Rham and L.~Heisenberg, {\it {Cosmology of the Galileon from Massive
  Gravity}},  {\em Phys. Rev.} {\bf D84} (2011) 043503,
  [\href{http://xxx.lanl.gov/abs/1106.3312}{{\tt arXiv:1106.3312}}].

\bibitem{deRham_Gabadadze_et_Tolley2011a}
C.~de~Rham, G.~Gabadadze, and A.~J. Tolley, {\it {Comments on
  (super)luminality}},  \href{http://xxx.lanl.gov/abs/1107.0710}{{\tt
  arXiv:1107.0710}}.

\bibitem{deRham_Gabadadze_et_Tolley2011b}
C.~de~Rham, G.~Gabadadze, and A.~J. Tolley, {\it {Ghost free Massive Gravity in
  the St\'uckelberg language}},  {\em Phys. Lett.} {\bf B711} (2012) 190--195,
  [\href{http://xxx.lanl.gov/abs/1107.3820}{{\tt arXiv:1107.3820}}].

\bibitem{deRham_Gabadadze_et_Tolley2011c}
C.~de~Rham, G.~Gabadadze, and A.~J. Tolley, {\it {Helicity Decomposition of
  Ghost-free Massive Gravity}},  {\em JHEP} {\bf 1111} (2011) 093,
  [\href{http://xxx.lanl.gov/abs/1108.4521}{{\tt arXiv:1108.4521}}].

\bibitem{D'Amico_et_al2011}
G.~D'Amico, C.~de~Rham, S.~L. Dubovsky, G.~Gabadadze, D.~Pirtskhalava, and
  A.~J. Tolley, {\it {Massive Cosmologies}},  {\em Phys. Rev.} {\bf D84} (2011)
  124046, [\href{http://xxx.lanl.gov/abs/1108.5231}{{\tt arXiv:1108.5231}}].

\bibitem{Koyama:2011wx}
K.~Koyama, G.~Niz, and G.~Tasinato, {\it {The Self-Accelerating Universe with
  Vectors in Massive Gravity}},  {\em JHEP} {\bf 1112} (2011) 065,
  [\href{http://xxx.lanl.gov/abs/1110.2618}{{\tt arXiv:1110.2618}}].

\bibitem{Berezhiani_et_al2011}
L.~Berezhiani, G.~Chkareuli, C.~de~Rham, G.~Gabadadze, and A.~J. Tolley, {\it
  {On Black Holes in Massive Gravity}},  {\em Phys. Rev.} {\bf D85} (2012)
  044024, [\href{http://xxx.lanl.gov/abs/1111.3613}{{\tt arXiv:1111.3613}}].

\bibitem{Burrage_et_al2011}
C.~Burrage, C.~de~Rham, L.~Heisenberg, and A.~J. Tolley, {\it {Chronology
  Protection in Galileon Models and Massive Gravity}},  {\em JCAP} {\bf 1207}
  (2012) 004, [\href{http://xxx.lanl.gov/abs/1111.5549}{{\tt
  arXiv:1111.5549}}].

\bibitem{de_Rham_et_Renaux-Petel2012}
C.~de~Rham and S.~Renaux-Petel, {\it {Massive Gravity on de Sitter and Unique
  Candidate for Partially Massless Gravity}},
  \href{http://xxx.lanl.gov/abs/1206.3482}{{\tt arXiv:1206.3482}}.

\bibitem{De_Felice_Gumrukcuoglu_etMukohyama2012}
A.~De~Felice, A.~E. Gumrukcuoglu, and S.~Mukohyama, {\it {Massive gravity:
  nonlinear instability of the homogeneous and isotropic universe}},
  \href{http://xxx.lanl.gov/abs/1206.2080}{{\tt arXiv:1206.2080}}.

\bibitem{D'Amico2012}
G.~D'Amico, {\it {Cosmology and perturbations in massive gravity}},
  \href{http://xxx.lanl.gov/abs/1206.3617}{{\tt arXiv:1206.3617}}.

\bibitem{Fasiello_et_Tolley2012}
M.~Fasiello and A.~J. Tolley, {\it {Cosmological perturbations in Massive
  Gravity and the Higuchi bound}},
  \href{http://xxx.lanl.gov/abs/1206.3852}{{\tt arXiv:1206.3852}}.

\bibitem{Langlois_et_Naruko2012}
D.~Langlois and A.~Naruko, {\it {Cosmological solutions of massive gravity on
  de Sitter}},  \href{http://xxx.lanl.gov/abs/1206.6810}{{\tt
  arXiv:1206.6810}}.

\bibitem{Motohashi:2012jd}
H.~Motohashi and T.~Suyama, {\it {Self-accelerating Solutions in Massive
  Gravity on Isotropic Reference Metric}},
  \href{http://xxx.lanl.gov/abs/1208.3019}{{\tt arXiv:1208.3019}}.

\bibitem{Comelli_et_al2011b}
D.~Comelli, M.~Crisostomi, F.~Nesti, and L.~Pilo, {\it {FRW Cosmology in Ghost
  Free Massive Gravity }},  {\em JHEP} {\bf 1203} (2012) 067,
  [\href{http://xxx.lanl.gov/abs/1111.1983}{{\tt arXiv:1111.1983}}].

\bibitem{Hassan_et_al2012a}
S.~F. Hassan, R.~A. Rosen, and A.~Schmidt-May, {\it {Ghost-free Massive Gravity
  with a General Reference Metric}},  {\em JHEP} {\bf 02} (2012) 026,
  [\href{http://xxx.lanl.gov/abs/1109.3230}{{\tt arXiv:1109.3230}}].

\bibitem{Hassan_et_Rosen2012a}
S.~F. Hassan and R.~A. Rosen, {\it {Bimetric Gravity from Ghost-free Massive
  Gravity}},  {\em JHEP} {\bf 02} (2012) 126,
  [\href{http://xxx.lanl.gov/abs/1109.3515}{{\tt arXiv:1109.3515}}].

\bibitem{Hassan_et_Rosen2012b}
S.~F. Hassan and R.~A. Rosen, {\it {Confirmation of the Secondary Constraint
  and Absence of Ghost in Massive Gravity and Bimetric Gravity}},  {\em JHEP}
  {\bf 04} (2012) 123, [\href{http://xxx.lanl.gov/abs/1111.2070}{{\tt
  arXiv:1111.2070}}].

\bibitem{Comelli_et_al2011a}
D.~Comelli, M.~Crisostomi, F.~Nesti, and L.~Pilo, {\it {Spherically Symmetric
  Solutions in Ghost-Free Massive Gravity}},  {\em Phys. Rev.} {\bf D85} (2012)
  024044, [\href{http://xxx.lanl.gov/abs/1110.4967}{{\tt arXiv:1110.4967}}].

\bibitem{Paulos_et_Tolley2012}
M.~F. Paulos and A.~J. Tolley, {\it {Massive Gravity Theories and limits of
  Ghost-free Bigravity models}},  {\em JHEP} {\bf 1209} (2012) 002,
  [\href{http://xxx.lanl.gov/abs/1203.4268}{{\tt arXiv:1203.4268}}].

\bibitem{Hinterbichler_et_Rosen2012}
K.~Hinterbichler and R.~A. Rosen, {\it {Interacting Spin-2 Fields}},  {\em
  JHEP} {\bf 1207} (2012) 047, [\href{http://xxx.lanl.gov/abs/1203.5783}{{\tt
  arXiv:1203.5783}}].

\bibitem{Comelli_et_al2012b}
D.~Comelli, M.~Crisostomi, F.~Nesti, and L.~Pilo, {\it {Degrees of Freedom in
  Massive Gravity}},  \href{http://xxx.lanl.gov/abs/1204.1027}{{\tt
  arXiv:1204.1027}}.

\bibitem{Baccetti_et_al2012a}
V.~Baccetti, P.~Martin-Moruno, and M.~Visser, {\it {Massive gravity from
  bimetric gravity}},  \href{http://xxx.lanl.gov/abs/1205.2158}{{\tt
  arXiv:1205.2158}}.

\bibitem{Baccetti_et_al2012b}
V.~Baccetti, P.~Martin-Moruno, and M.~Visser, {\it {Null Energy Condition
  violations in bimetric gravity}},  {\em JHEP} {\bf 1208} (2012) 148,
  [\href{http://xxx.lanl.gov/abs/1206.3814}{{\tt arXiv:1206.3814}}].

\bibitem{Baccetti_et_al2012c}
V.~Baccetti, P.~Martin-Moruno, and M.~Visser, {\it {Gordon and Kerr-Schild
  ansatze in massive and bimetric gravity}},  {\em JHEP} {\bf 1208} (2012) 108,
  [\href{http://xxx.lanl.gov/abs/1206.4720}{{\tt arXiv:1206.4720}}].

\bibitem{Hassan:2012wr}
S.~Hassan, A.~Schmidt-May, and M.~von Strauss, {\it {On Consistent Theories of
  Massive Spin-2 Fields Coupled to Gravity}},
  \href{http://xxx.lanl.gov/abs/1208.1515}{{\tt arXiv:1208.1515}}.

\bibitem{Hassan:2012gz}
S.~Hassan, A.~Schmidt-May, and M.~von Strauss, {\it {On Partially Massless
  Bimetric Gravity}},  \href{http://xxx.lanl.gov/abs/1208.1797}{{\tt
  arXiv:1208.1797}}.

\bibitem{Nojiri:2012zu}
S.~Nojiri and S.~D. Odintsov, {\it {Ghost-free $F(R)$ bigravity and
  accelerating cosmology}},  \href{http://xxx.lanl.gov/abs/1207.5106}{{\tt
  arXiv:1207.5106}}.

\bibitem{Volkov2012a}
M.~S. Volkov, {\it {Cosmological solutions with massive gravitons in the
  bigravity theory}},  {\em JHEP} {\bf 01} (2012) 042,
  [\href{http://xxx.lanl.gov/abs/1110.6153}{{\tt arXiv:1110.6153}}].

\bibitem{Strauss_et_al2011}
M.~v. Strauss, A.~Schmidt-May, J.~Enander, E.~Mortsell, and S.~F. Hassan, {\it
  Cosmological solutions in bimetric gravity and their observational tests},
  {\em JCAP} {\bf 03} (2012) 042,
  [\href{http://xxx.lanl.gov/abs/1111.1655}{{\tt arXiv:1111.1655}}].

\bibitem{Volkov2012c}
M.~S. Volkov, {\it {Exact self-accelerating cosmologies in the ghost-free
  bigravity and massive gravity}},
  \href{http://xxx.lanl.gov/abs/1205.5713}{{\tt arXiv:1205.5713}}.

\bibitem{Comelli_et_al2012a}
M.~Crisostomi, D.~Comelli, and L.~Pilo, {\it {Perturbations in Massive Gravity
  Cosmology}},  {\em JHEP} {\bf 1206} (2012) 085,
  [\href{http://xxx.lanl.gov/abs/1202.1986}{{\tt arXiv:1202.1986}}].

\bibitem{Khosravi_et_al2012}
N.~Khosravi, H.~R. Sepangi, and S.~Shahidi, {\it {On massive cosmological
  scalar perturbations}},  {\em Phys.Rev.} {\bf D86} (2012) 043517,
  [\href{http://xxx.lanl.gov/abs/1202.2767}{{\tt arXiv:1202.2767}}].

\bibitem{Berg_et_al2012}
M.~Berg, I.~Buchberger, J.~Enander, E.~Mortsell, and S.~Sjors, {\it {Growth
  Histories in Bimetric Massive Gravity}},
  \href{http://xxx.lanl.gov/abs/1206.3496}{{\tt arXiv:1206.3496}}.

\bibitem{Kuhnel:2012gh}
F.~Kuhnel, {\it {On Instability of Certain Bi-Metric and Massive-Gravity
  Theories}},  \href{http://xxx.lanl.gov/abs/1208.1764}{{\tt arXiv:1208.1764}}.

\bibitem{Weinberg:2008}
S.~Weinberg, {\em Cosmology}.
\newblock Oxford University Press, 2008.

\bibitem{Mukhanov:2005}
V.~Mukhanov, {\em Physical Foundations of Cosmology}.
\newblock Cambridge University Press, 2005.

\bibitem{Frieman:2008sn}
J.~Frieman, M.~Turner, and D.~Huterer, {\it {Dark Energy and the Accelerating
  Universe}},  {\em Ann.Rev.Astron.Astrophys.} {\bf 46} (2008) 385--432,
  [\href{http://xxx.lanl.gov/abs/0803.0982}{{\tt arXiv:0803.0982}}].

\bibitem{Martin:2012bt}
J.~Martin, {\it {Everything You Always Wanted To Know About The Cosmological
  Constant Problem (But Were Afraid To Ask)}},
  \href{http://xxx.lanl.gov/abs/1205.3365}{{\tt arXiv:1205.3365}}.

\bibitem{Copeland:2006wr}
E.~J. Copeland, M.~Sami, and S.~Tsujikawa, {\it {Dynamics of dark energy}},
  {\em Int.J.Mod.Phys.} {\bf D15} (2006) 1753--1936,
  [\href{http://xxx.lanl.gov/abs/hep-th/0603057}{{\tt hep-th/0603057}}].

\bibitem{Clifton:2011jh}
T.~Clifton, P.~G. Ferreira, A.~Padilla, and C.~Skordis, {\it {Modified Gravity
  and Cosmology}},  {\em Phys.Rept.} {\bf 513} (2012) 1--189,
  [\href{http://xxx.lanl.gov/abs/1106.2476}{{\tt arXiv:1106.2476}}].

\bibitem{Amendola:2012ys}
L.~Amendola and D.~F. Mota, {\it {Cosmology and fundamental physics with the
  Euclid satellite}},  \href{http://xxx.lanl.gov/abs/1206.1225}{{\tt
  arXiv:1206.1225}}.

\bibitem{Hossenfelder:2008bg}
S.~Hossenfelder, {\it {A Bi-Metric Theory with Exchange Symmetry}},  {\em
  Phys.Rev.} {\bf D78} (2008) 044015,
  [\href{http://xxx.lanl.gov/abs/0807.2838}{{\tt arXiv:0807.2838}}].

\bibitem{Koivisto:2011vq}
T.~S. Koivisto, {\it {On new variational principles as alternatives to the
  Palatini method}},  {\em Phys.Rev.} {\bf D83} (2011) 101501,
  [\href{http://xxx.lanl.gov/abs/1103.2743}{{\tt arXiv:1103.2743}}].

\bibitem{Tamanini:2012mi}
N.~Tamanini, {\it {Variational approach to gravitational theories with two
  independent connections}},  {\em Phys.Rev.} {\bf D86} (2012) 024004,
  [\href{http://xxx.lanl.gov/abs/1205.2511}{{\tt arXiv:1205.2511}}].

\bibitem{Westman:2012xk}
H.~Westman and T.~Zlosnik, {\it {Gravity, Cartan geometry, and idealized
  waywisers}},  \href{http://xxx.lanl.gov/abs/1203.5709}{{\tt
  arXiv:1203.5709}}.

\bibitem{BeltranJimenez:2012sz}
J.~Beltran~Jimenez and T.~S. Koivisto, {\it {The Bimetric variational principle
  for General Relativity}},  \href{http://xxx.lanl.gov/abs/1201.4018}{{\tt
  arXiv:1201.4018}}.

\bibitem{Komatsu:2010fb}
{\bf WMAP Collaboration} Collaboration, E.~Komatsu {\em et.~al.}, {\it
  {Seven-Year Wilkinson Microwave Anisotropy Probe (WMAP) Observations:
  Cosmological Interpretation}},  {\em Astrophys.J.Suppl.} {\bf 192} (2011) 18,
  [\href{http://xxx.lanl.gov/abs/1001.4538}{{\tt arXiv:1001.4538}}].

\bibitem{Beutler:2011hx}
F.~Beutler, C.~Blake, M.~Colless, D.~H. Jones, L.~Staveley-Smith, {\em
  et.~al.}, {\it {The 6dF Galaxy Survey: Baryon Acoustic Oscillations and the
  Local Hubble Constant}},  {\em Mon.Not.Roy.Astron.Soc.} {\bf 416} (2011)
  3017--3032, [\href{http://xxx.lanl.gov/abs/1106.3366}{{\tt
  arXiv:1106.3366}}].

\bibitem{Percival:2009xn}
{\bf SDSS Collaboration} Collaboration, W.~J. Percival {\em et.~al.}, {\it
  {Baryon Acoustic Oscillations in the Sloan Digital Sky Survey Data Release 7
  Galaxy Sample}},  {\em Mon.Not.Roy.Astron.Soc.} {\bf 401} (2010) 2148--2168,
  [\href{http://xxx.lanl.gov/abs/0907.1660}{{\tt arXiv:0907.1660}}].

\bibitem{Blake:2011en}
C.~Blake, E.~Kazin, F.~Beutler, T.~Davis, D.~Parkinson, {\em et.~al.}, {\it
  {The WiggleZ Dark Energy Survey: mapping the distance-redshift relation with
  baryon acoustic oscillations}},  {\em Mon.Not.Roy.Astron.Soc.} {\bf 418}
  (2011) 1707--1724, [\href{http://xxx.lanl.gov/abs/1108.2635}{{\tt
  arXiv:1108.2635}}].

\bibitem{Riess:1998cb}
{\bf Supernova Search Team} Collaboration, A.~G. Riess {\em et.~al.}, {\it
  {Observational evidence from supernovae for an accelerating universe and a
  cosmological constant}},  {\em Astron.J.} {\bf 116} (1998) 1009--1038,
  [\href{http://xxx.lanl.gov/abs/astro-ph/9805201}{{\tt astro-ph/9805201}}].

\bibitem{Perlmutter:1998np}
{\bf Supernova Cosmology Project} Collaboration, S.~Perlmutter {\em et.~al.},
  {\it {Measurements of Omega and Lambda from 42 high redshift supernovae}},
  {\em Astrophys.J.} {\bf 517} (1999) 565--586,
  [\href{http://xxx.lanl.gov/abs/astro-ph/9812133}{{\tt astro-ph/9812133}}].

\bibitem{Suzuki:2011hu}
N.~Suzuki, D.~Rubin, C.~Lidman, G.~Aldering, R.~Amanullah, {\em et.~al.}, {\it
  {The Hubble Space Telescope Cluster Supernova Survey: V. Improving the Dark
  Energy Constraints Above and Building an Early-Type-Hosted Supernova
  Sample}},  {\em Astrophys.J.} {\bf 746} (2012) 85,
  [\href{http://xxx.lanl.gov/abs/1105.3470}{{\tt arXiv:1105.3470}}].

\bibitem{Hassan:2012rq}
S.~Hassan, A.~Schmidt-May, and M.~von Strauss, {\it {Bimetric Theory and
  Partial Masslessness with Lanczos-Lovelock Terms in Arbitrary Dimensions}},
  \href{http://xxx.lanl.gov/abs/1212.4525}{{\tt arXiv:1212.4525}}.

\bibitem{Cowan:1998}
G.~Cowan, {\em Statistical Data Analysis}.
\newblock Oxford University Press, 1998.

\bibitem{Akrami:2009hp}
Y.~Akrami, P.~Scott, J.~Edsjo, J.~Conrad, and L.~Bergstrom, {\it {A Profile
  Likelihood Analysis of the Constrained MSSM with Genetic Algorithms}},  {\em
  JHEP} {\bf 1004} (2010) 057, [\href{http://xxx.lanl.gov/abs/0910.3950}{{\tt
  arXiv:0910.3950}}].

\bibitem{Akrami:2010cz}
Y.~Akrami, C.~Savage, P.~Scott, J.~Conrad, and J.~Edsjo, {\it {Statistical
  coverage for supersymmetric parameter estimation: a case study with direct
  detection of dark matter}},  {\em JCAP} {\bf 1107} (2011) 002,
  [\href{http://xxx.lanl.gov/abs/1011.4297}{{\tt arXiv:1011.4297}}].

\bibitem{Akrami:2011vh}
Y.~Akrami, {\it {Supersymmetry vis-\`a-vis Observation: Dark Matter
  Constraints, Global Fits and Statistical Issues}},
  \href{http://xxx.lanl.gov/abs/1111.0710}{{\tt arXiv:1111.0710}}.

\bibitem{D'Agostini:1995fv}
G.~D'Agostini, {\it {Probability and measurement uncertainty in physics: A
  Bayesian primer}},  \href{http://xxx.lanl.gov/abs/hep-ph/9512295}{{\tt
  hep-ph/9512295}}.

\bibitem{Trotta:2005ar}
R.~Trotta, {\it {Applications of Bayesian model selection to cosmological
  parameters}},  {\em Mon.Not.Roy.Astron.Soc.} {\bf 378} (2007) 72--82,
  [\href{http://xxx.lanl.gov/abs/astro-ph/0504022}{{\tt astro-ph/0504022}}].

\bibitem{Trotta:2008qt}
R.~Trotta, {\it {Bayes in the sky: Bayesian inference and model selection in
  cosmology}},  {\em Contemp.Phys.} {\bf 49} (2008) 71--104,
  [\href{http://xxx.lanl.gov/abs/0803.4089}{{\tt arXiv:0803.4089}}].

\bibitem{Liddle:2009xe}
A.~R. Liddle, {\it {Statistical methods for cosmological parameter selection
  and estimation}},  {\em Ann.Rev.Nucl.Part.Sci.} {\bf 59} (2009) 95--114,
  [\href{http://xxx.lanl.gov/abs/0903.4210}{{\tt arXiv:0903.4210}}].

\bibitem{Hobson:2010}
M.~Hobson, A.~Jaffe, A.~Liddle, P.~Mukherjee, and D.~Parkinson, {\em Bayesian
  Methods in Cosmology}.
\newblock Cambridge University Press, 2010.

\bibitem{2005NIMPA.551..493R}
W.~A. {Rolke}, A.~M. {L{\'o}pez}, and J.~{Conrad}, {\it {Limits and confidence
  intervals in the presence of nuisance parameters}},  {\em Nuclear Instruments
  and Methods in Physics Research A} {\bf 551} (Oct., 2005) 493--503,
  [\href{http://xxx.lanl.gov/abs/physics/0403059}{{\tt physics/0403059}}].

\bibitem{Neyman}
J.~Neyman, {\it {Outline of a Theory of Statistical Estimation based on the
  Classical Theory of Probability}},  {\em Phil.Trans.Royal.Soc.London.} {\bf
  A236} (1937) 333.

\bibitem{1998PhRvD..57.3873F}
G.~J. Feldman and R.~D. Cousins, {\it {Unified approach to the classical
  statistical analysis of small signals}},  {\em Phys.Rev.} {\bf D57} (Apr.,
  1998) 3873--3889, [\href{http://xxx.lanl.gov/abs/physics/9711021}{{\tt
  physics/9711021}}].

\bibitem{Gamerman:2006}
D.~Gamerman and H.~F. Lopes, {\em Markov Chain Monte Carlo: Stochastic
  Simulation for Bayesian Inference}.
\newblock Chapman and Hall/CRC Texts in Statistical Science, 2006.

\bibitem{Lewis:2002ah}
A.~Lewis and S.~Bridle, {\it {Cosmological parameters from CMB and other data:
  A Monte Carlo approach}},  {\em Phys.Rev.} {\bf D66} (2002) 103511,
  [\href{http://xxx.lanl.gov/abs/astro-ph/0205436}{{\tt astro-ph/0205436}}].

\bibitem{2004AIPC..735..395S}
J.~{Skilling}, {\it {Nested Sampling}},  in {\em American Institute of Physics
  Conference Series} (R.~{Fischer}, R.~{Preuss}, and U.~V. {Toussaint}, eds.),
  vol.~735 of {\em American Institute of Physics Conference Series},
  pp.~395--405, Nov., 2004.

\bibitem{SkillingNS2}
J.~Skilling, {\it {Nested sampling for general Bayesian computation}},  {\em
  Bayesian.Anal.} {\bf C1} (2006) 833.

\bibitem{Feroz:2007kg}
F.~Feroz and M.~Hobson, {\it {Multimodal nested sampling: an efficient and
  robust alternative to MCMC methods for astronomical data analysis}},  {\em
  Mon.Not.Roy.Astron.Soc.} {\bf 384} (2008) 449,
  [\href{http://xxx.lanl.gov/abs/0704.3704}{{\tt arXiv:0704.3704}}].

\bibitem{Feroz:2008xx}
F.~Feroz, M.~Hobson, and M.~Bridges, {\it {MultiNest: an efficient and robust
  Bayesian inference tool for cosmology and particle physics}},  {\em
  Mon.Not.Roy.Astron.Soc.} {\bf 398} (2009) 1601--1614,
  [\href{http://xxx.lanl.gov/abs/0809.3437}{{\tt arXiv:0809.3437}}].

\bibitem{Trotta:2008bp}
R.~Trotta, F.~Feroz, M.~P. Hobson, L.~Roszkowski, and R.~Ruiz~de Austri, {\it
  {The Impact of priors and observables on parameter inferences in the
  Constrained MSSM}},  {\em JHEP} {\bf 0812} (2008) 024,
  [\href{http://xxx.lanl.gov/abs/0809.3792}{{\tt arXiv:0809.3792}}].

\end{thebibliography}\endgroup


\providecommand{\href}[2]{#2}\begingroup\raggedright\endgroup

\end{document}